\documentclass[11pt]{article}

\usepackage{verbatim}
\usepackage{amsmath}
\usepackage{amsfonts}
\usepackage{amssymb}
\usepackage{bbm}
\usepackage{bbold}
\usepackage{latexsym}
\usepackage{lscape} 
\usepackage{hhline} 
\usepackage{epsfig}
\usepackage[usenames,dvipsnames]{color}
\usepackage{graphicx}
\usepackage{enumitem}
\usepackage[normalem]{ulem} 
\usepackage[table]{xcolor}
\usepackage{subfig}
\usepackage{multirow}
\usepackage{mathrsfs}
\usepackage{blkarray}
\usepackage{amscd}
\usepackage{cite}

\newcommand{\ket}[1]{| #1 \rangle}

\DeclareMathSymbol{\mg}{\mathrel}{symbols}{"1D}

\newcommand{\bes}{\begin{split}}
\newcommand{\ees}{\end{split}}

\def\ie{{\it i.e.~}}

\newcommand{\ga}{\alpha}
\newcommand{\gb}{\beta}

\newcommand{\gd}{\delta}

\newcommand{\gve}{\varepsilon}
\newcommand{\gf}{\phi}

\newcommand{\gm}{\mu}
\newcommand{\gn}{\nu}

\newcommand{\gl}{\lambda}
\newcommand{\gr}{\rho}
\newcommand{\gth}{\theta}
\newcommand{\gvth}{\vartheta}

\newcommand{\gt}{\tau}
\newcommand{\go}{\omega}

\newcommand{\gps}{\psi}
\newcommand{\get}{\eta}
\newcommand{\gch}{\chi}
\newcommand{\gG}{\Gamma}
\newcommand{\gD}{\Delta}

\newcommand{\gX}{\Xi}
\newcommand{\gL}{\Lambda}

\newcommand{\gTh}{\Theta}

\newcommand{\cC}{{\cal C}}
\newcommand{\cD}{{\cal D}}

\newcommand{\cN}{{\cal N}}

\newcommand{\cP}{{\cal P}}

\newcommand{\cR}{{\cal R}}
\newcommand{\cS}{{\cal S}}

\newcommand{\cV}{{\cal V}}

\newcommand{\ui}{{\underline i}}
\newcommand{\uj}{{\underline j}}

\newcommand{\Id}{\text{\small 1}\hspace{-3.5pt}\text{1}}

\newcommand{\ra}{\rightarrow}

\newcommand{\der}{\partial}

\newcommand{\dsp}{\displaystyle}

\newcommand{\undr}[1]{{\underline{#1}}}

\newcommand{\beq}{\begin{equation}}
\newcommand{\eeq}{\end{equation}}
\newcommand{\barr}{\begin{array}}
\newcommand{\earr}{\end{array}}
\newcommand{\equ}[1]{\begin{gather} #1 \end{gather}}
\newcommand{\equa}[1]{\begin{align} #1 \end{align}}
\newcommand{\items}[1]{\begin{itemize} #1 \end{itemize}}
\newcommand{\enums}[1]{\begin{enumerate} #1 \end{enumerate}}

\newcommand{\arry}[2]{\begin{array}{#1} #2 \end{array}}

\newcommand{\mtrx}[1]{\begin{matrix} #1 \end{matrix}}
\newcommand{\pmtrx}[1]{\begin{pmatrix} #1 \end{pmatrix}}

\newcommand{\sfrac}[2]{\mbox{$\frac{#1}{#2}$}}
\newcounter{oldcounter}

\newcommand{\bq}{{\bar q}}

\newcommand{\bw}{{\bar w}}

\newcommand{\byy}{{\bar y}}
\newcommand{\bz}{{\bar z}}
\newcommand{\bgf}{{\bar\phi}}

\newcommand{\bgl}{{\bar\lambda}}

\newcommand{\bgt}{{\bar\tau}}

\newcommand{\bgps}{{\bar\psi}}
\newcommand{\bget}{{\bar\eta}}

\newcommand{\Bga}{{\boldsymbol \alpha}}
\newcommand{\Bgb}{{\boldsymbol \beta}}
\newcommand{\Bgg}{{\boldsymbol \gamma}}
\newcommand{\Bgd}{{\boldsymbol \delta}}

\newcommand{\Bgx}{{\boldsymbol \xi}}

\newcommand{\BgX}{{\boldsymbol \Xi}}

\newcommand{\Intr}{\mathbbm{Z}}

\newcommand{\Real}{\mathbbm{R}}

\newcommand{\ba}[2]{\[\begin{array}{#2}\label{#1}}
\newcommand{\ea}{\end{array}\]}
\newcommand{\be}{\begin{equation}}
\newcommand{\ee}{\end{equation}}
\newcommand{\bea}{\begin{eqnarray}}
\newcommand{\eea}{\end{eqnarray}}

\newcommand{\brkt}[2]{\bigl[ ^{#1}_{#2} \bigr]}

\newcommand{\rep}[1]{\mathbf{#1}}

\newcommand{\sm}{{\,\mbox{-}}}

\newcommand{\Z}[1]{\mathbb{Z}_{#1}}
\newcommand{\Zclass}{$\mathbb{Z}$-class}

\begin{document}

\thispagestyle{empty}

\begin{flushright}
LMU-ASC 08/16 \\
LTH-1076\\ 
\end{flushright}
\vskip 1 cm
\begin{center}
{\Large {\bf 
Heterotic free fermionic and symmetric toroidal orbifold models 
} 
}
\\[0pt]

\bigskip
\bigskip {\large
{\bf P.~Athanasopoulos$^{a,}$}\footnote{
E-mail: panos@liverpool.ac.uk},
{\bf A.E.~Faraggi$^{a,}$}\footnote{
E-mail: alon.faraggi@liverpool.ac.uk},
{\bf S.~Groot Nibbelink$^{b,}$}\footnote{
E-mail: groot.nibbelink@physik.uni-muenchen.de},
{\bf V.M.~Mehta$^{c,}$}\footnote{
E-mail: viraf.mehta@thphys.uni-heidelberg.de}
\bigskip }\\[0pt]
\vspace{0.23cm}
${}^a$ {\it 
Department of Mathematical Sciences, University of Liverpool, Liverpool L69 7ZL, UK 
 \\[1ex] } 
${}^b$ {\it 
Arnold Sommerfeld Center for Theoretical Physics,  Ludwig-Maximilians-Universit\"at M\"unchen, 80333 M\"unchen, Germany
 \\[1ex]}
${}^c$ {\it 
Institute for Theoretical Physics, University of Heidelberg, 69120 Heidelberg, Germany
 } 
\\[1ex] 
\bigskip
\end{center}

\subsection*{\centering Abstract}

Free fermionic models and symmetric heterotic toroidal orbifolds both constitute exact backgrounds that can be used effectively for phenomenological explorations within string theory. 
Even though it is widely believed that for $\Intr_2\times\Intr_2$ orbifolds the two descriptions should be equivalent, a detailed dictionary between both formulations is still lacking.
This paper aims to fill this gap: We give a detailed account of how the input data of both descriptions can be related to each other. 
In particular, we show that the generalized GSO phases of the free fermionic model correspond to generalized torsion phases used in orbifold model building. 
We illustrate our translation methods by providing free fermionic realizations for all $\Intr_2\times\Intr_2$ orbifold geometries in six dimensions.

\newpage 
\setcounter{page}{1}
\tableofcontents

\section{Introduction}
\label{sc:Introduction}

The properties of elementary subatomic particles and interactions are well accounted for by the Standard Model (SM) of particle physics. This rosy picture is spoiled due to  the omission of gravity from the subatomic universe as treating gravity in a fully consistent quantum field theory framework has, thus far, proven to be extremely difficult.
 Furthermore, a fundamental dichotomy  exists between the contribution of the subatomic interactions to the vacuum  energy versus the constraint determined by  gravitational observations.  We may anticipate that the resolution of this basic conflict may materialize only when a conjugal union of gravity and quantum field theories is accomplished.  Alas, such a day is, for now, unforeseeable.  In the meantime, all we may strive for is to build inadequate models that harbor some affinity to the gauge and gravitational  phenomena as they are seen in terrestrial  and extra-terrestrial observatories.

Toward that end, we are guided by theoretical constraints, as well as by some prejudices motivated by the observed data. On the theoretical front, experience suggests that ideating elementary subatomic particles as points breaks down in the presence of gravitational interactions. A logical extension is to consider elementary particles as extended objects, with the one dimensional extension being the next step on the complexity ladder. On the observational side, we may take account of the  fact that the charges of elementary particles strongly hint to the realization of unified structures in nature.  Among those, SO(10) Grand Unified Theories (GUTs) are particularly appealing as their spinorial $\rep{16}$ representation accommodates a complete SM family, vastly  reducing the number of free parameters  needed to account for the Standard Model gauge charges.  Heterotic string theories give rise to spinorial representations in their perturbative spectrum, and therefore reproduce the SO(10) GUT structures that underly the SM, while at the same time giving rise to a perturbatively consistent framework for quantum gravity. Equipped with these features, heterotic string theories provide a well motivated contemporary arena to explore how the properties of the  elementary subatomic particles arise from  a fundamental synthesis of gravity and quantum mechanics.

The era of string phenomenology started with the seminal paper~\cite{Candelas:1985en}. By now a plethora of methods  have been devised to construct phenomenological string vacua.  These, in general, can be divided into target space constructions, in which the internal space  of the heterotic string is compactified on a six dimensional Calabi-Yau manifold, or on a toroidal orbifold~\cite{Dixon:1985jw,Dixon:1986jc,Ibanez:1987sn,Ibanez:1987pj}, and worldsheet constructions, in which  all the degrees of freedom needed to obtain a consistent string theory are represented as  internal, two dimensional fields propagating on the string worldsheet~\cite{Antoniadis1987a, AB,Kawai1987, Kawai1988,Gepner:1987vz}. A variety of phenomenological string models were constructed using target space and worldsheet techniques. The remarkable point, however, is that these two seemingly distinct approaches are in fact intimately related. This has been most  beautifully demonstrated in the case of compactifications on Calabi-Yau manifolds with SU(n) holonomy, which were shown to be equivalent to the gluing of interacting worldsheet conformal field theories  with central charge $c=9$ together~\cite{Gepner:1987vz}.  This observation led to a deep mathematical insight  into the properties of complex manifolds, and in particular to the development of mirror symmetry \cite{Greene:1990ud,Candelas:1989hd},  which provided useful insight into the arithmetic properties of Calabi-Yau manifolds~\cite{Candelas:1990rm}.   Among the most widely explored heterotic string constructions are those that utilize free bosonic and  fermionic conformal field theories. These include the  toroidal orbifold models~\cite{Dixon:1985jw,Dixon:1986jc}, that are viewed as compactifications on an internal toroidal space~\cite{narain_86_2,narain_86}, divided by some symmetry group of the internal tori. In the free fermionic models, all the extra degrees of freedom needed to cancel the worldsheet conformal anomaly are realized as free fermions propagating on the string worldsheet, at a specific  point in the moduli space~\cite{Antoniadis1987a, AB,Kawai1987}.  Deformations away from the free fermionic point that correspond to exact marginal deformations can be incorporated in the fermionic formalism in the form of worldsheet Thirring interactions \cite{Bagger:1986cd,Chang:1988ci}.

Phenomenological heterotic string models using orbifold tools~\cite{string_compactification_phys_rept,Choi2006} have lead to various interesting models on a large number of orbifold geometries. Refs.~\cite{Buchmuller:2005jr,Buchmuller:2006,Lebedev:2006kn,Lebedev:2007hv,Lebedev:2008un} constructed MSSM--like models on the toroidal ${\mathbb Z}_\text{6--II}$ orbifold. Similar constructions were possible on the $\Intr_2\times \Intr_4$ orbifold~\cite{Z2xZ4}, $\Intr_\text{12--I}$ orbifold~\cite{Kim:2006hv,Kim:2007mt} and $\Intr_8$ orbifolds~\cite{Nibbelink:2013lua}. (For a recent overview see e.g.~\cite{Nilles:2014owa}.) Using free fermionic techniques~\cite{Antoniadis1987a, AB,Kawai1987}, many phenomenologically interesting models~\cite{Antoniadis1989,Faraggi:1989ka,Faraggi:1991jr,Faraggi1992a,Antoniadis1990,Cleaver:1998sa,Cleaver2000,Assel2010,Bernard2012,Faraggi:2014hqa,Faraggi:2014ica} have been constructed since the mid-eighties.  

Free fermionic models are expected to correspond to particular $\Intr_2\times \Intr_2$ orbifolds of $\cN=4$ toroidal Narain lattices, as established  in quite a few particular cases~\cite{Faraggi1994,Faraggi2002,Kiritsis:1997ca,Berglund:1998eq,Berglund:1998rq,Donagi2004,Faraggi:2006bs,Florakis2011}. However, even though this correspondence has been discussed in general for a long time, a complete and detailed dictionary between the prominent orbifold constructions and the free fermionic formalism is not available in the literature. The goal of this paper is to cover precisely this gap: Give a detailed mapping of the input data from one formulation to the other and indicate where potential loopholes may appear. Such a dictionary is important to enable the identification of equivalent vacua in the two representations and facilitate the communication between the orbifold and free fermionic communities. Moreover, the development of methods to translate vacua from one approach to the other is particularly worthwhile, because the two approaches may yield complementary insights into phenomenological string model building.

\subsubsection*{Outline} 

We have organized this paper as follows: We begin in Sections~\ref{sc:Orbifolds} and~\ref{sc:FreeFermi} with brief but comprehensive reviews  of model building using both symmetric orbifolds and free fermionic constructions, respectively. 
In Section~\ref{sc:OrbitoFFF} we describe how one can translate any symmetric $\Intr_2\times\Intr_2$ orbifold model with arbitrary gauge shifts and discrete Wilson lines into the free fermionic language. 
Section~\ref{sc:FFFtoOrbi} describes the translation in the opposite direction: 
We give conditions when this translation is essentially simply the inverse of the description in the previous section and when one needs to use the Narain moduli space to read off the bosonic data, and, particularly, the Wilson lines. 
In Section~\ref{sc:Examples} we illustrate these procedures with various examples from both the orbifold and the free fermionic literature: 
We provide an explicit correspondence between free fermionic models and the lattice vectors that determine the crystallographic classification of $\Intr_2\times\Intr_2$ orbifolds which has never been given before. 
To show that our procedures can also be applied to more complicated models, we translate some phenomenologically interesting (MSSM-like) models constructed in the past in one formulation to the other. 
Finally, in Section~\ref{sc:Conclusion} we summarize our most important findings and give an outlook of possible extensions of this work.

\section{Symmetric heterotic orbifolds}
\label{sc:Orbifolds}

\subsection{Geometrical lattices underlying symmetric orbifolds}

One of the defining elements of any orbifold model is the underlying six-dimensional lattice that is defined through the identification
\equ{\label{eq:identificationX}
X^i\sim X^i+ 2\pi \gve^i{}_\ui n_\ui \quad i=1,\ldots,6\ .
}
where $n$ is a vector of integers and the lattice, 
\equ{ \label{TorusLattice} 
\gL = \big\{ \gve\, n = \gve_\ui \, n_\ui \big| n_\ui \in \Intr \big\}~,  
}
is spanned by a set of basis vectors $\gve_\ui$, $\ui=1,\ldots, 6$. The matrix $\gve$, with these basis vectors as its columns, can be thought of as a vielbein associated with the metric, 
\equ{ \label{TorusMetric} 
G = \gve^T \gve~, 
}
on the six-torus. This metric carries all the information about the lengths and the angles of the lattice basis vectors. We refer to the vectors $\gve_\ui$ as the lattice basis. The lattice basis is in general not the standard orthogonal Euclidean basis; we reserve the notation $e_i$ to denote the standard basis vectors of $\Real^6$: $(e_i)_j = \gd_{ij}$ and write $e_{12} = e_1 + e_2$, etc.

\begin{table}[!t]
\noindent 
\small
\begin{center}
\begin{tabular}{|c|c|p{9.5cm}|}
\hline\hline
\textbf{Sector} &\textbf{Label} &\textbf{Description} \\
\hline\hline 
SUSY  & $X_R^i$ & Bosonic internal coordinates  \\
(holomorphic) & $\psi_R^\mu, \psi_R^i$ & Real superpartners of the bosonic coordinates $x^\mu, X^i$ \\
\hline\hline
Non-SUSY & $ X_L^i$ & Bosonic internal coordinates \\
(anti-holomorphic) & $Y_L^{I}$ & Real bosons living on an internal torus $T^{16}$ that are responsible for the gauge degrees of freedom. 
\\\hline\hline
\end{tabular}\end{center}
\caption{\label{table:wsbosons} 
This table gives the states that freely propagate on the string worldsheet: $\gm=1,2$, $i=1,\ldots, 6$ and $I=1,\ldots, 16$, are four dimensional light-cone, six dimensional internal and sixteen left-moving bosonic indices, respectively. The right-moving sector, labeled by $R$, is supersymmetric, while the left-moving sector, labeled by $L$, is not. 
}
\end{table}

\subsection{Orbifold actions} 
\label{sc:OrbiActions} 

Let $\gG = \Intr_{N_1}\times \Intr_{N_2}\times \ldots$ be a finite Abelian group, often referred to as the point group. The  generators of this finite group on $\Real^6$ are denoted $\gth_1,\gth_2,\ldots$,  i.e.\ the action of a generic element of $\gG$ can be written as $\gth^k := \gth_1^{k_1} \gth_2^{k_2}\ldots$ with $k_1=0,\ldots N_1-1$, etc. The action of the point group has to be compatible with the lattice $\gL$ in the sense that
\equ{ \label{lattice_compatibility} 
\gth^k\, \gL = \gL~: 
\quad 
\gth^k \, \gve = \gve\, \gr^k~, 
\quad 
\gr^k = \gr_1^{k_1} \gr_2^{k_2}\ldots~, 
\quad 
\gr_s \in GL(n,\Intr)~. 
}
The order of $\gr_s$ is at most $N_s$, but may be lower. The elements $\gth_s$ generate the point group $\gG$ in the standard Euclidean basis. In the lattice basis, this group is generated by the matrices $\gr_s$. We normally first specify the point group in the Euclidean basis. If one also has a compatible lattice basis then one simply determines the point group generators in the lattice basis via $\gr_s = \gve^{-1} \gth_s \gve$. We denote the resulting symmetric orbifold  with point group $\gG$ as $T^6/\gG$.


The orbifold can be equivalently described as the quotient of $\Real^6/S$ where $S$ is the so-called space group. The space group $S$ combines the elements of the lattice $\gL$ and the point group $\gG$. It acts on the coordinates $X$ of the covering space $\Real^6$ as 
\equ{ \label{SpaceGroup} 
h = (\gth^k, L_h)\in S~:
\quad
X \mapsto h \circ X = \gth^k\, X +  2\pi\, L_h~, 
\quad 
L_h = \ell\, k +  \gve\, n~. 
}
The vector $\ell =(\ell_s)$ that appears in the last equation encodes the information about the translation part of the space group element $h$. In particular, there is a vector $\ell_s\in\Real^6$ associated with each generator $\gth_s$ of the point group, and the vector associated with a generic element $\gth^k$ will then be $ \ell\, k=k_1 \ell_1+k_2 \ell_2+ \cdots $.
This realization induces the following group multiplication of space group elements:
\equ{ \label{spacegroup_property} 
h'\, h = (\gth^{k'},\ell\, k' + \gve\, n') \, (\gth^k, \ell\, k + \gve\, n) 
= (\gth^{k'+k}, \gth^{k'}(\ell\, k+ \gve\, n) + \ell\, k'+ \gve\, n')~. 
}
To ensure that the orbifold elements have finite order, we need $N_s\, \ell_s \in \gL$. Depending on the choice of $\gth_s$ and $\ell_s$ for a given $\Intr_{N_s}$ factor, we distinguish between pure twist, pure shift and roto-translational orbifold actions: 
\[
\arry{|c||p{6cm}|}{
\hline 
\textbf{Orbifold action} & \textbf{Characterization} 
\\ \hline\hline 
\text{pure twist} & $\gth_s \neq 1~,~~ \ell_s = 0$
\\
\text{pure shift} &$ \gth_s = 1~,~~\ell_s \notin \gL$
\\
\text{roto-translation} & $\gth_s\neq 1~,~~ \ell_s \notin \gL$
\\
\text{true roto-translation} &  $\ell_s \notin \gL$ has components in directions in which $\gth_s\neq 1$ acts trivially.
\\\hline 
}
\]
In principle, for pure twist orbifolds we could allow for $\ell_s \in \gL$, but this can be absorbed by a redefinition of the vector $n\in\Intr^6$. A pure shift orbifold can equivalently be thought of as a torus compactification with a new lattice in which some of the basis vectors $e_i$ are replaced by the $\ell_s$ corresponding to the pure shift actions. 

The distinction between a twist and a roto-translation is not always a coordinate independent statement: When the shift part of a roto-translation points only in directions where it also acts as a rotation, then one can change the origin and this action can look like a pure twist. On the other hand, when the shift of a roto-translation also has directions which are left inert by the twist part, the shift in these directions cannot be removed. We call this a true roto-translation. Note that even when a given roto-translation can be turned into a pure shift, it often happens that, at the same time, other pure twist actions become roto-translations. In such cases the effects of the roto-translations are also physical; they cannot be removed by a coordinate redefinition.

In the following we will also need the important concept of fixed points and fixed tori, because this is where additional so-called twisted matter typically arises. An orbifold fixed set arises as a solution to the fixed point equation $g\circ X = X$: 
Pure twist and roto-translations have fixed tori or points, depending on the twist action. 
A roto-translation, that has the same twist action as a pure twist, has its fixed points/tori simply shifted with respect to those of the pure twist. 
True roto-translations never leave any point inert, hence have an empty fixed set. 
Two space groups $S_1$ and $S_2$ belong to the same \Zclass{} if generators $\gr_s$ and $\widetilde{\gr}_s$ of the corresponding point groups are related by
\equ{
U^{-1}\gr_s U=\tilde\gr_s~, 
}
with $U\in \text{GL}(6,\Intr)$. Two orbifolds with the same \Zclass{} means that they are defined on the same lattice. The structure of fixed points and/or tori is highly dependent both on the \Zclass{} of the lattice as well as on the orbifold action under consideration.

\subsection{Conditions for supersymmetry} 

In this work we focus on six dimensional orbifolds $T^6/\gG$ which preserve (at least) $\cN=1$ supersymmetry. Since the group $\gG$ is Abelian, we can simultaneously diagonalize all elements of $\gG$ using a complex basis,  labeled by $\ga=1,2,3$,  and write each element $\gth \in\gG$ in terms of the twist vector $v$ as 
\equ{
\gth^k = e^{2\pi i\, v_h }~,
\quad 
v_h = k_s\, v_s~, 
\quad 
v_s = \big(0,(v_s)_1, (v_s)_2, (v_s)_3\big)~,~ \text{etc.}~,
}
(where the sum over $s$ labels the different point group generators $\gth_s$)
for a space group element $h\in S$ with $N_1\, (v_1)_\ga\, ,\,  N_2 \, (v_2)_\ga\, , \ldots = 0 \text{ mod } 1$ to ensure that $\gth_s^{N_s} =1$.

A positive chiral target space spinor in ten dimensions can be represented by  vectors of the form $ \sfrac 12 (\undr{\pm1^4})$ (i.e.\ all four entries can either be $+1/2$ or $-1/2$) with an even number of minus signs. The action of $\gth$ on a spinor state $|s_0, s_1,s_2,s_3\rangle$ reads
\equ{ \label{SpinorAction} 
\gth^k\, |s_0, s_1,s_2,s_3\rangle = e^{2\pi i\, (v_h)_\ga {s_\ga}}\, 
|s_0, s_1,s_2,s_3\rangle~, 
}
($s_0, \ldots, s_3 = \pm1/2$) 
where the sum is over the three complexified internal directions. Therefore, if we assume that the components of the surviving four dimensional supersymmetry are represented by $\pm\sfrac12(1^4)$, we have to require that 
\equ{ \label{TwistSUSY}
\sum_\ga (v_s)_\ga  = 0 \text{ mod } 2~.
}
In the heterotic orbifold literature, mostly twists that make the sum strictly zero are used in order to obtain a unique representation of the twist vectors.

The worldsheet supersymmetry generator is given by 
\equ{ \label{susycurrent_bos} 
T_F =  \psi_\gm\, \der x^\mu + \psi^i\, \der X_R^i 
}
in terms of the four dimensional coordinate field $x^\gm$ and the fields given in Table~\ref{table:wsbosons}.

\subsection{Shift embedding and discrete Wilson lines} 

In the bosonic orbifold description the gauge degrees of freedom are described by real left-moving coordinate fields $Y_L$ that live on a sixteen dimensional torus $\Real^{16}/2\pi\gL_\text{gauge}$ where the lattice $\gL_\text{gauge}$ is either the root lattice $\gL_{8+8} = \gL_8\oplus\gL_8$ of E$_8\times$E$_8$ or $\gL_{16}$ of Spin(32)$/\Intr_2$, where
\equ{ \label{LRlattices}
\gL_\text{8n} = \bigoplus\limits_{t=0,1}
\big\{u_\text{sh} = u+ \sfrac t2\,\mathbf{1}_{8n}~\big|~ u \in \Intr^{8n}~,~  
 \mathbf{1}_{8n}^T u = 0 \text{ mod } 2 \big\}~, 
}
with $\mathbf{1}_d = (1^d)$ (the vector with $d$ entries equal to $1$) for $n=1,2$. It 
consists of the direct sum of the root ($t=0$) and spinorial ($t=1$) lattices.
In particular, $\gL_\text{8n}$ is even and self-dual.
We use $\ga_I$ to denote the simple roots of these algebras. 
In the E$_8\times$E$_8$ case, we label the two spin-structures $t_a$ for both $\gL_8$ lattices by $a=1,2$.
In most orbifold models the action of the space group on these gauge degrees of freedom is assumed to be via the so-called shift embedding:
\equ{ \label{ShiftEmbedding}
Y_L \mapsto  
h\circ Y_L = Y_L +2\pi\, V_h~, 
\quad 
V_h = k_s\, V_s + n_\ui\,A_\ui~, 
}
for any space group element $h$ defined in~\eqref{SpaceGroup}. 
The vectors $A_\ui$ are called discrete Wilson lines and compatibility with the group property~\eqref{spacegroup_property} of the space group elements implies that 
\equ{ \label{discreteWLs}
A\, \gr_s  \cong A~, 
}
where $A\cong A'$ means that $A-A'\in \gL_\text{gauge}$. These conditions often relate various discrete Wilson lines to each other and strongly restrict the order $M_\ui$ of the discrete Wilson lines $A_\ui$: 
\equ{ \label{OrderShiftsWLs}
N_s\, V_s \cong 0~, 
\qquad 
M_\ui\, A_\ui \cong 0~, 
}
The gauge shift vectors $V_s$ have the same order as the point group generators $\gth_s$.

\subsection{Narain moduli space}

The untwisted sector of orbifold models corresponds to a torus compactification which can conveniently be encoded in the Narain lattice description. This description starts from a Narain lattice \cite{narain_86_2,narain_86} of dimensions $(6,22)$ with Minkowskian signature defined by the metric 
\equ{\label{eq:metric1}
\get = \pmtrx{ -\Id_6 & 0 \\[1ex] 0 & \Id_{22}}~.
}
Points on the Narain lattice, 
\equ{ \label{NarainLattice} 
P = \pmtrx{ P_R \\[1ex] P_L } =E\, N~, 
\qquad 
N \in \Intr^{28}~, 
}
are the variables that appear in the untwisted sector partition function in the Hamilton representation 
\equ{\label{eq:pfN}
Z_\text{Narain} (\gt,\overline{\gt})= 
\frac 1{\get^6\bget^{22}}\,  
\sum_{P} q^{\frac 12\, P_R^2} \, \bar q^{\frac 12\, P_L^2}~, 
}
where $q=e^{2\pi i\gt}$  and the Dedekind-Eta function $\get = \get(\tau)$ are holomorphic functions of the Teichmueller parameter of the worldsheet torus $\gt$ and $\bar q=e^{-2\pi i\bgt}$ and $\bget = \bget(\bgt)$ of its conjugate $\bgt$.
This is the combined partition function of the six-torus and gauge lattice in the untwisted sector (with $k=0$).  
A basis for these lattice vectors is encoded in the columns of the so-called generalized vielbein 
\equ{ \label{eq:genV}
E 
=\frac 1{\sqrt 2}  \pmtrx{
\gve+ \gve^{-T} C^T& -\gve^{-T} & \gve^{-T} A^T \ga
\\[1ex]
\gve- \gve^{-T} C^T& \gve^{-T} & -\gve^{-T} A^T \ga
\\[1ex] 
\sqrt{2}\, A & 0 & \sqrt{2}\, \ga
}~. 
}
The generalized vielbein contains the lattice vectors $\gve_\ui$ of the six-torus introduced in~\eqref{TorusLattice}. The continuous Wilson lines $A_\ui$ get completely frozen to discrete ones when the combined orbifold actions act on all six torus directions.
Moreover, the anti-symmetric Kalb-Ramond tensor $B$ is contained inside the matrix\footnote{
In the literature there are various forms of~\eqref{eq:genV} and the definition of $C$ as they crucially depend on the string slope parameter $\ga'$; throughout this paper we set $\ga'=1$.
}: $C=B+\frac{1}{2} A^T A$. 
Finally, $\ga$ are the simple roots of a sixteen dimensional even-self-dual lattice and $g=\ga^T\ga$ the corresponding metric. For this, we can either choose the simple roots  of E$_8\times$E$_8$ or Spin(32)$/\Intr_2$: 
The simple roots of Spin(32)$/\Intr_2$ and the corresponding Cartan matrix read 
\equ{ \label{CartanSpin32}
\ga^{\;}_{16} = \scalebox{.6}{$\left(\arry{rrrrrr}{
1  & 0 &\cdots & &~\, 0&~\,\sfrac 12\\
-1  & 1 &\cdots & &0& \sfrac 12\\  
0  & -1 &\cdots&  & 0&\sfrac 12\\  
\vdots & \vdots& \ddots& &\vdots &\vdots \\  
0  & 0 &\cdots &1& 1 &\sfrac 12\\
0  & 0 &\cdots &-1& 1 &\sfrac 12\\
0  & 0 &\cdots &0 & 0 &\sfrac 12\\
}
\right) 
$}_{16\times 16}~, 
\qquad 
g^{\;}_{16} = \ga_{16}^T \ga^{\;}_{16} = \scalebox{.5}{$
\left(\arry{rrrr rrrr}{
2  & -1 &0&\cdots &0 &0 &0&~\,0\\
-1  & 2 &-1 & \cdots&0 &0 &0&0\\  
0  & -1 & 2&  \cdots&0 &0 &0&0\\  
\vdots & \vdots&\vdots& \ddots&\vdots & \vdots&\vdots&\vdots \\  
0  & 0 &\cdots & &2& -1&-1 &0\\
0  & 0 &\cdots & &-1& 2 &0&0\\
0  & 0 &\cdots & &-1& 0 &2&1\\
0  & 0 &\cdots & &0& 0 &1&4\\
}
\right)
$}_{16\times 16}~. 
}
The simple roots of E$_8\times$E$_8$ and the corresponding Cartan matrix read
\equ{ \label{CartanE88} 
\ga^{\;}_{8\times 8} = \pmtrx{ 
\ga^{\;}_{8} & 0   \\ 
0 & \ga^{\;}_{8}
}~, 
\qquad 
g^{\;}_{8\times 8} = \pmtrx{ 
g^{\;}_{8} & 0 \\ 
0 & g^{\;}_8
}~, 
}
given here in terms of those of E$_8$: 
\equ{
\ga_8 = \scalebox{.5}{$
\left(\arry{rrrr rrrr}{
1 & 0 & 0 & 0 & 0 & ~\,0 & -\sfrac12 & 0 \\
-1& 1 & 0 & 0 & 0 & 0 & -\sfrac12 & 0 \\
0 &-1 & 1 & 0 & 0 & 0 & -\sfrac12 & 0 \\
0 & 0 &-1 & 1 & 0 & 0 & -\sfrac12 & 0 \\
0 & 0 & 0 &-1 & 1 & 0 & -\sfrac12 & 0 \\
0 & 0 & 0 & 0 &-1 & 1 & -\sfrac12 & 1 \\
0 & 0 & 0 & 0 & 0 & 1 & -\sfrac12 &-1 \\
0 & 0 & 0 & 0 & 0 & 0 & -\sfrac12 & 0 \\
}
\right) 
$}_{8\times 8}~,
\qquad  
g_8 = \ga_{8}^T \ga^{\;}_{8} =  
\scalebox{.5}{$
\left(\arry{rrrr rrrr}{
 2 &-1 & 0 & 0 & 0 & 0 & 0 & 0 \\
-1 & 2 &-1 & 0 & 0 & 0 & 0 & 0  \\
 0 &-1 & 2 &-1 & 0 & 0 & 0 & 0 \\
 0 & 0 &-1 & 2 &-1 & 0 & 0 & 0 \\
 0 & 0 & 0 &-1 & 2 &-1 &0 & -1  \\
 0 & 0 & 0 & 0 &-1 & 2 & -1 & 0  \\
 0 & 0 & 0 & 0 &0 & -1 & 2 & 0  \\
 0 & 0 & 0 & 0 &-1 & 0 &0 & 2 
 }
 \right) 
 $}_{8\times 8}~. 
}
It is possible to transform from the E$_8\times$E$_8$ to the Spin(32)$/\Intr_2$ description, see e.g.~\cite{Ginsparg:1986bx}; in this work we will indicate explicitly which description we are using.

The partition function~\eqref{eq:pfN} is modular invariant by virtue of the following constraint on the generalized vielbein 
\equ{\label{eq:metric2}
E^T\get E=\hat\get~,  
\quad \text{where}\quad  
\hat\get = 
\pmtrx{
0 & \Id_6  & 0\\
\Id_6  & 0 & 0\\
0 & 0 & g }~: 
}
 In particular, under the modular transformation $\gt \ra \gt+1$ the partition function picks up a phase 
$\exp \pi i\, (P_R^2 - P_L^2)$ which is trivial by virtue of 
\equ{
-P_R^2+P_L^2=P^T\get P = N^T \hat\get N 
= 2\, m^Tn + p^T g\, p \in 2\, \Intr~,
}
parameterizing $N^T = (m^T, n^T, p^T)$ where $m,n\in \Intr^6$ and $p\in \Intr^{16}$.

The associated Narain partition function (\ref{eq:pfN}) can be expressed in terms of the generalized vielbein, 

\equ{\label{eq:bpf1}
Z_\text{Narain} = 
\frac 1{\get^6\bget^{22}}\,  
\sum_{N\in\Intr^{28}} q^{\frac 14\, N^T E^T\, (\mathbb{1} - \get)\, E N } \, \bar q^{\frac 14\,N^T E^T\, (\mathbb{1} + \get)\, E N }~.
}

\subsection{Orbifold partition functions}

The general form of an orbifold one-loop partition function is given as a sum over commuting space group elements 
\equ{ \label{FullPartitionFunction} 
Z(\gt,\overline{\gt}) = \sum_{[h,h']=0} c[^h_{h'}] \, Z[^h_{h'}](\gt,\overline{\gt})~,
}
where $c[^h_{h'}]$ are called generalized torsion phases and $Z[^h_{h'}]$ defines the partition function for a given sector, i.e. a set of boundary conditions, on the worldsheet torus, defined by the space group elements $h$ and $h'$. The elements $h$ are often referred to as the constructing elements. They define the different sectors in the theory and affect the $q,\bar{q}$ expansions of the partition function. The elements $h'$ are called projecting elements, as they only affect phases, i.e.\ the projection conditions in the partition function. 
We have restricted the sum to commuting constructing and projecting space group elements only; for non-commuting elements the corresponding partition function is simply zero.

The full one-loop partition function is required to be modular invariant, i.e.\ $Z(\gt+1) = Z(-1/\gt) = Z(\gt)$ (for brevity, we only indicate the $\gt$ dependence). 
The partition functions in the various sectors transform modular covariantly into each other, in the sense that 
\equ{ \label{ModCovPartitionFunctions} 
Z[^h_{h'}](-1/\gt) =  Z[^{h'}_{h}](\gt)~, 
\qquad 
Z[^h_{h'}](\gt+1) = Z[^h_{h'h}](\gt)~, 
}
without any additional phases (since we only sum over commuting elements the order of $h'$ and $h$ is irrelevant).

The partition function in a given sector, $(h;h')$, splits as a product of partition functions of the various worldsheet fields
\equ{ \label{PartitionFunctionPartitioned}
Z[^h_{h'}](\gt,\overline{\gt}) = Z_x(\gt,\overline{\gt})\, Z_X[^h_{h'}](\gt,\overline{\gt})\, Z_\psi[^h_{h'}](\gt)\, Z_{Y}[^h_{h'}](\overline{\gt})~. 
}
Let us briefly discuss the various factors in turn: The partition function $Z_x(\gt,\overline{\gt})$ is the partition function associated with the two non-compact coordinates $x^\gm$ in four dimensions in the light-cone gauge. The partition functions 
\equ{\label{eq:pfX} 
Z_X[^h_{h'}](\gt,\overline{\gt}) = 
Z_\parallel[^h_{h'}](\gt,\overline{\gt}) \, Z_\perp[^h_{h'}](\gt,\overline{\gt})
}
correspond to the compactified internal directions parameterized by $X^i$: 
Here we need to distinguish between the directions in which the orbifold twist $\gth^k$ acts non-trivially and those which are left inert. To project on these subspaces we can define the projections
\equ{
\cP{}^k_\parallel =  \frac 1{N_k}\, \sum_{r=0}^{N_k-1} (\gth^{k})^r~, 
\qquad 
\cP{}^k_\perp = \Id - \cP{}^k_\parallel~, 
} 
where $N_k$ is the order of $\gth^k$ (we will use similar notations to indicate other projected quantities). The dimensions of the corresponding subspaces are $D{}^k_\parallel$ and $D{}^k_\perp$, respectively, such that $D{}^k_\parallel+D{}^k_\perp=6$. 
The orbifold action $\gth^k$ has fixed points in the subspace on which $\cP{}^k_\perp$ projects, hence, in these directions, we only get contributions from the twisted excitations
\equ{ \label{eq:pfperp} 
Z_\perp[^h_{h'}](\gt,\overline{\gt}) = 
\Bigg| 
\frac{\get^{D{}^k_{\perp}/2}}
{ \gvth{}^{k}_{\perp}\brkt{\mathbf{1}_4/2 - v_h}{\mathbf{1}_4/2 - v_{h'}} }
\Bigg|^2
~. 
}
Here the notation $\gvth^{k}_\perp[{}^{v}_{v'}] = \prod \gvth[^{v_\ga}_{v'_\ga}]$ signifies that we only take the product of the genus-one Jacobi-theta function $\gvth[^a_{a^\prime}] =\gvth[^a_{a^\prime}](z=0; \gt)$, defined as
\equ{\label{eq:Jtheta}
\gvth[^a_{a^\prime}](z; \gt) =  
\sum_{n\in\Intr}\, 
q^{\sfrac 12\, (n- a)^2} \, 
e^{2\pi i\, (n- a)(z - a^\prime)}~, 
}
in the complexified directions where $\gth^k$ or $\gth^{k'}$ act non-trivially, i.e.\ {\em not} in the $\ga$ directions where $(v_h)_\ga = (v_{h'})_\ga = 0$. 
In the directions where the twist acts as the identity, we have the usual lattice sums of the Narain partition function~\eqref{eq:pfN} restricted to the appropriate lower dimensional sublattice.
For a symmetric orbifold, no further phases are needed to make these partition functions modular covariant.

The next partition function results from the superpartners $\gps=(\gps^\ga)$ of the coordinate fields $x^\gm, X^i$ in a complex basis: $\ga=0$ corresponds to the four dimensional light-cone coordinates $x^\mu$ and $\ga=1,2,3$ to the six internal directions in a complex basis. In a bosonized description it takes the form
\equ{ \label{FermiLatticeSumOrbi} 
Z_\gps[^{h}_{h'}](\gt) = 
e^{-2\pi i\, \sfrac 12 v_h{}^T v_{h'}}\, 
\frac 1{\get^4}\, 
\frac 12\, 
(-)^{s's+s'+s}
\sum_{p\in \Intr^4} 
q^{\frac 12 p_\text{sh}^2} \, 
e^{2\pi i\, s' \nu_R^T (p + s\, \nu_R)}\, 
e^{2\pi i\, v_{h'}^T p_\text{sh}}~, 
} 
where the vector $p_\text{sh} = p +s\, \nu_R + v_h$ has four entries. The vector $\nu_R =  \sfrac 12\, \mathbf{1}_4$ generates the right-moving spin structures labeled by $s,s'=0,1$. The phase factor $(-)^{s's+s'+s}$ ensures that $p+s\, \nu_R$ lives on the direct sum lattice of the four dimensional vectorial and spinorial lattices:
\equ{ \label{LRlattices4D}
\gL_4 = 
\big\{u ~\big|~ u \in \Intr^{4}~,~  
 \mathbf{1}_{4}^T u = 1 \text{ mod } 2 \big\} 
 \oplus
\big\{u+ \sfrac 12\,\mathbf{1}_{4}~\big|~ u \in \Intr^{4}~,~  
 \mathbf{1}_{4}^T u = 0 \text{ mod } 2 \big\}~. 
}
The next-to-last phase factor in~\eqref{FermiLatticeSumOrbi} implements the appropriate projection on the so-called right-moving lattice momentum $p$. The phase factor in front, often referred to as the vacuum phase, ensures that these partitions are modular covariant.

Finally, the partition function associated with the left-moving gauge lattice is given by
\equ{ \label{GaugeLatticePartitionFunction}
Z_{Y}[^h_{h'}](\overline{\gt}) = 
e^{2\pi i\, \sfrac 12 V_h{}^T V_{h'}} \, 
\frac {1}{\bget^{16} } \, 
\frac 12\, 
\sum_{P\in \Intr^{16}} 
 \bq^{\sfrac 12\, P_\text{sh}^2} \, 
 e^{-2\pi i\, t_u' \nu_{uL}^T (P + t_u\, \nu_{uL})}\,  
 e^{-2\pi i\, V_{h'}^T P_\text{sh}}~, 
 }
with 
\equ{ \label{PshL} 
P_\text{sh} = P + t_u\, \nu_{uL} + V_h
}
where  for the Spin(32)$/\Intr_2$ theory the index $u$ is obsolete and $\gn_L = \sfrac 12 (1^{16})$; while the index $u=1,2$ is summed over and $\nu_{1L} = \sfrac 12(1^8, 0^8)$ and $\nu_{2L} = \sfrac 12(0^8, 1^8)$ for the E$_8\times$E$_8$ theory. In the orbifold literature, the sums over the spin structures $s',s$ in~\eqref{FermiLatticeSumOrbi} and $t_u',t_u$ in~\eqref{GaugeLatticePartitionFunction} have often already been executed. One then writes $p_\text{sh} = p + v_h$ and $P_\text{sh} = P + V_h$ with $p\in \gL_4$, $P\in \gL_{16}$ or $\gL_8\oplus \gL_8$. To facilitate the comparison with the free fermionic formulation later, we choose to keep the sums over these spin structures explicit. The final phase factor in~\eqref{GaugeLatticePartitionFunction} implements the orbifold projection. Again, the vacuum phase factor ensures that these partition functions transform covariantly into each other. This lattice partition function can be obtained by assuming boundary conditions~\eqref{ShiftEmbedding}
for the left-moving coordinates $Y_L$ in the sector $h$ with spin structure(s) $t_u$.

The inclusion of the vacuum phases in front of the partition functions~\eqref{FermiLatticeSumOrbi} and~\eqref{GaugeLatticePartitionFunction} makes them all modular covariant. However, it is not necessarily guaranteed that the full resulting partition function~\eqref{FullPartitionFunction} has the proper orbifold and Wilson line projections built in, because of the factor of $1/2$ in these phases. To ensure this, we need to require that: 
\equ{ \label{ModInvConditions} 
\text{gcd}(N_s,N_t)\, (V_s{}^T V_{t} - v_s{}^T v_{t})~,~  
\text{gcd}(N_s, N_\ui)\, V_s{}^T A_\ui~,~ 
\text{gcd}(M_\ui,M_\uj)\, A_\ui{}^T A_{\uj} = 0 \text{ mod } 2~, 
}
(note there are no sums over repeated indices here). These conditions are commonly referred to as the modular invariance conditions.

\subsection{Generalized discrete torsion phases}
\label{sc:GenDiscreteTorsion}   

To ensure that the full partition function is modular invariant, the generalized torsion phases $c[^h_{h'}]$ satisfy the following conditions 
\equ{
c[^h_{h'}] = c[^{h'}_{h}] = c[^h_{h'h}]~. 
}
In particular, simply setting $c[^h_{h'}]=1$ is an allowed solution, which is the typical choice for heterotic orbifolds unless otherwise stated. In general, we may parameterize these phases as 
\equ{ \label{FullGenTorsion} 
c[^h_{h'}]  = 
c_\text{anti}[^h_{h'}]  \, c_\text{sym}[^h_{h'}] 
}
in terms of so-called generalized torsion phases. We distinguish between the symmetric and anti-symmetric phase factors: The anti-symmetric generalized torsion phases can be product expanded as 
\equ{ \label{GenTorsion}
c_\text{anti}[^h_{h'}] = 
c_{st}[^{k_s}_{k_t'}] \,  
c_{\ui\uj}[^{n_\ui}_{n_\uj'}] 
c_{s\ui}[^{k_s\, n_\ui}_{k_s' \, n_\ui'}]~,  
}
where appropriate products over different indices in the various factors are implied, e.g.\ over $t>s$. 
The factors, defined, for example, as 
\equ{ 
c_{st}[^{k_s}_{k_t'}] = e^{2\pi i\, c_{st}\, k_s k_t'}~, 
\qquad 
c_{s\ui}[^{k_s n_\ui}_{k_s' n_\ui'}] = e^{2\pi\, c_{s\ui} (k_s n_\ui'-k_s' n_\ui)}~, 
}
are characterized by the generalized torsion matrices $c_{st}$, $c_{s\ui}$, etc.; 
their entries are anti-symmetric when they have two identical type indices, e.g.\ $c_{st}=-c_{ts}$. 
The generalized torsion matrices are subject to the quantization conditions to ensure that with these generalized torsion phases included one still has proper (orbifold) projections. They read, for instance, as 
\equ{ \label{GeneralizedTorsionConditions} 
\text{gcd}(N_{s},N_t)\, c_{st}~,
\quad 
\text{gcd}(N_s, M_\ui)\, c_{s\ui}~,
\quad 
\text{gcd}(M_\ui, M_\uj)\,c_{\ui\uj} 
=0 \text{ mod } 1~, 
}
(no sums implied) and are characterized by the order of the respective elements to which the indices correspond. Here, and throughout this paper, we will use the indices of the torsion matrices to indicate which torsion phases we are actually referring to: For example, $c_{uv}$ refers to the possible torsion phase between the spin structure of the two E$_8$ factors; for the Spin(32)$/\Intr_2$ theory, it is absent.

Furthemore, specifically for order-two elements we can admit additional symmetric phases: 
\equ{ \label{SymPhases}
 c_\text{sym}[^h_{h'}] = 
 c_s[^{k_s}_{k_s'}]\, 
 c_\ui[^{n_\ui}_{n_\ui'}]~, 
 \quad 
 \text{where, for example:}
 \quad  
 c_s[^{k_s}_{k_s'}] = (-)^{c_s(k_s + k_s' + k_s'k_s)}~,  
}
and the only allowed values are $c_s, c_\ui, c_u = 0,1$. These phases are symmetric under the interchange of primed and non-primed quantities. The phases $c_s, c_u$ effectively select the spinorial lattice of the opposite chirality.

It should be emphasized that many of the generalized torsion phases introduced in~\eqref{GenTorsion} and~\eqref{SymPhases} are normally not considered in the orbifold literature. The discrete torsion discussed by Vafa-Witten~\cite{Vafa:1994} only corresponds to the phase $c_{st}$. 
In~\cite{Ploger:2007iq} no symmetric torsion phases were introduced, only the anti-symmetric ones and in the current version of the {\tt orbifolder} package~\cite{Nilles:2011aj} these symmetric torsion phases are not available. Moreover, one can introduce many additional symmetric and anti-symmetric generalized torsion phases that involve the spin structures $\gn_R$ and $\gn_{uL}$: 
\equ{ \label{AddGenTorsion} 
c_{add} = 
c_R[^s_{s'}] \, c_u[^{t_u}_{t_u'}] ~
c_{uv}[^{t_u}_{t_v'}]\, 
c_{Ru}[^{s\, t_u}_{s'\,t_u'}]\, 
c_{R\ui}[^{s\, n_\ui}_{s'\,n_\ui'}]\, 
c_{su}[^{k_s\, t_u}_{k_s'\, t_u'}] \,
c_{\ui u}[^{n_\ui\, t_u}_ {n_\ui' \,t_u'}]~. 
}

\subsubsection*{Brother models}

Having fixed the orbifold geometry, the gauge shift and discrete Wilson lines, and the generalized torsion phases, one might hope that a heterotic orbifold model is uniquely specified. Unfortunately, this specification is somewhat redundant: Naively, one would think that by adding combinations of lattice vectors, $\gD V_s, \gD A_\ui \in \gL_\text{gauge}$ to the defining gauge shifts $V$ and discrete Wilson lines $A$:
\equ{ \label{Brothers}
\widetilde V = V + \gD V~, 
\quad 
\widetilde A = A + \gD A~,
}
 would not change the model at all, as, for example, the resulting gauge group is typically unaffected by such changes. However, this is, in general, not true since adding such vectors 
leads to a whole family of so-called brother models~\cite{Ploger:2007iq}. Consequently, two heterotic orbifold brother models with gauge shift and Wilson lines satisfying~\eqref{ModInvConditions} which are related via~\eqref{Brothers}, can be viewed as two versions of the same orbifold model but with different generalized torsion phases~\cite{Ploger:2007iq}
\equ{ \label{BrotherPhases} 
\widetilde c[^h_{h'}]  = 
e^{-2\pi i\, \frac 12 \big( 
V_{h'}{}^T \gD V_h - \gD V_{h'}{}^T V_h + \gD V_{h'}{}^T \gD V_h
\big)}\, c[^h_{h'}]~. 
}
The first two terms in the exponential are manifestly anti-symmetric, while the last term is not. To see that this term is in fact also anti-symmetric, one should realize that this term is always integral because $\gD V_s$ and $\gD A_\ui$ are lattice vectors. In fact, for the diagonal part, i.e.\ $h'=h$, this term is even as $\gL_\text{gauge}$ is even. For the off-diagonal parts, $h' \neq h$, we may flip the signs of the contributions because they are half-integral taking the factor of $1/2$ out front in the exponential into account. Finally, the conditions~\eqref{ModInvConditions} ensure that the phase satisfies the quantization conditions of the generalized torsion~\eqref{GeneralizedTorsionConditions}.

\subsection{Massless spectrum}

Using the expressions for the partition functions for the various worldsheet fields, we can determine the complete spectrum of the orbifold theory. In the orbifold literature one often restricts oneself to the massless spectrum only in a {\em generic} point of the moduli space. This means that one considers the compactification on orbifolds with arbitrary radii (as long as they are not set equal by the orbifold action). For such generic values of the orbifold radii, there is no ``accidental'' gauge symmetry enhancement, i.e.\ the lattice sum in \eqref{eq:pfX} can be ignored as long as one is only interested in the massless spectrum. 

The massless spectrum of an orbifold theory, in the sector $h\in S$ at a generic point of its moduli space, reads 
\equ{
M_R^2 = \sfrac 12\, p_\text{sh}^2 + \gd c - \sfrac 12~, 
\qquad
M_L^2 = \sfrac 12\, P_\text{sh}^2 + \gd c - 1 + N_L~, 
}
where $N_L$ is the left-moving number operator and $p_\text{sh}$ and $P_\text{sh}$ the shifted  momenta, defined below \eqref{FermiLatticeSumOrbi} and \eqref{GaugeLatticePartitionFunction},  respectively. The level matched massless states, of course, correspond to $M_R^2 = M_L^2 =0$ (for supersymmetric orbifolds right-moving oscillator excitations will always lead to positive $M_R^2$, hence never constitute massless states).  Here we have defined the shift $\gd c$ in the zero point energy, given by 
\equ{
\gd c = \sfrac 12\, \go^T(1_4 - \go)~, 
}
where the entries of $\go_\ga = (v_h)_{\ga} \text{ mod } 1$ are such that $0 \leq \go_\ga < 1$. The spectrum is subject to the orbifold projection condition
\equ{ \label{SpectrumProjection} 
v_{h'}^T R - V_{h'}^T P_\text{sh} = 
\sfrac 12\, \Big( v_{h'}^T v_h - V_{h'}^T V_h \Big) 
\text{ mod } 1
}
for all projecting elements $h'$ of the space group $S$ that commute with the constructing elements $h$ (only the standard generalized torsion phase $c[^h_{h'}] =1$ is considered here for simplicity). Here we have defined 
\equ{ 
R^\ga = p_\text{sh}^\ga - N_L^{\ga}+ N_L^{\ga*}~, 
}
which involves the shifted right-moving momentum and the number operators $N_L^{\ga}$ and $N_L^{\ga*}$ counting the bosonc oscillators, e.g.\ $\overline{\partial} X^\ga$ and $\overline{\partial} X^{\ga*}$. Note that the conditions~\eqref{ModInvConditions} are essential for the projection conditions~\eqref{SpectrumProjection} to be well-defined.

\begin{table} 
\begin{center} 
\scalebox{.88}{
\(
\arry{|c|c|c|c||c|c|c|c|}{
\hline\hline 
\textbf{FRTV} & \textbf{DW} &\textbf{twists / roto-}& \textbf{Hodge} 
& 
\textbf{FRTV} & \textbf{DW} &\textbf{twists / roto-}& \textbf{Hodge}  
\\[-2ex]  
\textbf{label} &\textbf{label}  &\textbf{translations} &  \textbf{numbers} 
&
\textbf{label} &\textbf{label}  &\textbf{translations} &  \textbf{numbers} 
\\ \hline\hline 
\multicolumn{4}{|l||}{ \textbf{{\tt CARAT} } \boldsymbol{\Intr}\textbf{-class -- 1~\, : } \big\{e_{1},e_2,e_{3}, e_4,e_{5},e_6\big\} } 
&
\multicolumn{4}{l|}{ \textbf{{\tt CARAT} } \boldsymbol{\Intr}\textbf{-class -- 5~\, : }  \big\{\sfrac 12 e_{135}, e_2,e_3,e_{4},e_5,e_6\big\} }
\\ \hline
\text{(1 - 1)} &\text{(0 - 1)} & \big(\theta_1,0\big),\big(\theta_2,0\big) & (51,3)  
&
\text{(5 - 1)}&\text{(1 - 1)} & \big(\theta_1,0\big),\big(\theta_2,0\big) & (27,3)
\\
\text{(1 - 2)} &\text{(0 - 2)} & \big(\theta_1,\sfrac{1}{2} e_2\big),\big(\theta_2,0\big)	 & (19,19) 
&
\text{(5 - 2)}&\text{(1 - 3)}  & \big(\theta_1,\sfrac{1}{2} e_4\big),\big(\theta_2,0\big)	 & (11,11)
\\
\text{(1 - 3)}&\text{(0 - 3)}  &\big(\theta_1,\sfrac{1}{2} e_{26}\big),\big(\theta_2,0\big) & (11,11)
&
\text{(5 - 3)}&\text{(1 - 2)} & \big(\theta_1,\sfrac{1}{2} e_{23}\big),\big(\theta_2,0\big)  & (15,15)
\\
\text{(1 - 4)}& \text{(0 - 4)} &\big(\theta_1,\sfrac{1}{2} e_{26}\big),\big(\theta_2,\sfrac{1}{2} e_4\big)	 & (3,3)   
&
\text{(5 - 4)}&\text{(1 - 4)}  & \big(\theta_1,\sfrac{1}{2} e_4\big),\big(\theta_2,\sfrac{1}{2} e_5\big) & (7,7) 
\\  \cline{1-4} 
\multicolumn{4}{|l||}{ \textbf{{\tt CARAT} } \boldsymbol{\Intr}\textbf{-class -- 2~\, : }  \big\{\sfrac 12 e_{15},e_{2},e_3,e_{4},e_5,e_6\big\} }
&
\text{(5 - 5)}&\text{(1 - 5)} & \big(\theta_1,\sfrac{1}{2} e_{46}\big),\big(\theta_2,\sfrac{1}{2} e_5\big)  &  (3,3)
\\ \hline
\text{(2 - 1)}&\text{(1 - 6)} &\big(\theta_1,0\big),\big(\theta_2,0\big) & (31,7) 
&
\multicolumn{4}{l|}{ \textbf{{\tt CARAT} } \boldsymbol{\Intr}\textbf{-class -- 7~\, : }   \big\{\sfrac 12 e_{15},\sfrac 12 e_{26},\sfrac 12 e_{36},e_4,e_5,e_6\big\} }
\\  \cline{5-8} 
\text{(2 - 2)}&\text{(1 - 8)} & \big(\theta_1,\sfrac{1}{2} e_3\big),\big(\theta_2,0\big)	 & (15,15)
&
\text{(7 - 1)}&\text{(3 - 3)} & \big(\theta_1,0\big),\big(\theta_2,0\big)& (17,5)
\\
\text{(2 - 3)}&\text{(1 - 10)} & \big(\theta_1,\sfrac{1}{2} e_{36}\big),\big(\theta_2,0\big) & (11,11)
&
\text{(7 - 2)}&\text{(3 - 4)} & \big(\theta_1,0\big),\big(\theta_2,\sfrac{1}{2} e_6\big)  & (7,7)
\\  \cline{5-8} 
\text{(2 - 4)}&\text{(1 - 7)} & \big(\theta_1,0\big),\big(\theta_2,\sfrac{1}{2} e_5\big)	 & (11,11)
&
\multicolumn{4}{l|}{ \textbf{{\tt CARAT} } \boldsymbol{\Intr}\textbf{-class -- 8~\, : }  \big\{\sfrac 12 e_{15},\sfrac 12 e_{26},\sfrac 12 e_{35},\sfrac 12 e_{46},e_5,e_6\big\} } 
\\  \cline{5-8} 
\text{(2 - 5)}&\text{(1 - 9)} &\big(\theta_1,\sfrac{1}{2} e_3\big),\big(\theta_2,\sfrac{1}{2} e_5\big) & (7,7)
&
\text{(8 - 1)} &\text{(4 - 1)}  & \dsp\big(\theta_1,0\big),\big(\theta_2,0\big)  & (15,3)
\\ \cline{5-8} 
\text{(2 - 6)}&\text{(1 - 11)} &\big(\theta_1,\frac{1}{2}  e_{36}\big),\big(\theta_2,\sfrac{1}{2} e_5\big) & (3,3) &
\multicolumn{4}{l|}{ \textbf{{\tt CARAT} } \boldsymbol{\Intr}\textbf{-class -- 9~\, : }   \big\{\sfrac 12 e_{135},\sfrac 12 e_{26},e_{3},e_4,e_5,e_6\big\}  }
\\ \hline
\multicolumn{4}{|l||}{ \textbf{{\tt CARAT} } \boldsymbol{\Intr}\textbf{-class -- 3~\, : }  \big\{\sfrac 12 e_{15},e_{2},\sfrac 12 e_{35},e_{4},e_5,e_6\big\} }
&
\text{(9 - 1)}&\text{(2 - 3)} & \big(\theta_1,0\big),\big(\theta_2,0\big)  & (17,5)
\\ \cline{1-4} 
\text{(3 - 1)}&\text{(2 - 9)} &\big(\theta_1,0\big),\big(\theta_2,0\big) & (27,3)
&
\text{(9 - 2)}&\text{(2 - 5)}& \big(\theta_1,0\big),\big(\theta_2,\sfrac{1}{2} e_6\big)	 & (7,7)
\\
\text{(3 - 2)}&\text{(2 - 10)} &\big(\theta_1,\sfrac{1}{2} e_6\big),\big(\theta_2,0\big) & (11,11)
&
\text{(9 - 3)}&\text{(2 - 4)}& \big(\theta_1,\sfrac{1}{2} e_{23}\big),\big(\theta_2,0\big)  & (11,11)
\\ \cline{5-8} 
\text{(3 - 3)}&\text{(2 - 11)} & \big(\theta_1,\frac{1}{2} e_6\big),\big(\theta_2,\sfrac{1}{2} e_5\big)& (7,7)
&
\multicolumn{4}{l|}{ \textbf{{\tt CARAT} } \boldsymbol{\Intr}\textbf{-class -- 10 : }  \big\{\sfrac 12 e_{135},\sfrac 12 e_{26},e_{3},\sfrac 12 e_{46},e_5,e_6 \big\} }
\\ \cline{5-8} 
\text{(3 - 4)}&\text{(2 - 12)} & \big(\theta_1,\sfrac{1}{2} e_{46}\big),\big(\theta_2,\sfrac{1}{2} e_5\big) & (3,3)
&
\text{(10 - 1)}&\text{(3 - 5)}   & \big(\theta_1,0\big),\big(\theta_2,0\big)  & (15,3)
\\ \cline{1-4} 
\multicolumn{4}{|l||}{ \textbf{{\tt CARAT} } \boldsymbol{\Intr}\textbf{-class -- 4~\, : }  \big\{\sfrac 12 e_{15},\sfrac 12 e_{26},e_{3},e_{4},e_5,e_6\big\}} 
& 
\text{(10 - 2)}&\text{(3 - 6)} & \big(\theta_1,\sfrac{1}{2} e_{12}\big),\big(\theta_2,0\big)  & (9,9)
\\ \hline
\text{(4 - 1)}&\text{(2 - 13)}& \big(\theta_1,0\big),\big(\theta_2,0\big) &  (21,9) 
& 
\multicolumn{4}{l|}{ \textbf{{\tt CARAT} } \boldsymbol{\Intr}\textbf{-class -- 11 : }  \big\{\sfrac 12 e_{14},\sfrac 12 e_{26},\sfrac 12 e_{35},e_{4},e_5,e_6 \big\} }
\\ \cline{5-8} 
\text{(4 - 2)}&\text{(2 - 14)}  & \big(\theta_1,0\big),\big(\theta_2,\sfrac{1}{2} e_4\big)	 & (7,7)
& 
\multirow{2}{*}{\text{(11 - 1)}} & \multirow{2}{*}{$\dsp \mtrx{ \text{(3 - 1)} \\[-2ex] \rotatebox{90}{$\equiv$}  \\[-2ex] \text{(3 - 2)} }$}  & \multirow{2}{*}{$\dsp \big(\theta_1,0\big),\big(\theta_2,0\big)$} & \multirow{2}{*}{(12,6)} 
\\ \cline{1-4} 
\multicolumn{4}{|l||}{ \textbf{{\tt CARAT} } \boldsymbol{\Intr}\textbf{-class -- 6~\, : }  \big\{\sfrac 12 e_{15},\sfrac 12 e_{23},e_3,e_{4},e_5,e_6\big\}  }
& 
&&&
\\\hline
\text{(6 - 1)}&\text{(2 - 6)} & \big(\theta_1,0\big),\big(\theta_2,0\big)	& (19,7) 
& 
\multicolumn{4}{l|}{ \textbf{{\tt CARAT} } \boldsymbol{\Intr}\textbf{-class -- 12 : } \big\{\sfrac 12 e_{135},\sfrac 12 e_{246},e_{3},e_4,e_5,e_6 \big\} }
\\ \cline{5-8} 
\text{(6 - 2)}&\text{(2 - 7)} & \big(\theta_1,0\big),\big(\theta_2,\sfrac{1}{2} e_5\big)  & (9,9)
&
\text{(12 - 1)}&\text{(2 - 1)}  & \big(\theta_1,0\big),\big(\theta_2,0\big)  & (15,3)
\\
\text{(6 - 3)}&\text{(2 - 8)} & \big(\theta_1,\sfrac{1}{2}\,e_6\big),\big(\theta_2,\sfrac{1}{2} e_5\big)  & (5,5)
& 
\text{(12 - 2)}&\text{(2 - 2)}  & \big(\theta_1,\sfrac{1}{2} e_{56}\big),\big(\theta_2,0\big)  & (9,9)
\\ \hline\hline 
}
\)
}
\end{center}
\caption{ \label{tb:Z22classification}
Classification of all six-dimensional lattices that admit a $\Intr_2\times \Intr_2$ orbifold action  according to \cite{FRTV} and \cite{DW} with the hodge numbers $(h_{11},h_{21})$ indicated. 
We have grouped the geometries according to their {\tt CARAT} $\Intr$-classes and we give representative lattice choices for each of these $\Intr$-classes. Here $\theta_1$ and $\theta_2$ denote the two $\Intr_2$ reflections that leave the first and second two-torus fixed; $e_i$ denotes the $i$-th standard Euclidean basis vector and $e_{ij} = e_i + e_j$, etc. 
}
\end{table}

\subsection{Special features of $\boldsymbol{\Intr_2\times \Intr_2}$ orbifolds}

So far our discussion has been for general orbifolds; in this section we make some statements that are specific to $\Intr_2\times\Intr_2$ orbifolds which we will be using later.

\subsubsection*{Standard form of the $\boldsymbol{\Intr_2\times \Intr_2}$ orbifold twists} 

First of all, in this paper we will use the following conventions to represent $\Intr_2\times\Intr_2$ orbifolds. All $\Intr_2\times\Intr_2$ orbifolds contain two twist elements combined with possible translations, i.e.\ roto-translations. The point group parts of the orbifolding elements are taken to be
\equ{ \label{Z22twists} 
\gth_1 = \pmtrx{ \Id_2 & & \\ & - \Id_2 & \\ & & -\Id_2}~, 
\quad 
\gth_2 = \pmtrx{ -\Id_2 & & \\ &  \Id_2 & \\ & & -\Id_2}~,
\quad  
\gth_3 = \gth_1\gth_2 = \pmtrx{ -\Id_2 & & \\ & - \Id_2 & \\ & & \Id_2}~.  
}
They define reflections in four of the six torus directions in the standard Euclidean basis, leaving the first, second and third two-torus inert, respectively. 
Their actions on the spinors~\eqref{SpinorAction} are defined by the vectors 
\equ{ \label{Z2xZ2twists}
v_1 = \big(0, 0, \sfrac 12, -\sfrac 12)~, 
\quad 
v_2 = \big(0,-\sfrac 12, 0, \sfrac 12)~.  
}

\subsubsection*{Classification of $\boldsymbol{\Intr_2\times \Intr_2}$ orbifolds}

The possible $\Intr_2\times\Intr_2$ twist orbifolds were classified by Donagi and Faraggi in~\cite{Donagi2004}. The classification was extended to include roto-translations  by Donagi and Wendland in~\cite{DW}. A full classification of all symmetric toroidal orbifolds that preserve at least $\cN=1$ supersymmetry in four dimensions has been performed in~\cite{FRTV}: This classification includes, but is not restricted to, $\Intr_2\times \Intr_2$ or even Abelian orbifolds; most orbifolds turn out to possess non-Abelian point groups.

All these classifications are ultimately inspired by crystallography: The orbifold actions have to be compatible with a particular lattice; for given orbifold twists $\gth_s$ and lattice vectors $\gve_i$, one needs to be able to fix the matrices $\gr_s\in\text{GL}(6;\Intr)$ such that~\eqref{lattice_compatibility} is fulfilled. This, in turn, restricts the form of the metric $G$ on the six-torus. Moreover, this determines the number and positions of two-tori and points that the various orbifold actions leave fixed. All $\Intr_2\times\Intr_2$ orbifolds only possess fixed two-tori, which are either orbifolded by the second orbifold action or pairwise identified. All this information is encoded in the \Zclass{} (or arithmetic crystal class) of the six-dimensional lattice. The possible $\Z2\times\Z2$ compatible lattices have been classified up to six dimensions~\cite{Opgenorth:1998}. 
The required algorithms have been collected in the computer package {\tt CARAT}~\cite{CARAT}. This software provides a complete catalog of the \Zclass{es}.

The representations of both the lattice and the orbifold actions used in the classification are far from unique: For example, by scaling or permuting the torus directions and by shifting the origin on the six-torus, one obtains very different looking representations of the same orbifold. Moreover, the same lattice can be described in infinitely many bases.

We have given a compact representation of the $\Intr_2\times\Intr_2$ orbifolds in Table~\ref{tb:Z22classification}. The data in this table are as follows: 
The first two columns give ${\Intr_2\times \Intr_2}$ classifications following both Donagi,Wendland~\cite{DW} and Fischer~et al.~\cite{FRTV}. The various {\tt CARAT} \Zclass{}es following~\cite{CARAT} are given with a representative lattice for each. The third column indicates a representation of the various orbifold actions on these lattices. The final column of this table displays the Hodge numbers of the various $\Intr_2\times \Intr_2$ orbifolds. They can be determined as the number of generations and anti-generations when one uses the orbifold standard embedding, in which the orbifold shifts $V_s$ are taken to be equal to $v_s$ (completed with 13 zeros).

\section{Free fermionic models}
\label{sc:FreeFermi}

Next we review the free fermionic formulation\footnote{There exists an alternative fermionic description~\cite{Kawai1987,Kawai1988}; a mapping between these formalisms may be found in Appendix A of~\cite{Dienes:1995bx}.} as first outlined in~\cite{Antoniadis1987a,AB}. In this formalism, the internal sectors of the string are described by fermionic degrees of freedom. In general, there are $n_f$ right-moving (or holomorphic) fermions $f$ and $n_{\overline{f}}$ left-moving (or anti-holomorphic) fermions $\overline{f}$. In the case of heterotic string theories with four non-compact target space dimensions, again described by light-cone coordinates $x^\mu$ with superpartner $\gps^\mu$, conformal invariance requires that we have 
\begin{align}\label{confanom4d}
n_f =18~, 
\qquad 
n_{\overline{f}} =44~.
\end{align}
The holomorphic sector has worldsheet supersymmetry, which is non-linearly realised by the supercurrent
\begin{align}\label{eq:susycurrent} 
  T_F= \psi_\mu\, \partial x^\mu-\chi^i y^i w^i~, 
\end{align}
on the internal fermions $\chi^i, y^i, w^i$, $i=1,\ldots,6$. The $44$ real anti-holomorphic fermions are conventionally separated into two sets of real fermions $\overline{y}{}^i, \overline{w}{}^i$ and sixteen complex fermions $\overline\gl{}^I$, $I=1,\ldots, 16$. Often these fermions are further divided into three classes as indicated in Table~\ref{table:wsfermions}. 

\begin{table}[!t]
\noindent 
\small
\begin{center}
\begin{tabular}{|c|c|p{9.5cm}|}
\hline\hline
\textbf{Sector}& \textbf{Label} &\textbf{Description} \\
\hline \hline 
{SUSY} & $\psi^\mu, \chi^{i}$ & Real superpartners of the bosonic coordinate $x^\mu$ and the six compactified directions in the bosonic formulation\\
(holomorphic) & $y^{i},w^{i}$ & Real fermions that correspond to the bosons describing the six compactified directions\\
\hline\hline
{Non-SUSY} & $\overline{y}{}^{i},\overline{w}{}^{i}$ & Real fermions that correspond to the bosons describing the six compactified dimensions  in the orbifold formulation\\
(anti-holomorphic) & 
\multirow{3}{*}{
 $\overline{\lambda}{}^{I} = 
 \left\{ \arry{l}{
 \overline{\psi}{}^{1,\dots,5} \\ 
 \overline{\eta}{}^{1,2,3} \\[2ex] 
 \overline{\phi}^{1,\dots,8} } \right.$}
& Complex fermions that describe the visible gauge sector, corresponding to eight of the internal directions in $T^{16}$\\[2ex] 
& & Complex fermions that describe the hidden gauge sector, corresponding to the remaining eight internal directions in $T^{16}$ \\
\hline\hline
\end{tabular}\end{center}
\caption{ \label{table:wsfermions}
This table gives the fermionic states that freely propagate on the string worldsheet: 
$\gm=1,2$, $i=1,\ldots, 6$ and $I=1,\ldots, 16$, are four dimensional light-cone, six real internal and sixteen complex indices, respectively. 
The right-moving sector is supersymmetric, while the left-moving sector is not. 
}
\end{table}

\subsection{Basis vectors and the additive group}

A 48-component vector 
$\boldsymbol{\ga} = \big(\ga(\gps), \ga(\chi),\ga(y),\ga(w) \,\big|\, \ga({\overline{y}}), 
\ga({\overline{w}});\ga({\overline{\gl}})\big)$ 
characterizes a sector in a free fermionic model by defining a set of boundary conditions
\begin{align}\label{eq:fermichanges}
  f \mapsto -e^{i\pi \ga(f)}\, f~, 
  \qquad 
  \overline{f} \mapsto -e^{-i\pi\, \ga(\overline{f})}\, \overline{f}~, 
\end{align}
for all the fermions. 
The line $|$ between the components of the vector $\Bga$ separates the boundary conditions for holomorphic and anti-holomorphic fermions, $f=\gps^\mu,\chi^i,y^i,w^i$ and $\overline{f} = \overline{y}{}^i,\overline{w}{}^i;\overline{\gl}{}^I$, and the semi-colon distinguishes the latter between real fermions, $\overline{y}{}^i,\overline{w}{}^i$, and complex fermions $\overline{\gl}{}^I$. This convention means that when an entry $\ga(f) = 0$, the fermion is anti-periodic, i.e.\ with NS boundary conditions. 
The transformations~\eqref{eq:fermichanges} imply that combining boundary conditions leads to the addition rule: $(\Bga, \Bgb) \mapsto \Bga+\Bgb - \mathbf{1}$ with unit element: $\mathbf{1}$.

The reduced version $[\Bga]$ of a vector $\Bga$ has entries equal to those of $\Bga$ up to even integers such that all entries of $[\Bga]$ lie within the range
\equ{\label{eq:fermiphases}
\big( -1,+1 \big]~.
}
In particular, $[\Bga](f)$ is the entry of $\alpha$ for the fermion $f$, restricted to the above range for complex fermions, and it is simply $0$ or $1$ for real fermions.  
Often the basis vectors are chosen to lie within this restricted range. The difference between a vector and its reduced representation is denoted by
\equ{\label{eq:reduced}
  2r(\boldsymbol{\alpha})\equiv\boldsymbol{\alpha}-\left[ \boldsymbol{\alpha} \right].
}
Moreover, it is conventional to only indicate the fermions with non-vanishing entries:
For illustration, in Table~\ref{tb:BasisVectors} we have given a number of basis vectors that appear in many free fermionic models. They are described either by the names of the fermions that appear in them or equivalently by the values of all of their 48 entries.
We represent any such vector by $\Bga_{R}$ and $\Bga_L$ with components $\ga_R(f)$ and $\ga_L(\overline{f})$. The Lorentzian inner product between two vectors, $\boldsymbol{\alpha}$ and $\boldsymbol{\beta}$ is defined as  
\begin{align}\label{eq:lorentzprod}
 \boldsymbol{\alpha}\cdot\boldsymbol{\beta} 
 =  \Bga_R^T\Bgb_R - \Bga_L^T \Bgb_L 
 =  \sfrac 12\, \ga(f)^T \gb(f) -  \sfrac 12\, \ga(\overline{f})^T \gb(\overline{f}) - 
 \ga(\overline{\gl})^T \gb(\overline{\gl})~, 
\end{align}
with half-weighting for the real fermionic components  $f=\gps^\mu,\chi^i,y^i,w^i$ and $\overline{f} = \overline{y}{}^i,\overline{w}{}^i$.  

The collection of all such vectors defines a finite additive group, $\BgX \cong \mathbb{Z}_{N_1}\oplus\dots\oplus\mathbb{Z}_{N_K}$. This group 
\begin{align}\label{eq:addset}
  \BgX=\mathrm{span}\left\{\mathbf{B}_1,\dots,\mathbf{B}_K \right\}
\end{align}
is generated by the set $\mathsf{B} = \{\mathbf{B}_a\}$ of basis vectors, which are linearly independent and non-redundant, in the sense that each $\Bga \in \BgX$ can be written as $\Bga = \sum m_a\,  \mathbf{B}_a$, $m_a\in\Real$, such that 
\begin{align}\label{eq:A1}
  m_a\, \mathbf{B}_a   = 0 \text{ mod } 2 
\quad \Leftrightarrow \quad   
m_a = 0 \mbox{ mod } N_a
\end{align} 
for all $a=1,\ldots,K$, where the mod 2 for vectors is understood component wise. Here $N_a$ is the smallest integer satisfying $N_a\,\mathbf{B}_a=0 \text{ mod } 2$ and is called the order of $\mathbf{B}_a$. 

Furthermore, any set of boundary conditions, $\Bga$, has to be compatible with the worldsheet supersymmetry current $T_F$, \ie all terms in \eqref{eq:susycurrent} need to transform with the same phase: 
\begin{align}\label{eq:delta}
T_F  \mapsto -\gd_\Bga\, T_F~, 
\qquad 
  \delta_\Bga = e^{i\pi\alpha\left( \psi^\mu \right)}~.
\end{align}
This is determined by the $\gps^\mu$ component of $\Bga$, as it has been assumed that the non-compact Minkowski coordinates, $x^\mu$, do not transform under any element of $\BgX$. Consequently, all vectors in the additive group $\BgX$ must satisfy: 
\equ{ \label{PreserveSUSY}
\ga(\chi^i) + \ga(y^i) + \ga(w^i) = \ga(\gps^\gm) \text{ mod } 2~, 
}
for all $i=1,\ldots,6$. This implies that if $\ga(\gps^\gm) = 0$ then, for each $i$, the fermions $\{\chi^i, y^i,w^i\}$ may only appear in pairs in $\Bga$; when $\ga(\gps^\gm) =1$, then, for each $i$, either just one fermion or all three out of these sets have to be present in $\Bga$.

In order to ensure that the resulting partition function for the fermions is modular invariant, yet non-vanishing, it is crucial that all fermions can have both R and NS sectors. This means that the collection of all vectors in the additive set $\BgX$ should affect all fermions. 
This is automatically guaranteed because the unit element $\mathbf{1}$ of the boundary condition addition rule is part of the additive set~\cite{AB}.

\begin{table}[!t]
\noindent 
\small
\begin{center}
\scalebox{1}{
\begin{tabular}{cc}
\begin{tabular}{|c|l|}
\hline\hline
$\mathbf{B}_a$ & \textbf{Basis vector }$\mathbf{B}_a$\textbf{ components in fermions}  \\
\hline \hline 
$\mathbf{1}$ & $\big\{\psi^\mu,\chi^{1\dots 6};y^{1\dots6},w^{1\dots6 }\,|\,\overline{y}^{1 \dots 6},\overline{w}^{1 \dots 6},\overline{\psi}{}^{1 \dots 5},\overline{\eta}^{123},\overline{\phi}{}^{1 \dots 8}\big\}$ 
\\
$\mathbf{S}$ & $\big\{\psi^\mu,\chi^{1\dots 6}\big\}$ 
\\
$\boldsymbol{\xi}_1$ &  
$\big\{\overline{\psi}{}^{1 \dots 5},\overline{\eta}{}^{123}\big\}$ 
\\ 
$\boldsymbol{\xi}_2$ & 
$\big\{\overline{\phi}^{1 \dots 8}\big\}$ 
\\ 
$\boldsymbol{\xi}$ & 
$\boldsymbol{\xi}_1 + \boldsymbol{\xi}_2 = \big\{\overline{\psi}^{1 \dots 5},\overline{\eta}{}^{123},\overline{\phi}{}^{1 \dots 8}\big\}$ 
\\ 
$\mathbf{e}_i$ & 
$\big\{y^i,w^i\,|\,\overline{y}^i,\overline{w}^i  \big\}$ 
\\ 
$\mathbf{b}_1$ & $\big\{\chi^{3456};y^{3456} \,|\, \byy^{3456}; \overline{\eta}^{23}\big\}$ 
\\
$\mathbf{b}_2$ & $\big\{\chi^{1256};y^{12},w^{56} | \byy^{12},\bw^{56}; \overline{\eta}^{13}\big\}$ 
\\\hline\hline
\end{tabular}
\hspace{-2ex}  & \hspace{-2ex} 
\begin{tabular}{|c || r rrr r rr |}
\hline \hline 
$\mathbf{B}_b \cdot \mathbf{B}_a$ & $\mathbf{1}$ & $\mathbf{S}$ & $\Bgx_1$ & $\Bgx_2$ & $\Bgx$ & $\mathbf{e}_i$ & $\mathbf{b}_s$ \\ \hline\hline 
$\mathbf{1}$ & -12 & 4& -8 & -8 & -16 & 0 & 0 \\ 
$\mathbf{S}$ & 4 & 4 & 0 & 0 & 0 & 0 & 2 \\ 
$\Bgx_1$ & -8 & 0 & -8 & 0 & -8 & 0 & -2 \\ 
$\Bgx_2$ & -8 & 0 & 0 & -8 & - 8 & 0 & 0 \\ 
$\Bgx$ &  -16 & 0 & -8 & -8 & -16 & 0 & -2\\
$\mathbf{e}_j$  & 0 & 0 & 0 & 0 & 0 & 0 & 0 \\
$\mathbf{b}_t$ & 0 & 2 & -2 & 0 & -2 & 0 & 0 
\\ 
 &&&&&&&
\\ \hline\hline 
\end{tabular} 
\end{tabular}
}
\end{center}
\caption{ \label{tb:BasisVectors}
The left part of this table gives a number of important basis vectors that appear in many free fermionic models. 
The vector $\mathbf{1}$ is necesarily part of the additive set $\BgX$.
The vector $\mathbf{S}$ is associated with target space supersymmetry. 
The right part gives their multiplication table using the product defined in~\eqref{eq:lorentzprod}.
}
\end{table}

\subsection{The free fermionic partition function}

The full partition function of a free fermionic model  \cite{AB},
\equ{  \label{FullPartitionFunctionFF} 
Z(\gt,\overline{\tau}) = \sum_{\Bga^\prime,\Bga \in \gX} C[^\Bga_{\Bga^\prime}] \, Z[^\Bga_{\Bga^\prime}](\gt,\overline{\tau})~,
}
is given by a sum over the additive set $\BgX$ of partition functions defined by the boundary conditions $\Bga$ and $\Bga^\prime$ when parallel transported around the non-contractible loops of the torus amplitude,
%
%
\equ{\label{eq:FFZ}
Z[^\Bga_{\Bga^\prime}](\gt,\overline{\tau}) = 
Z_x(\gt,\overline{\tau})
\! %
\left[
 \frac{\Theta[^{\ga(y)}_{\ga'(y)}] \Theta[^{\ga(w)}_{\ga'(w)}] \Theta[^{\ga(\gps)}_{\ga^\prime(\gps)}] \Theta[^{\ga(\chi)}_{\ga^\prime(\chi)}]}{\get^{20}}( \tau)
\right]^{\frac 12}
\!%
 \left[
 \frac{\overline{\Theta}[^{\ga(\byy)}_{\ga'(\byy)}] \overline{\Theta}[^{\ga(\bw)}_{\ga'(\bw)}]}{\bget^{12}}( \bgt)
 \right]^{\frac 12}   
 \frac{\overline{\Theta}[^{\ga(\overline{\gl})}_{\ga^\prime(\overline{\gl})}]}{\overline{\get}^{16}}(\overline{\tau})~,
}
in terms of the Mumford theta functions $\gTh[^\ga_{\ga^\prime}](\gt) = \gTh[^\ga_{\ga^\prime}](0; \gt)$: 
\equ{ \label{ThetaAB} 
\gTh[^\ga_{\ga^\prime}](z; \gt) =  
e^{-\pi i\, \sfrac 12\, \ga^T\ga^\prime}\, 
\sum_{n\in\Intr^d}\, 
q^{\sfrac 12\, (n+ \sfrac12 \ga)^2} \, 
e^{2\pi i\, (n+\sfrac 12 \ga)^T(z+\sfrac 12 \ga^\prime)}
~. 
}
The $Z_x(\gt,\overline{\tau})$ factor corresponds to the non-compact bosons $x^\gm$ and is therefore the same as in \eqref{PartitionFunctionPartitioned}.  The absolute value sign appears since we combine the contributions of the right-moving fermions, $y,w$ with those of their left-moving partners $\overline{y},\overline{w}$ and, as they are real fermions, this term is not squared. $\psi$ and $\chi$ are also real fermions but cannot be combined with any left-movers.

Modular invariance of the full partition function restricts both the choice of basis vectors of the additive group, $\BgX$, as well as the generalized GSO phases. All pairs of basis vectors $\mathbf{B}_a,\mathbf{B}_b$ need to satisfy the following conditions (no sums implied here and the dot product is defined in \eqref{eq:lorentzprod}): 
\begin{subequations} \label{eq:ABnorms} 
\begin{align}\label{eq:A3}
   \text{lcm}(N_a,N_b)
   \, \mathbf{B}_a\cdot \mathbf{B}_b & = 0 \mbox{ mod } 4~,
\end{align}
hence in particular 
$N_a\, \mathbf{B}_a^2=0\mbox{ mod } 4$. Moreover, when $N_a$ is even, an even stronger condition has to be imposed, namely,
\begin{align}\label{eq:A4}
  N_a\, \mathbf{B}_a^2=0\mbox{ mod } 8~. 
\end{align}
\end{subequations} 
This means that for models with only basis elements of order 2, $ \mathbf{B}_a^2=0\mbox{ mod } 4$.
Finally, real fermions which are simultaneously periodic under any three boundary condition basis vectors must come in pairs \cite{Kawai1988}.

\subsection{Conditions on generalized GSO phases}

In addition, there are constraints on the generalized GSO phases coming from modular invariance \cite{AB}:
\begin{subequations}\label{eq:modinvphases}
\equ{  \label{modinvphases1a0}
   C[^\Bga_{\Bga^\prime}] = C^*[^{-\Bga}_{~\Bga^\prime}]~,
    \\[1ex] \label{modinvphases1a}
    C[^ \Bga_{\Bga^\prime }] = -e^{\frac 14 i\pi\,\Bga\cdot \Bga}\,C[ ^\Bga_{\Bga^\prime-\Bga+\mathbf{1} }]~,
    \\[1ex] \label{modinvphases1b0}
        C[ ^\Bga_{\Bga^\prime }] = e^{\frac 12 i\pi\, \Bga\cdot\Bga^\prime }\,C^*[ ^{\Bga^\prime}_{\Bga }]~,
    \\[1ex]   \label{modinvphases1b}
    C[ ^\Bga_{\Bgb+\Bgg}] = \delta_\Bga \, C[^ \Bga_{\Bgb}]\,  C[^ \Bga_{\Bgg}]
   \\[1ex]\label{modinvphases2}
    C[ ^\Bga_{\Bga^\prime }]\, C[ ^{\Bgb}_{\Bgb^\prime }] = 
    \delta_\Bga \, \delta_{\Bgb}\, 
    e^{-\frac 12 i\pi\, \Bga\cdot\Bgb }\,
    C[^ \Bga_{\Bga^\prime+\Bgb}]\, 
    C[ ^{\Bgb}_{\Bgb^\prime+\Bga }]~,
}
\end{subequations}
at the one- and two-loop level. The general solution to these conditions can be parameterized as follows~\cite{AB}: 
\equ{ \label{eq:phaseparamFF1}
 C[^\Bga_{\Bga^\prime}] = \big(\gd_\Bga\big)^{\sum_a n^\prime_a -1}\,  \big(\gd_{\Bga^\prime}\big)^{\sum_a n_a -1}\, 
 e^{-\pi i\, r(\Bga)\cdot \Bga^\prime}\, 
 \prod_{a,b}  C\left[^{\mathbf{B}_a}_{\mathbf{B}_b}\right]^{n_a n^\prime_b}~,  
}
for two arbitrary vectors $\Bga = \sum n_a\, \mathbf{B}_a\,$, $\Bga^\prime = \sum n^\prime_b\, \mathbf{B}_b \in \BgX$, with $r(\Bga)$ defined in \eqref{eq:reduced}. It is important to note that \eqref{eq:phaseparamFF1} gives $C[^\mathbf{0}_\mathbf{0}] = 1$. This tells us that all generalized GSO phases are fixed in terms of the phases $C[^{\mathbf{B}_a}_{\mathbf{B}_b}]$ for all the basis vectors generating the additive group $\BgX$. 
The phases that can be chosen freely are those of the upper triangular part of the GSO phase matrix $C$ including the diagonal ($b \geq a$); the phases in the lower triangular part ($b<a$) are fixed by~\eqref{modinvphases1b0}.
 
It might sometimes happen that some vector $\Bga$ does not lie in the reduced range defined in~\eqref{eq:fermiphases}. One can bring it into this range by adding a vector $\Bgd$ with only even entries. The generalized GSO phases are, in general, not invariant under such changes, but transform as 
\equ{ \label{TrivialPhaseChanges}
C[^{\Bga\, + \Bgd}_{\Bga' + \Bgd'}] = e^{\sfrac 12 \pi i\, \Bgd \cdot \Bga'}\, C[^{\Bga}_{\Bga'}]~, 
}
provided that $\Bgd,\Bgd'$ have only even entries, as can be inferred from~\eqref{FullPartitionFunctionFF}  and~\eqref{ThetaAB}. This means that two sets of basis vectors, which only differ in  vectors with only even entries, describe fully equivalent models provided that one transforms their generalized GSO phases via~\eqref{TrivialPhaseChanges}. It also shows that there is no loss of generality when enforcing all basis vectors to have entries that lie inside the range~\eqref{eq:fermiphases}.

\subsection{Massless spectrum}

The spectrum in the $\Bga \in \BgX$ sector of a free fermionic model is built upon the left- and right-moving vacua, $|0\rangle^\Bga_R \otimes |0\rangle^\Bga_L$. When a fermion, $f$ or $\overline{f}$, is strictly periodic, i.e.\ $\ga(f)=1$ or $\ga(\overline{f})=1$, then this fermion has a zero mode. 
In all models, properties of the fermions are always defined pairwise, hence we can use complex fermions from which we can construct spin up/down generators. 
A single complex fermion zero mode leads to two degenerate vacua represented as $\ket{\pm}$; when we have a collection of fermionic zero modes we write $\ket{\pm, \ldots, \pm}$. Consequently, their vacua are associated with spinorial representations in target space. In particular, when  the fermions $\gps^\mu$ have periodic boundary conditions, their zero modes form the light-cone version of the four dimensional Clifford algebra and hence define target space fermions. Thus, whether the sector $\Bga$ corresponds to bosons or fermions in target space is determined by the quantity $\gd_\Bga$ defined in \eqref{eq:delta}. Making use of \eqref{eq:phaseparamFF1} we then obtain 
\begin{align}\label{statistics}
 \delta_\Bga^{-1}   = C[^\mathbf{0}_\Bga] = C[_\mathbf{0}^\Bga]
  =
  \begin{cases} 
  ~~\,1 & \mbox{ spacetime bosons~,} 
  \\[1ex] 
  -1& \mbox{ spacetime fermions~.}
  \end{cases} 
\end{align}
Both bosonic and fermionic oscillator excitations may act on the vacuum of such sectors.
The oscillator modes associated with the boson $x^\mu$ have always non-zero, integral frequencies. The smallest non-zero fermionic frequencies are
\begin{align}\label{eq:fermifreq}
  \nu(f) = \sfrac 12\, \big( 1+\alpha_R( f) \big)\ , 
  &&
  \nu(\overline{f}) = \sfrac 12\, \big( 1+\alpha_L(\overline{f} ) \big)~, 
\end{align}
for real fermions, $f$ and $\overline{f}$, while for the complex fermions, $\overline{\gl}$, and their complex conjugates we have 
\begin{align}
  \nu(\overline{\gl}) = \sfrac 12\, \big( 1+\alpha_L(\overline{\gl} ) \big)~, 
  &&
    \nu(\overline{\gl}^*) = \sfrac 12\, \big( 1 -\alpha_L(\overline{\gl} ) \big)~. 
\end{align}
The left- and right-moving masses of such states are given by 
\begin{align}\label{eq:Virasoro}
  M^2_R = \sfrac{1}{8}\, \Bga_R^2- \sfrac 12 +\sum_f \nu(f) + N_R~, 
  &&
  M^2_L = \sfrac{1}{8}\, \Bga_L^2  -1 + \sum_{\overline{f}} \nu(\overline{f}) + N_L~, 
\end{align}
where $N_{R/L}$ are the number operators associated with bosonic oscillators on the right-/left-moving sides. Level-matching requires that these left- and right-moving masses are equal. Moreover,  if we are only interested in massless states, both the left- and right-moving masses in \eqref{eq:Virasoro} need to vanish. Hence, only for the values $\Bga_{R}^2 \leq 4$ and $\Bga_L^2 \leq 8$ are massless states possible. 

On the states in each sector, $\boldsymbol{\alpha}\in\Xi$, the generalized GSO projections, 
\begin{align}\label{eq:hilbertspace}
  e^{i\pi \, \mathbf{B}_a \cdot F} \ket{\mbox{state}}_{\boldsymbol{\alpha}}= {\delta_\Bga}C^*[^\Bga_{\mathbf{B}_a}]\ket{\mbox{state}}_{\boldsymbol{\alpha}}\ ,
\end{align}
are imposed for all basis elements $\mathbf{B}_a$, where

\equ{ \label{eq:b.F}
\mathbf{B}_a \cdot F = 
\sum_{f} \mathbf{B}_a \cdot F(f) 
-
\sum_{\overline{f}} \mathbf{B}_a \cdot F(\overline{f})~.
}
Here we work in a complex basis for all fermions; the fermion number operator $F$  is defined such that $F(f) = - F(f^*) = 1$. $F$ vanishes on any NS-vacuum as well as on the ``true'' R-vacuum $|+1^n\rangle$, which we define as $f_0^{i*} |+1^n\rangle = 0$ when it corresponds to $n$ complexified fermions with periodic boundary conditions;  $f_0^{1} | +1^n \rangle = | -1, 1^{n-1}\rangle$, etc. (Note that $n=10$ for the right-moving Ramond vacuum and $n=28$ for the left-moving Ramond vacuum.) Only the states that survive the generalized GSO projections are physical, i.e.\ correspond to states in the four dimensional target space.

\subsection{Conditions for supersymmetry}

The generator of target space supersymmetry is denoted by $\mathbf{S}$; its explicit form can be found in Table~\ref{tb:BasisVectors}. Different forms for $\mathbf{S}$ are, in principle, possible, but it was shown in~\cite{Dreiner1989} that they never lead to models with less than $\mathcal{N}=2$ supersymmetry and will, therefore, not be considered further here. To preserve modular invariance, fermions with identical transformation properties always come in pairs, hence we can make use of a complex notation for the fermions as well. 

Whenever $\mathbf{S}$ is part of the set of basis vectors $\{\mathbf{B}_a\}$, we know that associated with any sector $\Bga$ there will be a sector $\Bga+\mathbf{S}$. Since~\eqref{statistics} decides whether a sector corresponds to target space bosons or fermions and $\mathbf{S}$ involves $\gps^\gm$, it follows that if $\Bga$ is bosonic then $\Bga+\mathbf{S}$ is fermionic and vice versa. The supersymmetry element $\mathbf{S}$ then leads, via~\eqref{eq:hilbertspace}, to the projection, that imposes the following for the signs $s$: 
\equ{
\sum_\ga s_\ga = 
\begin{cases}
\text{even} \\
\text{odd} 
\end{cases}
\quad \text{for} \qquad 
C[^{\mathbf{S}}_{\mathbf{S}}] = \mp 1~. 
}
Either choice corresponds to $\mathcal{N}=4$ spacetime supersymmetry in four dimensions, but of opposite chirality in ten dimensions; conventionally one takes for positive chirality that the spinors' sums are even, so that $C[^{\mathbf{S}}_{\mathbf{S}}] = -1$.

In order to break $\mathcal{N}=4$ down to $\mathcal{N}=1$ supersymmetry in four dimensions, the set of basis vectors $\{\mathbf{B}_a\}$ must contain elements that overlap with the vector $\mathbf{S} = \{\psi^\mu, \chi^i\}$. In light of~\eqref{eq:A3}, their overlaps always involve an even number of complexified combinations of the fermions in $\mathbf{S}$.
 To fix conventions, we choose the surviving four dimensional gravitino, 
\equ{
  \ket{s}_R^\mathbf{S} \otimes \overline{\partial} x^\gm_{\sm1} \ket{0}_L^\mathbf{S}, 
}
to have components $s=\pm(1^4)$. This then requires that the generalized GSO phases involving $\mathbf{S}$ have to be chosen such that 
\equ{ \label{SusyConditions}
C[^{\mathbf{S}}_{\mathbf{B}_a}] = C[^{\mathbf{S}}_{\mathbf{1}}]
= C[^{\mathbf{S}}_{\mathbf{S}}] = -1~,
}
to preserve at least $\mathcal{N}=1$ supersymmetry.  In particular, for basis vectors that do not overlap with $\mathbf{S}$ the opposite sign for GSO phases would kill all gravitino states. The second equality holds even when $\mathbf{1}$ is not part of the basis by~\eqref{modinvphases1a}.

\section{Converting symmetric $\boldsymbol{\Intr_2\times\Intr_2}$ orbifolds to free-fermionic models}
\label{sc:OrbitoFFF}

In this section, we describe how one can associate a free fermionic model with the input data of a given symmetric orbifold model. This conversion takes an orbifold model, defined at a generic point, to a specific point in the geometrical moduli space; namely a point that actually admits a free fermionic description.

Heterotic symmetric orbifolds are defined as orbifolds of either the E$_8\times$E$_8$ or the Spin(32)$/\Intr_2$ string. A generic  $\Intr_2\times\Intr_2$ symmetric orbifold model is defined by the two $\Intr_2$ orbifold elements $\gth_s$ that can act as pure twists or as roto-translations on the geometry, accompanied by specific embeddings in the gauge degrees of freedom as encoded by the gauge shifts $V_s$. In addition, there are the Wilson lines $A_\ui$, associated with the translations in the various lattice directions, $\gve_\ui$, that define the underlying torus or lattice. Finally, the model might possess some generalized torsion phases. This is the input data we need to translate into a collection of free fermionic basis vectors $\textsf{B}$ and generalized GSO phases.

To define such a set of basis vectors, we need to take into account both the Wilson lines as well as the free fermionic requirement that the $\mathbf{1}$ is in the additive set. To this end, we first observe that having an order $M_\ui$ Wilson line, $A_\ui$, associated with a certain translation $\gve_\ui$, can be thought of as a $\Intr_{M_\ui}$ pure shift orbifold: On a torus with a radius $M_\ui$ times that of the original one, we see that applying the translational element with the Wilson line $M_\ui$ times acts as standard periodicity of the bigger torus. For this reason, we will take this bigger six-torus as our starting point and assume that it has an orthonormal lattice with unit radius. Hence, we first define a standard set of basis vectors, $\textsf{B}_0,$ that describes the E$_8\times$E$_8$ or the Spin(32)$/\Intr_2$ theory on this orthonormal unit six-torus: 
\equ{ \label{Orbi->FFFbasisMin}
\textsf{B}_0 = \big\{ \mathbf{S}, \Bgx_u, 
\mathbf{e}_1,\ldots, \mathbf{e}_6 \big\}~, 
}
with $\Bgx_u = \Bgx$ (or $\Bgx_1, \Bgx_2$) for the Spin(32)$/\Intr_2$ (or E$_8\times$E$_8$) case, respectively.

Next we extend this set to include basis vectors  $\mathbf{\widetilde{b}}_s$ and $\Bgb_\ui$ that correspond to the orbifold elements, $\gth_s$, and the Wilson lines, $A_\ui$, respectively. The resulting canonical basis set, 
\equ{ \label{Orbi->FFFbasis}
\textsf{B} = \text{B}_0 \cup 
 \big\{ 
 \mathbf{\widetilde{b}}_1,  \mathbf{\widetilde{b}}_2,  
\Bgb_1,\ldots, \Bgb_6 
\big\}
= 
 \big\{ 
 \mathbf{S}, \Bgx_u, \mathbf{e}_1,\ldots, \mathbf{e}_6, 
\mathbf{\widetilde{b}}_1,  \mathbf{\widetilde{b}}_2,  
\Bgb_1,\ldots, \Bgb_6 
\big\}~, 
}
contains up to 16 (or 17) elements for the Spin(32)$/\Intr_2$ (or E$_8\times$E$_8$) case. Any element $\Bga$ in the additive set $\BgX$, associated with a given orbifold model, can therefore be expanded as 
\equ{ \label{OrbiFFFExpansion} 
\Bga =
s\, \mathbf{S} 
+ t_u\, \Bgx_u
+ m_i\, \mathbf{e}_i 
+ k_s\, \mathbf{\widetilde{b}}_s
+ n_\ui \, \Bgb_\ui~. 
}
For the set of basis vectors in~\eqref{Orbi->FFFbasis}, we need a prescription for a choice of the generalized GSO phase matrix.

\subsection{Defining the free fermionic basis vectors}

\subsubsection*{Choice of ten dimensional heterotic theory}

Depending on whether the orbifolded string theory is the Spin(32)$/\Intr_2$ or the E$_8\times$E$_8$ theory, the set of basis vectors $\textsf{B}$ contains:
\equ{ 
\text{Spin(32)$/\Intr_2$ :}\quad \mathbf{S}, \Bgx \in \textsf{B}~, 
\quad\text{or}\quad 
\text{E$_8\times$E$_8$ :}\quad \mathbf{S}, \Bgx_1, \Bgx_2 \in \textsf{B}~. 
}

\subsubsection*{Encoding Wilson lines}

Next, we turn to an order $M_\ui$ Wilson line, $A_\ui$ associated with a lattice translation $\gve_\ui$. Any of the lattice translations can be decomposed in the standard Euclidean basis $e_i$ as: $\gve_\ui = (n_{\ui})_i\, e_i/M_\ui$, where we treat $n_\ui$ as integral vectors. The associated fermionic basis vector, $\Bgb_\ui$, can then be taken to be: 
\equ{ \label{WLtranslationFFF} 
\Bgb_\ui = \big\{ 0^{8}; \sfrac {n_\ui}{M_\ui},  \sfrac {n_\ui}{M_\ui} \,|\, \sfrac { n_\ui}{M_\ui},  \sfrac { n_\ui}{M_\ui}\big\}\big(2A_\ui\big)~. 
}
The notation here means that no $\psi^\mu, \chi^i$ fermions are involved and only the pairs of fermions $y^i,w^i$ and $\byy^i,\bw^i$, in the Euclidean directions in which $\gve_\ui$ is pointing, appear. The latter part indicates that one completes the basis vector by two times the value of the discrete Wilson line in the orbifold formulation. As an illustrative example, the order-two Wilson lines, $A_\ui = (0^7, 1)(0^8)$ in the $\gve_\ui = \sfrac 12\, e_i$ direction in the E$_8\times$E$_8$ theory, become 
\(
\Bgb_i  = \{0^8; y^i,w^i \,|\, \byy^i,\bw^i\} (0^7, 1, 0^8)\,.
\) 
Also, the spin structure vector, say $\nu_{L}$ for the Spin(32)$/\Intr_2$ theory defined under~\eqref{PshL}, which is a shift only in the gauge lattice, can be translated to a free fermionic basis vector using~\eqref{WLtranslationFFF} to give $\Bgx$ (similarly, $\nu_{1L}$ and $\nu_{2L}$ correspond to $\Bgx_1$ and $\Bgx_2$, respectively). 
Note that we did not include an extra factor of $2$ in the $y,w$ and $\byy,\bw$ parts of~\eqref{WLtranslationFFF} since this element represents an order $M_\ui$ vector w.r.t.\ the orthonormal lattice that was already generated by $\mathbf{e}_1, \ldots, \mathbf{e}_6$.

\subsection{Orbifold elements in the free fermionic formulation} 
\label{sc:OrbioldActionsinFFF} 

In the same way, we can associate the basis vectors $\mathbf{b}_1$ and $\mathbf{b}_2$ with the orbifold elements $\gth_1$ and $\gth_2$. Here the following complication arises: As discussed in Subsection~\ref{sc:OrbiActions} there are different types of orbifold actions and their characterization is partially parameterization dependent. As can be inferred from the bosonization relation: 
\equ{ \label{Simple_Bosonization}
-i\, y^i w^i \simeq i\, \der X_R^i~,  
}
in order to represent twists or shifts, but not roto-translations, the fermionic basis vectors can be chosen as 
\equ{ \label{TwisttranslationFFF} 
\mathbf{\widetilde{b}}_1 = \big\{ \chi^{34},-\chi^{56}; z^{34},z^{56} \,|\, \bz^{34},\bz^{56} \big\} 
\big(2V_1\big)~, 
\qquad 
\mathbf{\widetilde{b}}_2 = \big\{ \!-\!\chi^{12},\chi^{56}; z^{12},z^{56} \,|\, \bz^{12},\bz^{56} \big\} 
\big(2V_2\big)~,
\quad 
}
where the signs in front of the complexified fermions, e.g.\ $\chi^{12} = \chi^1 + i\chi^2$,
have been chosen such that they are compatible with the sign choices for the $\Intr_2\times\Intr_2$ actions on the spinor in the bosonic formulation in~\eqref{Z2xZ2twists}. (We use the same notation for the complexified $z$'s as well.) The non-removable parts of the shifts in the true roto-translations can be taken into account by including the corresponding fermion pairs $y^i,w^i$ and $\byy^i,\bw^i$ in their associated fermionic basis vectors in the same fashion as we did for the Wilson line elements, as in~\eqref{WLtranslationFFF}. Furthermore, each $z^i$, $i=1,\ldots,6$, equals either $y^i$ or $w^i$ and $\bz^i$ either $\byy^i$ or $\bw^i$. Thus, a similar ambiguity is present in the fermionic description when defining the twist actions.

This seems to imply that there is also an ambiguity of how to associate definite fermionic basis vectors with their corresponding orbifold twist actions. To shed light on this issue, we compare the partition functions of the bosonic and fermionic descriptions of the orbifold twisted sectors. When doing so one notices some seemingly  unrelated differences:  
\items{
\item 
  In the bosonic description only commuting, constructing and projecting, elements give contributions to the partition function, while by definition all boundary conditions encoded in the additive set $\BgX$ are allowed. Hence, the number of sectors on the worldsheet torus does not seem to be the same in both descriptions. 
\item 
Secondly, the bosonic twisted partition function, given in~\eqref{PartitionFunctionPartitioned}, involves $\gvth$-functions in the denominator as can be seen from~\eqref{eq:pfperp}. In contrast, the fermionic partition function~\eqref{eq:FFZ} always has $\gvth$-functions in the numerator only. Moreover, for the geometrical part, the fermionic description involves twice as many $\gvth$-functions as the bosonic description, since each right-(left-)moving bosonic coordinate $X_R^i$ corresponds to two fermions $y^i,w^i$.
} 
But these issues are closely related and can, in fact, help us understand whether the twist-like $\Intr_2$ elements are mutual twists or roto-translations: 

Suppose the two twist-like elements $\mathbf{\widetilde{b}}_1$ and $\mathbf{\widetilde{b}}_2$ both contain a specified $y^i$ or $w^i$. The part of the partition function in which one is the constructing and the other is the projecting element will vanish identically since this overlap leads to a (square root of) $\gvth[^{1/2}_{1/2}]=0$. This means that this sector does not give any contribution to the partition function; precisely as if we have two non-commuting space group elements. Hence, in the direction(s) where the overlapping $y^i$ or $w^i$ appear, one of the elements corresponds to a pure twist while the other acts as a roto-translation. Consequently, if the sector defined by one element is to have a proper projection from the other, then there should not be any overlap of any of the $y$s and $w$s.

We can see the same effect when we reverse the process: For commuting constructing and projecting space group elements, $h$ and $h'$, the geometrical twisted partition function is given in~\eqref{eq:pfperp}. Using the identity 
\equ{
\arry{c}{\dsp \frac{\get}{ \gvth\left[\substack{{\sfrac{1-a}2}\\[.5ex]{\sfrac{1-a'}2}}\right]}}   
= 
\arry{c}{\dsp \frac{ \gvth\left[\substack{{\sfrac{a}2}\\[.7ex]{0}\\[.3ex]}\right] \, \gvth\left[\substack{{0}\\[.5ex]{\sfrac{a'}2}}\right]}{2\,\get^2}}~, 
}
for any $a,a' = 0,1$, excluding $(a,a')=(0,0)$, we can rewrite this partition function with twice the number of $\gvth$-functions in the numerator, just like one has in the fermionic formulation, for the $\gvth$-functions associated with the fermions $y$ and $w$. Moreover, precisely as we noticed above, for elements that do not lead to a $\gvth[^{1/2}_{1/2}]$ in the partition function, the characteristics in these $\gvth$-functions do not overlap. 

Using these considerations it is always possible to find the appropriate choice of $y$s and $w$s (and their conjugates) in the two orbifold basis vectors $\mathbf{\widetilde{b}}_1, \mathbf{\widetilde{b}}_2$. In practice, figuring out the correct choices for given orbifold geometries can be quite tricky. Therefore, in Table~\ref{tb:FFFrealizationsDW} in the example section, we provide specific choices of free fermionic basis vectors that can represent all 35 $\Intr_2\times\Intr_2$ orbifold geometries of Table~\ref{tb:Z22classification}.

\subsubsection*{Some properties of the resulting set of basis vectors}

If we translate orbifold twists, shifts and Wilson lines to basis vectors of the corresponding models, we will always obtain basis vectors which will satisfy the modular invariance conditions~\eqref{eq:ABnorms} in the free fermionic formulation, since the orbifold input satisfied~\eqref{ModInvConditions}. 
By adding appropriate multiples of 2 to some of the entries of these basis vectors, they can be brought to the specific range~\eqref{eq:fermiphases} as long as one remembers to modify the generalized GSO phases accordingly, once they have been determined.

It should be noted that the notion of order of the resulting basis vectors in the free fermionic model will be two times that of the orbifold theory for those orbifold shifts $V_s$ or Wilson lines $A_\ui$ that are built from spinorial roots. For example, $A_1 = (0^8)(\frac 14{}^8)$ has order two in the orbifold language since $2 \, A_1 \in \gL_{8\times 8}$ while the corresponding 
\(
\Bgb_1 = \{ y^1,w^1 | \byy^1,\bw^1; \sfrac 12\, \bgf^{1\ldots 8}\}
\)
has order four.  The reason for this difference is that in the free fermionic construction the order of the vectors is counted with respect to the orthogonal lattice while on the orbifold side it is counted with respect to the E$_8\times$E$_8$ or Spin(32)$/\Intr_2$ lattice.

We would also like to emphasize that when converting an orbifold to a free fermionic model we are forcing the theory to move to a very particular point in the moduli space, namely a free fermionic point. 
By the rules of the dictionary presented here this is automatically guaranteed. In particular, the vector $\mathbf{1}$ is always in the additive set.
Moreover, we should mention that we can always find different lattice representations in the same \Zclass{} which are free fermionic points as well. Instead of starting from the basis vectors $\mathbf{e}_1, \ldots, \mathbf{e}_6$ that define the standard Euclidean basis, we can also use more miminal (i.e.\ with less basis vectors) to define other free fermionic realizations of the various orbifold geometries. Examples, for the different $\Intr_2\times\Intr_2$ orbifold geometries of Table~\ref{tb:Z22classification} are presented in Table~\ref{tb:FFFrealizationsDW}.

\subsection{Determining the associated generalized GSO phases}

The next step is to determine the generalized GSO phases from the partition function in the bosonic formulation. To do so, it is crucial to take into account all phases that appear in the partition functions on both the orbifold and the free fermionic sides. These phases in the orbifold description of Section~\ref{sc:Orbifolds} get contributions from the bosonized superpartners of the coordinate fields~\eqref{FermiLatticeSumOrbi}, the gauge lattice~\eqref{GaugeLatticePartitionFunction}, generalized torsion phases~\eqref{GenTorsion} and, finally, the additional symmetric phases~\eqref{SymPhases}. These phases should be compared with the generalized GSO phases in~\eqref{FullPartitionFunctionFF} taking into account the phases~\eqref{ThetaAB} included in the $\gvth$-functions, $\gTh$. 
An important fact here is that the projection phase structure in both theories is not fully identical:
In the free fermionic formulation, the projection phase, i.e.\ the final phases in~\eqref{ThetaAB}, are fully factorized in the exponential. On the orbifold side, however, the phases in the exponential are not factorized: there are two projection phases in both~\eqref{FermiLatticeSumOrbi}  and~\eqref{GaugeLatticePartitionFunction}: the last implement the orbifold and Wilson line projections while the next-to-last implement the various lattice constraints due to the spin structures.

Taking these observations into account, while comparing the various phases, we conclude that 
\equ{  \label{PhaseMatching} 
 (-)^{s's+s'+s}\, 
e^{- 2\pi i\, \sfrac 12 \big\{v_h{}^T v_{h'} - V_h{}^T V_{h'}\big\}}\, 
c[^h_{h'}] = 
e^{ -\pi i\, \sfrac 12 \Bga\cdot \Bga^\prime}\, 
e^{2\pi i\, (s' \nu_R^Tv_h - t'_u\nu_{uL}^TV_h)}\, 
C\big[^\Bga_{\Bga^\prime}\big]~, 
}
by simply setting the bosonic and fermionic phases equal, provided that we use the expansion in~\eqref{OrbiFFFExpansion} for the vectors $\Bga$ and $\Bga'$. The second phase on the right-hand-side takes into account the fact that on the orbifold side the fully factorized exponentials are not present. Inserting the various definitions we find
\equ{ 
C\big[^\Bga_{\Bga^\prime}\big] = (-)^{s's+s'+s} \,
e^{\pi i\, (v_h^Tv_h' - V_h^TV_h' ) } \, 
e^{-2\pi i\, t_u\, \nu_{uL}^T V_h'}\, 
c[^h_{h'}]~, 
}
where we have used that $\nu_R^Tv_s = 0$ strictly for all supersymmetric orbifolds.

If we make the identifications~\eqref{OrbiFFFExpansion}, we see that all the remaining phases also agree identically, hence, we can read off the generalized GSO phases of the free fermionic formulation from the orbifold input. For all phases involving $\mathbf{S}$ we find~\eqref{SusyConditions}. For the remaining phases involving $\mathbf{e}_i$, we conclude that they are simply
\begin{subequations}\label{eq:IdentifiedPhasesOrbi->FFF}
\equ{
 C[^{\mathbf{e}_i}_{\mathbf{B}_a}] = 1~, 
}
for all $\mathbf{B}_a \neq \mathbf{S}$. In addition, we find 
\equ{
C\left[^{\mathbf{\widetilde{b}}_1}_{\mathbf{\widetilde{b}}_2}\right] = 
e^{\pi i\,  ( v_1^T v_2 - V_1^TV_2)}\, 
e^{2\pi i\, c_{st}}~, 
\quad 
C\left[^{\Bgb_\ui}_{\Bgb_\uj}\right] = 
e^{-\pi i \, A_\ui^TA_\uj}\, 
e^{2\pi i\, c_{\ui\uj}}~, 
\quad 
C\left[^{\mathbf{\widetilde{b}}_s}_{\Bgb_\ui}\right] = 
e^{-\pi i \, V_s^TA_\ui}\, 
e^{2\pi i\, c_{s\ui}}~. 
}

As stressed in Subsection~\ref{sc:GenDiscreteTorsion}, all other possible generalized discrete torsion phases are (mostly implicitly) taken to be trivial, i.e.\ $c=0$, in the orbifold literature. Since any free fermionic construction is not complete without also specifying their values, we indicate the remaining phases here. We obtain  
\equ{ 
C\left[^{\mathbf{\widetilde{b}}_s}_{\mathbf{\widetilde{b}}_s} \right] 
= e^{\pi i\, (v_s^2 - V_s^2)} \, (-)^{c_s}~, 
\qquad  
C\left[^{\Bgb_\ui}_{\Bgb_\ui} \right] = e^{- \pi i\, A_\ui^2} \,  (-)^{c_\ui}~, 
\qquad 
C\left[^{\Bgx_u}_{\Bgx_u} \right] = (-)^{c_u}~, 
}
for the symmetric phases and 
\equ{
C\left[^{\mathbf{\Bgx}_1}_{\mathbf{\Bgx}_2}\right] = e^{2\pi i\, c_{uv}}~, 
\qquad 
C\left[^{\mathbf{\widetilde{b}}_s}_{\Bgx_u}\right] = e^{2\pi i\, c_{s u}}~, 
\qquad 
C\left[^{\Bgb_\ui}_{\Bgx_u}\right] = e^{2\pi i\, c_{\ui u}}~, 
\\ 
C\left[_{\mathbf{\widetilde{b}}_s}^{\Bgx_u}\right] = 
e^{- 2\pi i\, \nu_{uL}^T V_s}\, 
e^{-2\pi i\, c_{s u}}~, 
\qquad 
C\left[_{\Bgb_\ui}^{\Bgx_u}\right] = 
e^{- 2\pi i\, \nu_{uL}^T A_\ui}\, 
e^{-2\pi i\, c_{\ui u}}~, 
}
\end{subequations}
for the anti-symmetric phases.

\section{Converting free fermionic models to symmetric orbifolds}
\label{sc:FFFtoOrbi}

In this section we describe explicitly how to convert a free fermionic model to a symmetric orbifold model.  In the proceeding subsection, the various steps are discussed in detail. 
In Section \ref{sc:Examples} we then go through a number of examples to illustrate the general procedure. 

Since the task of converting models is --in its fine-print-- rather involved, we first present a brief, non-technical outline of the steps involved. The interested reader is encouraged to read the general discussion here and the examples in Section~\ref{sc:Examples} in parallel and, whenever necessary, consult the other subsections to find extensive explanations of the steps used.

\enums{ 
\item[{\bf 1.}] {\bf Convert to a basis that admits an orbifold interpretation} 
\\[1ex] 
As considered and described in Section~\ref{sc:FreeFermi}, a free fermionic model is defined by a set of basis vectors  $\mathsf{B} = \{\mathbf{B}_a\}$, generating an additive set $\BgX$, together with generalized GSO-phases that both satisfy a large set of consistency conditions.

The basis of a generic free fermionic model contains vectors whose role in the description of an orbifold geometry is rather obscure. For the subsequent identification of the properties of the orbifold model, it is necessary to go to a set of basis vectors that can be distinguished by the roles they play: 
\items{
\item supersymmetry vector $\mathbf{S}$\,,
\item twist-like vectors $\mathbf{\widetilde{b}}_s$\,, $s=1,2$\,,
\item Narain-like vectors $\boldsymbol{\gb}_x$\,, 
\item spin-structure vectors $\Bgx_u$\,. 
}
The twist-like generators, $\mathbf{\widetilde{b}}_1, \mathbf{\widetilde{b}}_2$\,, encode the two independent $\Intr_2$ reflections, possibly combined with simultaneous shifts, i.e.\ the orbifold twists or roto-translations. The Narain-like basis vectors, $\boldsymbol{\gb}_x$\,, are characterized by the requirement that they do not act on the fermions $\{\psi^\mu, \chi^i\}$\,. Often one can identify one or two spin-structure basis vectors:  either $\Bgx$ or $\Bgx_1,\Bgx_2$.  

\item[{\bf 2a.}] {\bf Directly determine the orbifold twists, shifts and Wilson lines} 
\\[1ex] 
If the spin-structure vectors, $\Bgx$ or both $\Bgx_1$ and $\Bgx_2$, can be identified, then one can directly interpret the free fermionic model as an orbifold of the Spin(32)$/\Intr_2$ or E$_8\times$E$_8$ theories, respectively. If the set of remaining Narain-like vectors is not redundant, then one can directly read off the orbifold shifts and Wilson lines. 
\item[{\bf 2b.}] {\bf Identify the geometrical Narain data} 
\\[1ex] 
Unfortunately, often the spin-structure vectors are not present in the additive set $\BgX$, or only one of the two $\Bgx_u$'s is. In this case, we can only  determine the orbifold data by comparison with the Narain description. This is possible because the Narain-like vectors, $\boldsymbol{\gb}_x$\,, define the untwisted sector of the orbifold. Their partition function can be represented as a lattice sum and from this we can, in principle, read off the geometrical parameters $G, B, A$ that define a Narain torus compactification. 
\item[{\bf 3.}] {\bf Determine the generalized discrete torsion phases} 
\\[1ex] 
We read off which generalized torsion phases are switched on for given generalized GSO phases. These relations are important since they affect the projection conditions on the spectra.
\item[{\bf 4.}] {\bf Classify the orbifold geometry}
\\[1ex] 
Once the six-torus background is specified, we can identify the orbifold geometry which the free fermionic model corresponds to. To this end, we need to identify the space group associated with the two twist-like elements $\mathbf{\widetilde{b}}_s$ and the torus lattice identified above. The combination of these data fixes the \Zclass{} of the bosonic model. In particular, it determines whether $\mathbf{\widetilde{b}}_s$ should be thought of as $\Intr_2$-twists and/or roto-translations. This will affect the number and type of fixed points of the orbifold and, consequently, the underlying geometry of the resolved manifold.
}
\noindent 
Before we go into the details, a couple of comments are in order:

When a complete set of spin-structure vector(s) can be identified, we suggest to use the direct route 2a to identify the Wilson lines. Of course, in that case, one can still follow the other route 2b: This gives more information as it does not only specify the topological data of the orbifold theory, but it also determines the value of all free moduli at the free fermionic point, where the free fermionic model is defined.

Especially via route 2b, one is confronted with the fact that the choice of twist-like vectors and Narain-like vectors out of the additive set is not unique. The representation of Wilson lines, or of the Narain lattice in general, is dependent on the choice of duality frame. In addition, one could keep some shift orbifold actions explicit in the description or absorb them, possibly including the associated generalized torsion phases, in a redefinition of the Narain lattice.
To make the matching of free fermionic models with orbifold models as transparent as possible, it is often preferable to translate all generalized GSO phases of a free fermionic model to generalized torsion phases in the corresponding orbifold model. However, we will also encounter examples where this is simply not directly possible or where it would lead to other complications.  
Different choices could lead to seemingly different orbifold models that are associated with one and the same free fermionic model; consequently, these different orbifold models are equivalent descriptions of the same physics.

Whether a basis vector is of type $\mathbf{S}$, $\mathbf{\widetilde{b}}_s$ or $\Bgb_x$ is determined by how it acts on the right-moving fermions only. Therefore, it is not automatically guaranteed that the twist-like elements $\mathbf{\widetilde{b}}_s$ have identical action on a certain set of left-moving fermions such that a symmetric orbifold interpretation is possible. Similarly, a Narain-like element might act as a twist on the left-moving coordinates, hence such Narain-like elements do not characterize the underlying Narain lattice of the construction. This is a subtle question because the pairing of the left-moving fermions with the right-moving $y$'s and $w$'s that correspond to the right-moving coordinates via~\eqref{Simple_Bosonization} is, in fact, arbitrary; for different choices the interpretation of the model might be very different.

Similarly, Step 3 might also be a show stopper for the matching: In principle, the free fermionic description allows for more choice of generalized GSO phases than the orbifold description. As stressed in Section~\ref{sc:Orbifolds}, it is conventional in the orbifold literature to fix certain phases once and for all, even though not all these choices are strictly necessary. However, we have included there additional generalized torsion phases that should correspond to the additional freedom of generalized GSO phases on the free fermionic side.

\subsection{Convert to a basis that admits an orbifold interpretation}\label{sc:convertToOrbiBasis} 

The first step in identifying an orbifold model that corresponds to a given free fermionic model is to bring the basis vectors into a form that makes interpreting them from the bosonic side easier.

\subsubsection*{Characterize different types of basis elements}

As discussed in the previous section, any free fermionic model under consideration in this paper 
possesses the supersymmetry vector $\mathbf{S}$ defined in Table~\ref{tb:BasisVectors} as an element of the additive set $\BgX$; conventionally, even as one of the basis vectors.
For such models we can find two independent vectors $\mathbf{\widetilde{b}}_1$ and $\mathbf{\widetilde{b}}_2$ such that both of these vectors and their sum, $\mathbf{\widetilde{b}}_{3} = \mathbf{\widetilde{b}}_1+\mathbf{\widetilde{b}}_2$, all act on some of the $\chi^i$ but not on $\gps^\gm$: 
\equ{
\mathbf{S} \cap \mathbf{\widetilde{b}}_s \neq \emptyset~,
\qquad 
\gd_{\mathbf{\widetilde{b}}_s} = 1~. 
}
These basis vectors,  $\mathbf{\widetilde{b}}_s$, are twist-like vectors since they act on the geometry at least as reflections and hence correspond to the orbifold elements as can be inferred from the bosonization relation~\eqref{Simple_Bosonization}. This can be obtained by comparing the supersymmetry currents in the bosonic and fermionic descriptions, given in~\eqref{susycurrent_bos}  and \eqref{eq:susycurrent}, respectively, upon identifying the notation $\gps^i = \chi^i$. 

For the remaining generators of the additive set, we construct linear combinations, $\Bgb_x$, such that none of them acts on the fermions $\{\psi^\mu,\chi^i\}$, i.e. 
\equ{ \label{NoOverlapS}
\Bgb_x \cap \mathbf{S} = \emptyset~. 
}
We refer to these vectors as Narain-like vectors. In this new basis, 
\equ{ \label{FFFBasisExpansion} 
\Bga =
s\, \mathbf{S} + \sum_{a\neq S} n_a\, \mathbf{B}_a
=
s\, \mathbf{S} 
+ k_s\, \mathbf{\widetilde{b}}_s
+ n_x \, \Bgb_x~, 
}
(with $s, k_s=0,1$ and $n_a$ up to the order of the various elements $\mathbf{B}_a$) only the supersymmetry generator $\mathbf{S}$ has $\gd_\mathbf{S} = -1$. Notice that the two basis vectors $\mathbf{\widetilde{b}}_s$ are not uniquely defined because we can always combine them with arbitrary linear combinations of the basis vectors $\Bgb_x$. A useful choice is to pick these linear combinations such that the overlap of the vectors $\mathbf{\widetilde{b}}_1$ and $\mathbf{\widetilde{b}}_2$ on the $y$'s and $w$'s is as small as possible.

\subsubsection*{Symmetric orbifold interpretation}

Before we continue, we need to check that the fermionic model admits an interpretation as a symmetric orbifold at all: The free fermionic basis elements translated into the bosonic language should either act as a twist-like action or as a shift action on both left- and right-moving coordinates. This is not guaranteed by the definitions of the twist-like and Narain-like basis vectors above as their characterizations involved their $\big\{\gps^\gm,\gch^i\big\}$-content only. 

To understand the relation between fermionic and bosonic boundary conditions, it is helpful to make use of the bosonization relation~\eqref{Simple_Bosonization}. Since the supercurrent~\eqref{eq:susycurrent} has to be preserved by all basis elements of a free fermionic model, we infer that for any Narain-like element $\Bgb_x$ the fermions $y^i,w^i$'s should always appear in pairs for any $i=1,\ldots,6$: Narain-like elements could act as translations on the coordinate fields but never as a twist, hence we see from~\eqref{Simple_Bosonization} that precisely in these cases $X_R$ does not change sign. For symmetric orbifolds, admissible Narain-like basis vectors should also contain $\byy^j, \bw^j$ pairwise.

Similarly, in any twist-like element, $\mathbf{\widetilde{b}}_s$, either $y^i$ or $w^i$ is present (but never both at the same time) whenever it contains $\chi^i$; when it does not, the $y^i,w^i$'s should appear pairwise. From~\eqref{Simple_Bosonization} we see that, in this case, $X_R$ at least changes sign, and so the interpretation of a twist-like element is justified. We demand that for a symmetric orbifold interpretation the same $\byy^i$'s and $\bw^i$'s should appear in the twist-like basis elements.

These criteria for having a symmetric orbifold interpretation are  up to renaming of the left-moving real and complex fermions, since splitting in real $\byy$ and $\bw$ and complex $\bgl$ fermions in Table~\ref{table:wsfermions} is somewhat arbitrary. For a free fermionic model to admit a symmetric orbifold interpretation, there should be some choice for this such that these statements all hold.

By a reordering of the indices $i$ we can ensure that we have chosen the twist-like elements such that 
\equ{ \label{IdentifyThetas}
\mathbf{\widetilde{b}}_1 \supset \big\{ \chi^{3,4}, \chi^{5,6} \big\}~, 
\qquad 
\mathbf{\widetilde{b}}_2 \supset \big\{\chi^{1,2}, \chi^{5,6} \big\}~. 
}
Again, using the invariance of the supercurrent~\eqref{eq:susycurrent} this implies that $\mathbf{\widetilde{b}}_1$ and $\mathbf{\widetilde{b}}_2$ act as twist-like actions on the bosonic coordinates with point group actions given by~\eqref{Z22twists}. In the following, we are considering only free fermionic models that admit a symmetric orbifold interpretation and that the basis vectors $\mathbf{b}_s$ and $\Bgb_x$ have been brought to the form defined here.

It is also possible to obtain some elements $\Bgb_x$ that do not involve any $y$ and $w$ fermions; such elements may be associated with the gauge spin structures $\gn_{uL}$ in the bosonic language: If the model includes $\Bgx_1$ and $\Bgx_2$ then we can think of it as an orbifold of the ten dimensional heterotic E$_8\times$E$_8$ theory, and when it only includes $\Bgx$, of the Spin(32)$/\Intr_2$ theory.  
It can also happen that there is no linear combination of the Narain-like basis vectors which equals $\Bgx$; in particular it might be that only one of the two $\Bgx_1,\Bgx_2$ is present. 
Given that the moduli space of Narain compactifications is connected, in such cases the free fermionic models  correspond to orbifold theories at points in the moduli space other than the E$_8\times$E$_8$ or Spin(32)$/\Intr_2$ points. Some examples are given in Table~\ref{tb:NarainModels} in Section~\ref{sc:Examples}.

If the additive set $\BgX$ includes a set of spin-structure vectors, i.e.\ either $\Bgx$ or both $\Bgx_1$ and $\Bgx_2$, and some further requirements are met, see below, we can continue either via route 2a or 2b. If this is not the case, only route 2b is available to us.

\renewcommand\thesubsection{\thesection.\arabic{subsection}a}
\subsection{Directly determine the orbifold twists, shifts and Wilson lines}

In this subsection we assume that we have a set of basis vectors
\equ{
\mathsf{B} = \big\{ \mathbf{B}_a\big\} = \big\{ \mathbf{S}, \mathbf{\widetilde{b}}_s, \Bgx_u, \Bgb_x\big\}~, 
}
 that admit a symmetric orbifold interpretation and has at most six remaining Narain-like basis vectors $\Bgb_x$. In addition, we demand that they are strictly symmetric, i.e.\ each of them contains the same $y^i,w^i$ as $\byy^i,\bw^i$-pairs. Finally, we require that they remain linearly independent when we restrict them to their geometrical action, characterized by the $y,w$-pairs only. 

If these conditions are not satisfied by the basis vectors in question, then the methods described in this subsection cannot be applied. One could try to modify the input data of the free fermionic model, such that the new set of basis vectors do satisfy these conditions. Of course, alternatively, one can use the more general procedures of the next Subsection corresponding to route 2b.

\subsubsection*{Free fermionic basis vectors and even lattice constraints}

The defining data of an orbifold model, in particular the orbifold twists, shifts and Wilson lines, are assumed to satisfy some additional conventions: The gauge shifts and Wilson lines multiplied by their order should be lattice vectors in the appropriate gauge lattices. The orbifold twists were chosen to leave a standard choice for the four dimensional supersymmetry generators invariant. These conditions are technically enforced by requiring that the twists $v_s$ satisfy~\eqref{TwistSUSY}  and the shifts $V_s$ and the Wilson lines $A_x$ multiplied by their orders are $\gL_\text{gauge}$ lattice vectors (see the requirements~\eqref{OrderShiftsWLs}). In addition, the orbifold input data needs to satisfy the generalized modular invariant conditions~\eqref{ModInvConditions}. The conventions on the free fermionic basis vectors $\mathbf{B}_a$ are slightly different: their entries have to fulfill~\eqref{eq:A1} and are 
conventionally chosen to lie in the range~\eqref{eq:fermiphases}. 

The additional specific lattice conditions on the orbifold input data translate in the free fermionic language as follows:  The standard choice for supersymmetry under~\eqref{TwistSUSY} requires that:
\equ{\label{eq:basisVecs} 
\mathbf{S} \cdot \mathbf{\widetilde{b}}_s = 0~
,}
(the conditions~\eqref{OrderShiftsWLs} are automatically fulfilled by~\eqref{eq:ABnorms}). If we have basis vectors that do not satisfy~\eqref{eq:basisVecs}, then we can modify them as 
\equ{ \label{BasisVectorOrbiForm}
\mathbf{\widetilde{b}}_s^\text{orbi} = \mathbf{\widetilde{b}}_s + \Bgd_s~, 
}
where $\Bgd_s$ are vectors with only even entries in the $\chi^i$-directions, such that some signs in $\chi^i$-entries of $\mathbf{\widetilde{b}}_s^\text{orbi}$ are flipped to satisfy~\eqref{eq:basisVecs}:
For example, we can take $\Bgd_1 = \{-2\chi^{34}\}$ and $\Bgd_2 = \{-2\chi^{12}\}$ so that $\mathbf{\widetilde{b}}_1^\text{orbi} \supset \{-\chi^{34},\chi^{56}\}$ and 
$\mathbf{\widetilde{b}}_2^\text{orbi} \supset \{-\chi^{12},\chi^{56}\}$.
This does not modify the free fermionic model at all, provided that one modifies the generalized GSO phases accordingly using~\eqref{TrivialPhaseChanges}.  
In the orbifold language, this corresponds to the twists 
\equ{
v_1 = (0,0,-\sfrac 12, \sfrac 12)~, 
\qquad 
v_2 = (0, -\sfrac 12,0, \sfrac 12)~. 
}
Up to possible brother phases~\eqref{BrotherPhases} this corresponds to the most common choice~\eqref{Z2xZ2twists} in the orbifold literature.

\subsubsection*{Characterizing the symmetric orbifold input data}

We can now immediately read off the orbifold input: The orbifold twists and shifts are given by 
\begin{subequations} \label{DataIdentificationFFF->Orbi}
\equ{ \label{IdentificationShifts}
v_s = \sfrac 12\, \widetilde{b}^\text{orbi}_s(\chi)~, 
\qquad 
V_s = \sfrac 12\, \widetilde{b}_s(\bgl)~, 
}
taking care when going from a real to a complex basis for the fermions $\chi^i$. 
Moreover, we can identify the Wilson lines 
\equ{ \label{IdentificationWLs}
A_x = \sfrac 12\, \gb_x(\bgl)~, 
}
\end{subequations} 
associated with translations in the directions $\gve_x = \sfrac 12\, \gb_x(y) = \sfrac 12\, \gb_x(w)$.

\addtocounter {subsection} {-1}
\renewcommand\thesubsection{\thesection.\arabic{subsection}b}

\subsection{Identify the geometrical Narain data} 
\label{sc:IdentifyNarainData}

The Narain lattice corresponding to a free fermionic model can be determined in the following fashion. Not the whole fermionic partition function~\eqref{eq:FFZ} admits a Narain lattice interpretation, therefore we only focus on the part of this partition function generated by the fermions $y^i, w^i, \byy^i, \bw^i, \bgl^I$. Moreover, only the non-twist part of the fermionic partition function~\eqref{eq:FFZ} should be considered, since the Narain description applies to torus compactifications. Hence, we further restrict to the basis vectors with $\Bgb = n_x\, \Bgb_x$ (i.e.\ setting $s=k_s=0$): 
\equ{ \label{NarainPartitionFun}
Z_\text{Narain} = \frac 1{N}\, \sum_{n,n'}\,  
 \frac{\Theta[^{\gb(y)}_{\gb'(y)}]}{\get^{6}} \, 
 \frac{  \overline{\Theta}[^{\gb(\overline{y})}_{\gb'(\overline{y})}] \, \overline{\Theta}[^{\gb(\overline{\gl})}_{\gb'(\overline{\gl})}]}{\overline{\get}^{22}}~, 
}
where $N$ is the product of the orders of the elements $\Bgb_x$.
Here, we used that, for the non-twist elements, $\gb(w) = \gb(y)$ and similarly for their conjugates. Using the sum representation~\eqref{ThetaAB}, this is immediately written in the form of a Narain lattice sum~\eqref{eq:bpf1} and hence one can read off a basis for the Narain lattice. An example illustrating this procedure in detail is given in Subsection~\ref{sc:NarainTorusModels}.

\subsubsection*{Narain standard form}

With either of the above methods, one obtains a basis for the Narain lattice. The collection of basis vectors may be interpreted as the generalized vielbein $E^\prime$. However, when we compute 
\equ{
E^{\prime T} \get E^\prime = \hat\get^\prime~, 
}
we generically do not find the metric $\hat\get$ generated in \eqref{eq:metric2}, but a matrix $\hat\get^\prime$ that is related to this via a transformation $M\in \text{GL}(28; \Intr)$: 
\equ{ \label{StandardLorentzMetric}
\hat\get = M^T \hat\get^\prime M~. 
}
It is important to realize that the determination of the Narain moduli strongly depends on the form of $\hat\get^\prime$. Hence, it is not sufficient to know the generalized vielbein $E'$ in some arbitrary basis, but it is crucial to find a matrix $M$ that brings it to a standard form. Unfortunately, as far as we are aware, no generic algorithm is known about how to determine such a transformation. However, this is not a problem of encoding a free fermionic model in the orbifold description, but rather an issue of how to practically work with Narain moduli spaces.

\renewcommand\thesubsection{\thesection.\arabic{subsection}}
\subsection{Determine the generalized torsion phases}

We have seen in the previous subsections that we can distinguish two types of free fermionic constructions: those that can be thought of as orbifolds of the Spin(32)$/\Intr_2$ or E$_8\times$E$_8$ theories and the others. This distinction is also important for how concretely one can describe the translation of the generalized GSO phases to the generalized torsion phases on the bosonic side.

\subsubsection*{Orbifolds of the Spin(32)$/\Intr_2$ or E$\boldsymbol{_8\times}$E$\boldsymbol{_8}$ theories}

Modulo the fact that one, in general, needs to add even entries to some of the basis vectors, i.e.~\eqref{BasisVectorOrbiForm}, we see that the translation of the free fermionic to the orbifold data in~\eqref{DataIdentificationFFF->Orbi} is essentially identical to that in the opposite direction, see~\eqref{WLtranslationFFF} and~\eqref{TwisttranslationFFF} (up to a factor of 1/2 in~\eqref{WLtranslationFFF}, which we included since all vectors $\mathbf{e}_i$ were taken to be in the basis vector set. Substituting the translations into each other, one gets the original input data back). Hence, to determine translation of the phases, we can also simply invert the phase relations~\eqref{eq:IdentifiedPhasesOrbi->FFF}.

Since free fermionic data do not necessarily satisfy~\eqref{eq:basisVecs}, we may need some sign flips in $\mathbf{\widetilde{b}}_s$. Via~\eqref{TrivialPhaseChanges}, we have 
\begin{subequations} \label{eq:IdentifiedPhasesFFF->Orbi}
\equ{
e^{2\pi i\, c_{st}} = 
e^{-\sfrac 14 \pi i\,  (\mathbf{\widetilde{b}}_1 - \Bgd_1) \cdot (\mathbf{\widetilde{b}}_2+\Bgd_2)}\, 
C\left[^{\mathbf{\widetilde{b}}_1}_{\mathbf{\widetilde{b}}_2}\right]~. 
}
In addition, we obtain: 
\equ{
e^{2\pi i\, c_{\ui\uj}}  = 
e^{-\sfrac 14 \pi i \, \Bgb_\ui\cdot \Bgb_\uj}\, 
C\left[^{\Bgb_\ui}_{\Bgb_\uj}\right]~, 
\quad 
e^{2\pi i\, c_{s\ui}} = 
e^{-\sfrac 14 \pi i \, \mathbf{\widetilde{b}}_s \cdot \Bgb_\ui}\, 
C\left[^{\mathbf{\widetilde{b}}_s}_{\Bgb_\ui}\right]~, 
}
\equ{ 
(-)^{c_s} = e^{ -\sfrac 14 \pi i\, \mathbf{\widetilde{b}}_s^2} \, 
C\left[^{\mathbf{\widetilde{b}}_s}_{\mathbf{\widetilde{b}}_s} \right]~, 
\qquad  
(-)^{c_\ui} = e^{- \sfrac 14 \pi i\, \Bgb_\ui^2} \,  
C\left[^{\Bgb_\ui}_{\Bgb_\ui} \right]~, 
\qquad 
(-)^{c_u} = 
C\left[^{\Bgx_u}_{\Bgx_u} \right]~, 
}
\equ{
e^{2\pi i\, c_{uv}} = C\left[^{\mathbf{\Bgx}_1}_{\mathbf{\Bgx}_2}\right] ~, 
\qquad 
e^{2\pi i\, c_{s u}} = C\left[^{\mathbf{\widetilde{b}}_s}_{\Bgx_u}\right]~, 
\qquad 
e^{2\pi i\, c_{\ui u}} = C\left[^{\Bgb_\ui}_{\Bgx_u}\right]~. 
}
\end{subequations}

\subsubsection*{General Narain orbifolds} 

If one has determined the Narain lattice associated with the Narain-like elements following route 2b, then one has absorbed some of the original generalized GSO phases into the Narain lattice. This will typically mean that the geometrical part of the lattice has changed, i.e.\ the $\gve$ in the generalized vielbein~\eqref{eq:genV} is not the same as the one we started with. Therefore, the Wilson lines that are read off from it, are related, in a complicated way, to the original ones, hence unfortunately, it is very difficult to describe the relation between the original phases of the free fermionic model and the remaining ones after rewriting the underlying torus compactification in the Narain form. In light of this, the most systematic approach seems to be to simply 
scan a variety of generalized torsions for the translated orbifold model.

\subsection{Identifying the orbifold geometry}
\label{sc:IdentifyOrbifoldAction}

Above, we obtained a basis of generators of the additive set which are divided into Narain-like and twist-like elements. The twist-like elements, $\mathbf{b}_1$ and $\mathbf{b}_2$, can either be interpreted as pure twists or roto-translations. However, reversing the logic presented in Subsection~\ref{sc:OrbioldActionsinFFF}, we are able to determine how to interpret their actions geometrically. 

Consequently, any free fermionic model that admits an interpretation as a symmetric $\Intr_2\times \Intr_2$ orbifold model should correspond to one of the geometries given in Table~\ref{tb:Z22classification}. When the orbifold actions and the six-torus lattice $\gve$ have been identified, the corresponding  $\Intr_2\times \Intr_2$ orbifold can be determined by referring to the program {\tt CARAT}. In particular, using this code, one determines the \Zclass{} of the lattice, simply by calculating the matrices $\gve^{-1}\gth_1\gve$ and $\gve^{-1}\gth_2\gve$ and using the {\tt CARAT} command: {\tt Name}.

\section{Examples}
\label{sc:Examples}

\subsection{Narain torus compactification models}
\label{sc:NarainTorusModels} 

\subsubsection*{The SO(12)$\times$SO(32) model}

Our review of free fermionic models in Section~\ref{sc:FreeFermi} indicated that all free fermionic models contain at least the vectors: $\big\{ \mathbf{1}, \mathbf{S}\big\}$. For simplicity, the first example we consider here is the free fermionic model obtained from this set augmented with the vector $\Bgx$ given in Table~\ref{tb:BasisVectors}, i.e.\ is defined by the set of basis vectors $\big\{  \mathbf{1},\mathbf{S},\Bgx\big\}$. The resulting model possesses $\cN=4$ supersymmetry in four dimensions and has an SO(12)$\times$SO(32) gauge group.

To translate this free fermionic model to the bosonic description, the first step is to define the orbifold interpretable basis. To this end, we make a change of basis such that the new basis vectors do not have any overlap: $\big\{ \mathbf{S}, \mathbf{e}_{123456},\Bgx\big\}$: $\Bgx$ is already a Narain-like basis vector. Since we have the basis vector $\mathbf{S}$ explicitly, the other element which does not contain $\gps^\mu$ and has no overlap with $\Bgx$ is 
\equ{ 
\mathbf{e}_{123456} = \mathbf{1} - \mathbf{S} -\Bgx= \big(0^8, 1^{12}\,|\,1^{12}; 0^{16}\big)~. 
}
As there is no overlap with $\mathbf{S}$, this is also a Narain-like basis vector. In addition, due to there being no overlap between the basis vectors $\mathbf{e}_{123456}$ and $\Bgx$, the resulting Narain part of the partition function~\eqref{NarainPartitionFun} factorizes as 
 \equ{
{Z}_\text{Narain} =
\frac{1}{4\, \get^6 \bget^{22}}\, \sum_{s',s=0,1} 
\Theta[^{s}_{s'}]^6 \overline{\Theta}[^{s}_{s'}]^{6}
\, \sum_{t',t=0,1} 
 \overline{\Theta}[^{t}_{t'}]^{16}
 }
Using the sum representation of the $\Theta$ function~\eqref{ThetaAB}, we can read off the projection conditions on the summation variables, $m'',n'' \in \Intr^6$ and $p''\in \Intr^{16}$, to obtain
\equ{\label{eq:ZNarainStep2}
Z_\text{Narain} = \frac1{4\,\get^6\bget^{22}}  \, 
\sum_{\substack{
s=0,1,~m'',n'' \in \Intr^6, 
\\
\sum (m''_i+n''_i)=0~\text{mod}~2}}
\bar q^{\frac 12 \sum_{i}(m''_i + \frac s2)^2} \, 
q^{\frac 12 \sum_{j}(n''_j+\frac s2 )^2}
\, 
 \sum_{\substack{
t=0,1,~p''\in\Intr^{16}
\\
\sum p''_k=0~\text{mod}~2}}
q^{\frac 12 \sum_{k}(p''_k + \frac t2)^2}~. 
}
We define new variables $m'$, $n'$ and $p'$ as
\equ{
m'_i=m''_i +\sfrac s2~, \quad  n'_i=n''_i+ \sfrac s2~, \qquad p'_k=p''_k +\sfrac t2~. 
}
Note that for $s = 0$ or $1$ variables $m'_i$'s and $n'_i$'s are all integral or all half-integral. 
The same holds for the new variables $ p'_k$'s. Furthermore, the sum restrictions imply that 
\equ{
\sum (m_i''+n_i'') = \text{even}~, 
\qquad 
\sum p_i'' = \text{even}~.  
}
These conditions together tell us that $(m',n') \in \cD_{12}$ and $p \in \cD_{16}$. 
Here the lattice $\cD_D$ in $D$ dimensions is defined as 
\equ{ \label{eq:DnNew}
\cD_D = \cR_D + \cS_D~, 
}
where we introduced the SO($2D$) root and spinor lattices
\begin{subequations} \label{SO2Dlattices}
\equ{
\cR_D = \big\{ n \in \Intr^D \,\big|\, \sum n = \text{even} \big\}~, 
\qquad 
\cS_D = \big\{ n + \sfrac 12\, \mathbf{1}_D \,\big|\, \sum n = \text{even}\big\}~. 
\\[1ex]  
\cV_D = \big\{ n \in \Intr^D \,\big|\, \sum n = \text{odd} \big\}~, 
\qquad 
\cC_D = \big\{ n + \sfrac 12\, \mathbf{1}_D \,\big|\, \sum n = \text{odd}\big\}~. 
}
\end{subequations} 
In particular, $\cD_8$ is the E$_8$ root lattice. Hence, we can write the lattice sum as 
\equ{ \label{eq:pfD32}
{Z}_\text{Narain} =   \frac1{4\,\get^6\bget^{22}}  \, 
\sum_{( {m}', {n}' )\in\cD_{12}} 
\bar q^{\frac 12 {m}'{}^2} \, 
q^{\frac 12 {n}'{}^2}
\, 
\sum_{{p}^{\,\prime} \in \cD_{16}} 
q^{\frac 12 {p}'{}^2}~. 
}
To identify this partition function~\eqref{eq:pfD32} with the Narain partition function given in~\eqref{eq:pfN}, one needs to find a change of variables, $N' = (m',n',p') =E\, N$, that solves the constraints and allows us to write the sum over all of $\Intr^{28}$ instead of the restricted set $\cD_{12}\oplus\cD_{16}$. This change of variables is precisely of the form of the Narain momentum vector~\eqref{NarainLattice}, hence the matrix $E$ can be taken in the form of the generalized vielbein~\eqref{eq:genV}. For the case at hand, a possible choice for this is given by
\equ{\label{ModSO32} 
\gve = \gve_{so}~, 
\qquad 
G = \gve^T \gve~, 
\qquad 
B = B_G~, 
\qquad 
A = 0_{16\times 6}~,
\qquad
\ga=\ga^{\;}_{16} 
}
using the notation defined below.

\subsubsection*{Other toroidal Narain models}    

\begin{table}
\[
\arry{|l||c|c|c|}{
\hline\hline
\textbf{Basis vectors} & \textbf{Gauge group} & \textbf{Six-torus lattice} & \textbf{Narain moduli} 
\\ \hline\hline 
\big\{ \mathbf{S}, \mathbf{e}_{1\ldots6}+\Bgx \big\} & \text{SO(44)} & 
\multirow{4}{*}{$\big\{ \sfrac 12 e_{1\ldots6},e_2 \ldots, e_6 \big\}$}&
\gve_{\mathbb{1}},\, B_{\mathbb{1}},\, A_{16},\, \ga^{\;}_{16}
\\ 
\big\{\mathbf{S}, \mathbf{e}_{1\ldots6}, \Bgx\big\} & \text{SO(12)}\times\text{SO(32)} & &
\gve_{so},\, B_{G},\, A=0,\, \ga^{\;}_{16}
\\ 
\big\{ \mathbf{S}, \mathbf{e}_{1\ldots6}+\Bgx_1, \Bgx_2\big\} & \text{SO(24)}\times\text{E}_8 & &
\gve_{\mathbb{1}},\, B_{\mathbb{1}},\, A_{8},\, \ga^{\;}_{8\times 8}
\\ 
\big\{\mathbf{S}, \mathbf{e}_{1\ldots6}, \Bgx_1, \Bgx_2\big\} & \text{SO(12)}\times\text{E}_8\times\text{E}_8 & &
\gve_{so},\, B_{G},\, A=0,\, \ga^{\;}_{8\times 8}
\\ \hline 
\big\{ \mathbf{S}, \mathbf{e}_1,\ldots, \mathbf{e}_6, \Bgx \big\} & \text{U(1)}^6\times\text{SO(32)} &
\multirow{2}{*}{$\big\{ \sfrac 12 e_1,\ldots, \sfrac 12 e_6 \big\}$}&
\gve_{\mathbb{1}},\, B=0,\, A=0,\, \ga^{\;}_{16}
\\ 
\big\{ \mathbf{S}, \mathbf{e}_1,\ldots, \mathbf{e}_6, \Bgx_1, \Bgx_2\big\} & \text{U(1)}^6\times\text{E}_8\times\text{E}_8 & &
\gve_{\mathbb{1}},\, B=0,\, A=0,\, \ga^{\;}_{8\times 8}
\\ \hline\hline 
}
\]
\caption{ \label{tb:NarainModels}
This table summarizes the most prominent free fermionic models that can be interpreted as Narain compactifications. The explicit moduli were derived for the standard choice of the GSO phases~\eqref{eq:StandardPhasesExample}. The notation for the Narain moduli fields is defined in Subsection~\ref{sc:NarainTorusModels}.
} 
\end{table}

To describe the previous and some other free fermionic models which correspond to purely Narain compactifications, we define: the six dimensional vielbeins, 
\equ{\label{eq:eso}
\gve_{\mathbb{1}}=\frac{1}{\sqrt2} 
\Id_6 , \quad
\gve_{so} = \frac{1}{\sqrt2} 
\scalebox{.5}{$\pmtrx{  
 1 &  0 & 0 & 0 & 0 & 0  \\ 
-1 & 1 & 0 & 0 & 0 & 0 \\ 
 0 &-1& 1 & 0 & 0 & 0 \\ 
 0 & 0  &-1 & 1 & 0 & 0 \\ 
 0 & 0 & 0 & -1 & 1 & 1 \\
 0 & 0 & 0 & 0 & -1 & 1
}$}_{6\times6}~; 
}
Kalb-Ramond B-fields, 
\equ{\label{eq:B}
B_{\mathbb{1}}=\frac{1}{\sqrt2} 
\scalebox{.6}{$ \pmtrx{ 
0 & -1 & \cdots &-1 \\ 
1 & \ddots &\ddots & \vdots \\  
\vdots & & \ddots & -1 \\ 
1 & \cdots & 1 & 0 
}_{6\times 6}$}~, \quad
B_{G} = \begin{cases} G_{ij} &\mbox{if } i<j \\
0 &\mbox{if } i=j\\
-G_{ij} & \mbox{if } i>j \end{cases}~;
}
and Wilson lines, 
\equ{\label{eq:A}
A_i=\scalebox{.8}{$\pmtrx{ 
 \text{\huge0} \\
1\ 1\ 1\ 1\ 1\ 1\\  
 \text{\huge0}
}_{16\times 6}
\begin{matrix} 
 \phantom{\text{\huge0}} \\
\leftarrow \mbox{i-th row}\\  
 \phantom{\text{\huge0}}
\end{matrix}
$}~.
}
Using these definitions, we can express the moduli of a number of pure Narain free fermionic models given in Table~\ref{tb:NarainModels}. They have been derived following the procedure in the previous subsection.  
For all of them we have made the standard choice of GSO phases, given by 
\equ{ \label{eq:StandardPhasesExample}
C[^\mathbf{S}_\mathbf{S}] = C[^\mathbf{S}_{\mathbf{B}_a}] = -1~, 
\qquad 
C[^{\mathbf{B}_a}_{\mathbf{B}_b}] = 1~, 
}
for all basis vectors $\mathbf{B}_{a},\mathbf{B}_b \neq \mathbf{S}$. 
Certain phases do not change the gauge group, but only the lattices. A simple example of this effect is  to set $C[^{ \boldsymbol{\xi}_2}_{ \boldsymbol{\xi}_2}]=-1$ leading to a change of the spinor lattice to the co-spinor lattice $\cD_8$ in~\eqref{SO2Dlattices} for the second $E_8$ factor.

\subsection{A simple free fermionic $\boldsymbol{\Intr_2\times\Intr_2}$ model}

We will start our analysis of free fermionic models that include orbifold twists by considering the free fermionic model with basis vectors 
 \begin{align}\label{eq:basisEg1}
\Big\{ \mathbf{S}, \mathbf{b}_1,\mathbf{b}_2,\mathbf{e}_{1\dots6}, \Bgx_1, \Bgx_2 \Big\}~,
\end{align}
introduced in Table~\ref{tb:BasisVectors}. The upper triangular part of the generalized GSO phase matrix, including the diagonal is taken to be: 
\equ{ \label{StandardPhasesExample} 
C[^{\mathbf{B}_a}_{\mathbf{B}_b}] = 
  \scalebox{.8}{$\begin{blockarray}{ccccccc}
 {\mathbf{B}_a} \backslash {\mathbf{B}_b}\!\!\!\! &  \mathbf{S}&\mathbf{b}_1&\mathbf{b}_2 & \Bgx_1 & \Bgx_2& \!\!\mathbf{e}_{1\ldots6}\\ 
 \begin{block}{l(rrrrrr)}
~\mathbf{S} 	& -1 &  -1 & -1 & -1 & -1 &  -1 \\
~\mathbf{b}_1 	& \color{gray}1 & -1 &  1 &  1 &  1 &  1 \\
~\mathbf{b}_2 	& \color{gray}1 & \color{gray}1 & -1 &  1 &  1 &  1 \\
~\Bgx_1       	& \color{gray}1 & \color{gray}-1 & \color{gray}-1 & 1 &  1 &  1 \\
~\Bgx_2       	& \color{gray}1 & \color{gray}1 & \color{gray}1 & \color{gray}1 &  1 &  1 \\
~\mathbf{e}_{1\ldots6}\!\!\!  & \color{gray}-1 & \color{gray}1 & \color{gray}1  & \color{gray}1 & \color{gray}1 &  1 \\
 \end{block}
\end{blockarray}$}~.
}
To emphasize that the entries in the lower triangular part cannot be chosen arbitrarily, but are fixed via~\eqref{modinvphases1b0}, we have drawn these entries in a grey colour. 

In this model, the interpretation of the basis vector elements is immediate: $\mathbf{S}$ is the target space supersymmetry element; $\mathbf{b}_1, \mathbf{b}_2$, the twist-like elements; and $\Bgx_1,\Bgx_2,\mathbf{e}_{1\ldots6}$, Narain-like elements. Since the twist-like elements involve the fermions $\chi^i$ as dictated in~\eqref{IdentifyThetas}, we can associate $\mathbf{b}_s$ with the orbifold twists $\gth_s$ defined in~\eqref{Z22twists}. Moreover, since these twists do not have any $y$ or $w$ overlap, we know we can interpret them both as generating pure twists, as discussed in Subsection~\ref{sc:IdentifyOrbifoldAction}.

In more detail, by the multiplication in Table~\ref{tb:BasisVectors} we notice that the inner products 
\equ{ 
 \mathbf{b}_s \cdot \mathbf{S} = 2 \text{ mod } 4~. 
}
Hence, the twist-like elements do not satisfy~\eqref{eq:basisVecs}. Therefore, when we want to read off the associated orbifold twists and gauge shifts according to~\eqref{IdentificationShifts}, we need to flip some signs (see~\eqref{BasisVectorOrbiForm}): 
\begin{subequations} 
 \label{StandardEmbedding} 
\equ{
\mathbf{b}_1~:
\quad 
v_1 = \sfrac 12\, \big(~\,0,-1,\,~1\big)~, 
\quad 
V^\text{SE}_1 = \sfrac 12\, \big(0^5,\,~0,\,~1,\,~1\big)\big(0^8\big)~, 
\\[1ex] 
\mathbf{b}_2~:
\quad 
v_2 = \sfrac 12\, \big(\!\!-\!1,\,~0,\,~1\big)~, 
\quad 
V^\text{SE}_2 = \sfrac 12\, \big(0^5,\,~1,\,~0,\,~1\big)\big(0^8\big)~, 
}
\end{subequations} 
which we can see with the help of \eqref{TrivialPhaseChanges}, do not modify the phases. Hence, we can keep using \eqref{StandardPhasesExample} in its current form. Since the model includes the basis vectors $\Bgx_1, \Bgx_2$, we can interpret it as an orbifold of the E$_8\times$E$_8$ theory. Moreover, since $V^\text{SE}_s$ contains $v_s$, this model corresponds to the standard embedding. Consequently, we can use the number of $\mathbf{16}$-plet generations and anti-generations to determine the Hodge numbers of the orbifold geometry.

The orbifold phases can be read from the matrix in~\eqref{StandardPhasesExample} using \eqref{eq:IdentifiedPhasesFFF->Orbi}. We find that all the orbifold torsion phases are trivial, i.e. 
\equ{ 
c_s = c_\ui = c_u = 0~, 
\qquad 
c_{st} = c_{\ui\uj} = c_{s\ui} = c_{uv} = c_{su}  = c_{\ui u} = 0~. 
}
In particular, the spin-structure projections are the standard ones used in the orbifold literature. Since, all the other possible generalized torsion phases~\eqref{FullGenTorsion} are also zero, this model can be directly understood as a standard orbifold model. 
Furthermore, the non-twist-like basis vectors, $\big\{\mathbf{S},\mathbf{e}_{1\ldots6} , \boldsymbol{\xi}_1, \boldsymbol{\xi}_2\big\}$, are the same as the set of basis vectors on the fourth row of Table~\ref{tb:NarainModels}. Hence, given that the relevant phases are also chosen identically, we can immediately read off the moduli from that row of the table.

To summarize, we have found that this simple free fermionic model corresponds to the standard $\Intr_2\times\Intr_2$ pure twist orbifold on the $SO(12)$ lattice with the standard embedding. This corresponds to the DW(1~-~1) geometry.

\subsection{Free fermionic realizations of the $\boldsymbol{\Intr_2\times\Intr_2}$ orbifold geometries}

In this subsection, we would like to give explicit examples of free fermionic models corresponding to each of the $\Intr_2\times\Intr_2$ orbifold geometries. The results of this analysis have been collected in Table~\ref{tb:FFFrealizationsDW}. (They are independent of the gauge structure and therefore apply to both the E$_8\times$E$_8$ and the Spin(32)$/\Intr_2$ cases.) 
In principle, we can directly use the results of Section~\ref{sc:OrbitoFFF} to translate each of these geometries in the free fermionic language. This way one obtains a large set of basis vectors which can be computationally inconvenient. In Table~\ref{tb:FFFrealizationsDW} we give free fermionic realizations of each of the $\Intr_2\times\Intr_2$ geometries that are minimal in their number of basis vectors. 

To determine these results we started from the explicit parameterization of the orbifold geometries given in~\cite{DW}: In particular, the periodicity of the target space two-tori in terms of a modular parameter is taken to be $2\gt$ (not to be confused with the Teichmueller parameter of the worldsheet torus defined under~\eqref{eq:pfN}). 
Whenever possible, we modified the shift elements indicated there such that they can be represented by free fermionic translational elements $\mathbf{e}_i$, $\mathbf{e}_{ij}$, etc., so that the sum of all these elements is identical to $\mathbf{e}_{123456}$ (combined with $\mathbf{S}$, $\Bgx_1$ and $\Bgx_2$, this ensures that $\mathbf{1}$ is part of the additive set). 
To that effect, we sometimes change $1$ or $\gt$ to $1+\gt$ throughout an orbifold geometry: i.e.\ both in the shift elements as well as in the twists/roto-translations. 
For all geometries, we extend the resulting elements such that we get a set of shift elements that sum to $\mathbf{e}_{123456}$.

We took the standard $\Intr_2\times\Intr_2$ action to be the one that leads to chirality in the standard embedding in the first E$_8$ of the E$_8\times$E$_8$ theory. This means that the twist-like elements in this case are simply $\mathbf{b}_1$ and $\mathbf{b}_2$, given in Table~\ref{tb:BasisVectors}.  The related non-chiral geometries in the same class have one or both twist elements replaced by roto-translations. These roto-translations can be represented in the fermionic language by combining the twist elements with the appropriate translational basis vectors $\mathbf{e}_i$.
We have tried to choose the free fermionic representations of the lattice and the twists/roto-translations such that they are all manifestly order two free fermionic elements. It was only for the DW geometry (2~-~12) that we were unable to find such a representation and resorted to a seemingly order four twist $\mathbf{b}_1 + \sfrac 12\, \mathbf{e}_2$.

The standard choice of generalized GSO phases we use in Table~\ref{tb:FFFrealizationsDW} is given by:  
\equ{ \label{PanosChoice} 
C[^{\mathbf{B}_a}_{\mathbf{B}_b}] = 
\scalebox{.8}{$
\begin{blockarray}{ccccc}
{\mathbf{B}_a} \backslash {\mathbf{B}_b}\!\!\!\! &\mathbf{S} & \mathbf{B}_1&\mathbf{B}_2 & \Bgb_x \\ 
 \begin{block}{l(rrrr)}
~~\mathbf{S} 		& -1 & -1 & -1 & -1  \\
~~\mathbf{B}_1 	&  \color{gray}1 & -1 &  1 &  1  \\
~~\mathbf{B}_2		&  \color{gray}1 &  \color{gray}1 & -1 &  1  \\
~~\Bgb_y 			& \color{gray} -1 &  \color{gray} \gd_y & \color{gray} \gd_y &  1  \\
 \end{block}
\end{blockarray}
$}~.
}
Here we define 
\equ{ 
\gd_y = 
\begin{cases}
-1 & \Bgb_y = \Bgx_1~, \\[1ex] 
+1 & \text{otherwise}~. 
\end{cases}
}
Moreover, $\mathbf{B}_s$, $s=1,2$, stands for the twist elements given in the next-to-last column of Table~\ref{tb:FFFrealizationsDW}, $\Bgb_x, \Bgb_y$ for $\Bgx_1, \Bgx_2$ and the shift elements given in the last column of that table.

\begin{table}[p!]  
\begin{center} 
\scalebox{.7}{
\(
\arry{| c | c || ll | ll |}{
\hline\hline  
\textbf{DW} & \textbf{Hodge} & \multirow{2}{*}{\textbf{Twists\,/\,roto-translations}} & \multirow{2}{*}{\textbf{Shifts elements}} & \multicolumn{2}{c |}{\textbf{Free fermionic basis vector realization}}
\\ 
\textbf{Label} & \textbf{\#} &  &  & \multicolumn{2}{c |}{\textbf{in the standard embedding: $\mathbf{S},\Bgx_1,\Bgx_2$ and}} 
\\ \hline\hline 
\text{(0 - 1)} & (51,3)  & \arry{l}{(0+,0-,0-)\,,~ (0-,0+,0-)} & \text{none} & 
  \mathbf{b}_1, \mathbf{b}_2, &
 \mathbf{e}_{12},  \mathbf{e}_{34},  \mathbf{e}_{56}
\\
\text{(0 - 2)} & (19,19) & \arry{l}{(0+,0-,0-)\,,~ (0-,0+,1-)} & \text{none} &
  \mathbf{b}_1, \mathbf{b}_2+ \mathbf{e}_5, &
 \mathbf{e}_{12},  \mathbf{e}_{34},  \mathbf{e}_{56}  
\\
\text{(0 - 3)} & (11,11) & \arry{l}{(0+,0-,0-) \,,~ (0-,1+,1-)} &  \text{none} &
  \mathbf{b}_1, \mathbf{b}_2 + \mathbf{e}_{35}, & 
\mathbf{e}_{12},  \mathbf{e}_{34},  \mathbf{e}_{56}
\\
\text{(0 - 4)} & (3,3)  & \arry{l}{(1+,0-,0-) \,,~ (0-,1+,1-)} & \text{none} &
 \mathbf{b}_1+ \mathbf{e}_{1}, \mathbf{b}_2 + \mathbf{e}_{35}, &
  \mathbf{e}_{12},  \mathbf{e}_{34},  \mathbf{e}_{56}
 \\ \hline\hline 
%
\text{(1 - 1)} & (27,3) & \arry{l}{(0+,0-,0-)\,,~ (0-,0+,0-)} & (\gt,\gt,\gt) & 
  \mathbf{b}_1, \mathbf{b}_2, & \mathbf{e}_{123456} 
\\
\text{(1 - 2)} & (15,15) & \arry{l}{(0+,0-,0-)\,,~ (0-,0+,\gt-)} & (\gt,\gt,\gt) & 
 \mathbf{b}_1, \mathbf{b}_2 + \mathbf{e}_{56}, & \mathbf{e}_{123456}
\\
\text{(1 - 3)} & (11,11) & \arry{l}{(0+,0-,0-)\,,~ (0-,0+,1-)} & (\gt,\gt,\gt) &
 \mathbf{b}_1, \mathbf{b}_2 + \mathbf{e}_5, & \mathbf{e}_{123456} 
\\
\text{(1 - 4)} & (7,7)  & \arry{l}{(0+,0-,0-)\,,~ (0-,1+,1-)} & (\gt,\gt,\gt) &
  \mathbf{b}_1, \mathbf{b}_2+ \mathbf{e}_{35}, & \mathbf{e}_{123456}
\\
\text{(1 - 5)} & (3,3)  & \arry{l}{(1+,0-,0-)\,,~  (0-,1+,1-)} & (\gt,\gt,\gt) & 
  \mathbf{b}_1+\mathbf{e}_1, \mathbf{b}_2 + \mathbf{e}_{35}, & \mathbf{e}_{123456}
\\ \hline 
\text{(1 - 6)} & (31,7) & \arry{l}{(0+,0-,0-)\,,~ (0-,0+,0-)} & (\gt,\gt,0) & 
 \mathbf{b}_1, \mathbf{b}_2, & \mathbf{e}_{1234}, \mathbf{e}_{56} 
\\
 \text{(1 - 7)} & (11,11)  & \arry{l}{(0+,0-,0-) \,,~ (0-,0+,1-)} & (\gt,\gt,0) & 
 \mathbf{b}_1, \mathbf{b}_2+\mathbf{e}_5, & \mathbf{e}_{1234}, \mathbf{e}_{56}   
\\
\text{(1 - 8)} & (15,15) & \arry{l}{(0+,0-,0-) \,,~ (0-,1+,0-)} & (\gt,\gt,0) & 
  \mathbf{b}_1, \mathbf{b}_2 + \mathbf{e}_3, & \mathbf{e}_{1234}, \mathbf{e}_{56}  
\\
\text{(1 - 9)} & (7,7) & \arry{l}{(0+,0-,0-) \,,~ (0-,1+,1-)} & (\gt,\gt,0) & 
\mathbf{b}_1, \mathbf{b}_2 + \mathbf{e}_{35}, &  \mathbf{e}_{1234}, \mathbf{e}_{56}   
\\
\text{(1 - 10)} & (11,11) & \arry{l}{(1+,0-,0-) \,,~ (0-,1+,0-)} & (\gt,\gt,0) & 
 \mathbf{b}_1+\mathbf{e}_1, \mathbf{b}_2+\mathbf{e}_3, & \mathbf{e}_{1234}, \mathbf{e}_{56}
\\
\text{(1 - 11)} & (3,3) & \arry{l}{(1+,0-,0-) \,,~ (0-,1+,1-)} & (\gt,\gt,0) & 
\mathbf{b}_1+\mathbf{e}_1, \mathbf{b}_2+ \mathbf{e}_{35}, &\mathbf{e}_{1234},  \mathbf{e}_{56} 
\\ \hline\hline  
\text{(2 - 1)} & (15,3) & \arry{l}{(0+,0-,0-) \,,~ (0-,0+,0-)} & \arry{l}{(1,1,1) \,,~ (\gt,\gt,\gt)}  & 
  \mathbf{b}_1, \mathbf{b}_2, & \mathbf{e}_{135}, \mathbf{e}_{246} 
\\
\text{(2 - 2)} & (9,9) & \arry{l}{(0+,0-,0-) \,,~ (0-,0+,1-)} & \arry{l}{(1,1,1) \,,~ (\gt,\gt,\gt)}  & 
  \mathbf{b}_1, \mathbf{b}_2+\mathbf{e}_5, & \mathbf{e}_{135}, \mathbf{e}_{246} 
\\ \hline 
\text{(2 - 3)} & (17,5) & \arry{l}{(0+,0-,0-) \,,~ (0-,0+,0-)} & \arry{l}{(1,1,1) \,,~ (\gt,\gt,0)}  & 
  \mathbf{b}_1, \mathbf{b}_2, & \mathbf{e}_{1356}, \mathbf{e}_{24}
\\
\text{(2 - 4)} & (11,11) & \arry{l}{(0+,0-,0-) \,,~ (0-,0+,1-)} & \arry{l}{(1,1,1) \,,~ (\gt,\gt,0)}  & 
  \mathbf{b}_1, \mathbf{b}_2+\mathbf{e}_{56}, & \mathbf{e}_{1356}, \mathbf{e}_{24}  
\\
\text{(2 - 5)} & (7,7) & \arry{l}{(0+,0-,0-) \,,~ (0-,0+,\gt-)} & \arry{l}{(1,1,1) \,,~ (\gt,\gt,0)}  & 
  \mathbf{b}_1, \mathbf{b}_2+\mathbf{e}_6, & \mathbf{e}_{1356}, \mathbf{e}_{24}  
\\ \hline 
\text{(2 - 6)} & (19,7) & \arry{l}{(0+,0-,0-) \,,~ (0-,0+,0-)} & \arry{l}{(1,1,1) \,,~ (\gt,1,0)}  & 
  \mathbf{b}_1, \mathbf{b}_2, & \mathbf{e}_{156}, \mathbf{e}_{234}
\\
\text{(2 - 7)} & (9,9) & \arry{l}{(0+,0-,0-) \,,~ (0-,0+,\gt-)} & \arry{l}{(1,1,1) \,,~ (\gt,1,0)}  & 
  \mathbf{b}_1, \mathbf{b}_2+\mathbf{e}_6, & \mathbf{e}_{156}, \mathbf{e}_{234}
\\
\text{(2 - 8)} & (5,5) & \arry{l}{(0+,0-,0-) \,,~ (0-,\gt+,\gt-)} & \arry{l}{(1,1,1) \,,~ (\gt,1,0)}  & 
  \mathbf{b}_1, \mathbf{b}_2+\mathbf{e}_{46}, & \mathbf{e}_{156}, \mathbf{e}_{234}
\\ \hline 
\text{(2 - 9)} & (27,3) & \arry{l}{(0+,0-,0-) \,,~ (0-,0+,0-)} & \arry{l}{(0,1,1) \,,~ (1,0,1)}  & 
  \mathbf{b}_1, \mathbf{b}_2, & \mathbf{e}_{12}, \mathbf{e}_{134}, \mathbf{e}_{156}
\\
\text{(2 - 10)} & (11,11) & \arry{l}{(0+,0-,0-) \,,~ (0-,0+,\gt-)} & \arry{l}{(0,1,1) \,,~ (1,0,1)}  &
  \mathbf{b}_1, \mathbf{b}_2+ \mathbf{e}_6, &   
\mathbf{e}_{12}, \mathbf{e}_{134}, \mathbf{e}_{156}
\\
\text{(2 - 11)} & (7,7) & \arry{l}{(0+,0-,0-) \,,~ (0-,\gt+,\gt-)} & \arry{l}{(0,1,1) \,,~ (1,0,1)}  &
  \mathbf{b}_1, \mathbf{b}_2+ \mathbf{e}_{46}, & \mathbf{e}_{12}, \mathbf{e}_{134}, \mathbf{e}_{156}
\\
\text{(2 - 12)} & (3,3) & \arry{l}{(\gt+,0-,0-) \,,~ (0-,\gt+,\gt-)} & \arry{l}{(0,1,1) \,,~ (1,0,1)}  &
  \mathbf{b}_1+ \frac12\mathbf{e}_2, \mathbf{b}_2+ \mathbf{e}_{46}, &   \mathbf{e}_{12}, \mathbf{e}_{134}, \mathbf{e}_{156}
\\ \hline 
\text{(2 - 13)} & (21,9) & \arry{l}{(0+,0-,0-) \,,~ (0-,0+,0-)} & \arry{l}{(1,1,0) \,,~ (\gt,\gt,0)} &
  \mathbf{b}_1, \mathbf{b}_2, & \mathbf{e}_{13}, \mathbf{e}_{24}, \mathbf{e}_{56} 
\\
\text{(2 - 14)} & (7,7) & \arry{l}{(0+,0-,0-) \,,~ (0-,0+,1-)} & \arry{l}{(1,1,0) \,,~ (\gt,\gt,0)}  & 
\mathbf{b}_1, \mathbf{b}_2+\mathbf{e}_5, &  \mathbf{e}_{13}, \mathbf{e}_{24}, \mathbf{e}_{56}
\\ \hline\hline  
\text{(3 - 1)} & (12,6) & \arry{l}{(0+,0-,0-) \,,~ (0-,0+,0-)} & 
  \arry{l}{(0,\gt,1),(\gt,1,0) \,,~ (1,0,\gt)} 
& 
  \mathbf{b}_1, \mathbf{b}_2, & \mathbf{e}_{45}, \mathbf{e}_{23}, \mathbf{e}_{16}   
\\ \hline 
\text{(3 - 3)} & (17,5) & \arry{l}{(0+,0-,0-) \,,~ (0-,0+,0-)} & \arry{l}{(1,1,0),(\gt,\gt,0) \,,~ (1,\gt,1)}  &
  \mathbf{b}_1, \mathbf{b}_2, & \mathbf{e}_{134}, \mathbf{e}_{124}, \mathbf{e}_{1456} 
\\ 
\text{(3 - 4)}  & (7,7) & \arry{l}{(0+,0-,0-) \,,~ (0-,0+,\gt-)} & \arry{l}{(1,1,0),(\gt,\gt,0) \,,~ (1,\gt,1)}  &
  \mathbf{b}_1, \mathbf{b}_2+ \mathbf{e}_6, & \mathbf{e}_{134}, \mathbf{e}_{124}, \mathbf{e}_{1456} 
\\ \hline 
\text{(3 - 5)} & (15,3) & \arry{l}{(0+,0-,0-) \,,~ (0-,0+,0-)} & \arry{l}{(0,1,1),(1,0,1) \,,~ (\gt,\gt,\gt)}  &
  \mathbf{b}_1, \mathbf{b}_2, & \mathbf{e}_{35}, \mathbf{e}_{15}, \mathbf{e}_{246} 
\\
\text{(3 - 6)} & (9,9) & \arry{l}{(0+,0-,0-) \,,~ (0-,0+,\gt-)} & \arry{l}{(0,1,1),(1,0,1) \,,~ (\gt,\gt,\gt)} &
  \mathbf{b}_1, \mathbf{b}_2+\mathbf{e}_{56}, &  \mathbf{e}_{35}, \mathbf{e}_{15}, \mathbf{e}_{246} 
  \\ \hline\hline  
\text{(4 - 1)} & (15,3) & \arry{l}{(0+,0-,0-) \,,~ (0-,0+,0-)} & \arry{l}{(0,\gt,1),(\gt,1,0) \,,~ (1,0,\gt)\,,~ (1,1,1)}  &
  \mathbf{b}_1, \mathbf{b}_2, & \mathbf{e}_{45}, \mathbf{e}_{23}, \mathbf{e}_{16}, \mathbf{e}_{135} 
\\\hline\hline  
}
\)
}
\end{center}
\caption{ \label{tb:FFFrealizationsDW}
Free fermionic realizations of all inequivalent $\Intr_2\times\Intr_2$ orbifold geometries \cite{DW} are suggested.
} 
\end{table}

\newpage 

\subsection{The NAHE set}
\label{sc:NAHEmodel} 

Maybe the most famous  free fermionic construction is the so-called NAHE set, which was first introduced in~\cite{Antoniadis1988,Antoniadis1988a,Faraggi1993}. This set has been the basis for many phenomenological explorations of free fermionic models. It reads: 
\begin{align}\label{eq:originalNAHEbasis}
  \Big\{ \mathbf{1},\mathbf{S},\mathbf{b}'_1,\mathbf{b}'_2,\mathbf{b}'_3 \Big\}~, 
\end{align}
The vectors $\mathbf{1}$ and $\mathbf{S}$ were defined in Table~\ref{tb:BasisVectors}; the vectors $\mathbf{b}'_s$ are given by: 
\begin{subequations}\label{eq:originalNAHE} 
\begin{align}
\mathbf{b}'_1&=\Big\{\psi^\mu,\chi^{1,2}, y^{3,\dots,6}\,|\,\overline{y}^{3,\dots,6},\overline{\psi}^{1,\dots,5},\overline{\eta}^{1}\Big\}~,
\\[1ex] 
\mathbf{b}'_2&=\Big\{\psi^\mu,\chi^{3,4}, y^{1,2},w^{5,6}\,|\,\overline{y}^{1,2},\overline{w}^{5,6},\overline{\psi}^{1,\dots,5},\overline{\eta}^{2}\Big\}~,
\\[1ex] 
\mathbf{b}'_3&=\Big\{\psi^\mu,\chi^{5,6}, w^{1,\dots,4}\,|\,\overline{w}^{1,\dots,4},\overline{\psi}^{1,\dots,5},\overline{\eta}^{3}\Big\}~. 
\end{align}
\end{subequations} 
These can be expanded as 
\equ{
\mathbf{b}'_1 = \mathbf{b}_1 + \mathbf{S} + \Bgx_1~, 
\quad 
\mathbf{b}'_2 = \mathbf{b}_2 + \mathbf{S} + \Bgx_1~, 
\quad 
\mathbf{b}'_3 = \mathbf{b}_1 + \mathbf{b}_2 + \mathbf{e}_{1\ldots6} + \mathbf{S} + \Bgx_1~,
}
in terms of the basis vectors given in Table~\ref{tb:BasisVectors}. 
In accordance with~\eqref{SusyConditions} the generalized GSO  projection phases are chosen such that
\equ{ \label{GSOphasesNAHE} 
C[^{\mathbf{B}_a}_{\mathbf{B}_b}] = 
\scalebox{.7}{$
\begin{blockarray}{cccccc}
{\mathbf{B}_a} \backslash {\mathbf{B}_b}\!\!\!\! &  \mathbf{1}&\mathbf{S}&\mathbf{b}'_1&\mathbf{b}'_2 & \mathbf{b}'_3\\ 
 \begin{block}{l(rrrrr)}
 \mathbf{1}	& 1&  -1& -1 & -1 &-1   \\
\mathbf{S} 	&  \color{gray}-1& -1 & -1 & -1 &-1  \\
\mathbf{b}'_1 	&  \color{gray}-1 &  \color{gray}-1  & -1 & -1 & -1 \\
\mathbf{b}'_2 	&  \color{gray}-1 &  \color{gray}-1  &  \color{gray}-1 & -1 & -1 \\
\mathbf{b}'_3 	&  \color{gray}-1 &  \color{gray}-1  &  \color{gray}-1 &  \color{gray}-1 & -1\\
 \end{block}
\end{blockarray}
$}~.
}
With these input parameters, the gauge group is 
SO(10)$\times$SO(6)$^3\times$E$_8$: 
In particular, the SO(16) gauge fields correspond to the states
\(
\overline{\phi}^{A} \overline{\phi}^{B}\, 
\ket{0}_L^{\mathrm{NS}}\otimes\psi^\mu\ket{0}_R^{\mathrm{NS}}\, 
\).
Additional gauge bosons arise in the
\(
\Bgx_2 = 
\mathbf{1}+\mathbf{b}'_1+\mathbf{b}'_2+\mathbf{b}'_3
\) 
sector; transforming in the $\mathbf{128}$ representation of SO(16). This enhances the gauge group to E$_8$. 
The charged matter consists of 48 generations of $\mathbf{16}$-plets of $SO(10)$; 16 originating in each of the $\mathbf{b}'_{i}$.

Since 
\begin{gather}
  \mathbf{b}'_1\cap\mathbf{S} =\Big\{ \psi^\mu,\chi^{1,2} \Big\}~, 
  \qquad  
  \mathbf{b}'_2\cap\mathbf{S}= \Big\{ \psi^\mu,\chi^{3,4} \Big\}~,
\end{gather}
the $\mathcal{N}=4$ spacetime SUSY generated by $\mathbf{S}$ is indeed reduced to $\mathcal{N}=1$.  The phases $C[_{\mathbf{b}'_s}^{\mathbf{S}}]= -1$ are chosen such that the remaining gravitino is not projected out.

We begin the translation of this NAHE model to the orbifold language by taking linear combinations of the basis vectors, so that it is clear which basis vectors are Narain-like and which impose the $\Intr_2$ orbifold actions. We can identify two Narain-like vectors via 
\equ{  
\Bgb = \mathbf{b}'_1 + \mathbf{b}'_2 + \mathbf{b}'_3 - \mathbf{S} = \mathbf{e}_{1\ldots6} + \Bgx_1~, 
\qquad 
\Bgx_2 = \mathbf{1} -  \mathbf{b}'_1 - \mathbf{b}'_2 - \mathbf{b}'_3~. 
} 
In addition, we define the twist-like elements 
\begin{subequations}\label{eq:originalNAHEmod1} 
\begin{align}
\mathbf{B}_1&=\mathbf{S}+\mathbf{b}'_1 = 
\big\{\chi^{3,4,5,6}, y^{3,\dots,6}\,|\,\overline{y}^{3,\dots,6},\overline{\psi}^{1,\dots,5},\overline{\eta}^{1}\big\}~,
\\[1ex] 
\mathbf{B}_2&=\mathbf{S}+\mathbf{b}'_2=
\big\{\chi^{1,2,5,6}, y^{1,2},w^{5,6}\,|\,\overline{y}^{1,2},\overline{w}^{5,6},\overline{\psi}^{1,\dots,5},\overline{\eta}^{2}\big\}~, 
\end{align}
\end{subequations} 
which are associated with the twists $\gth_1$ and $\gth_2$, respectively. Since they do not involve pairs of $y$'s and $w$'s and they do not overlap, they can be thought of as pure twist elements with the shift gauge embeddings:
\equ{
 V_1 = \sfrac 12\, (1^5, 1,2,0)(0^8)~, 
\qquad 
 V_2 = \sfrac 12\, (1^5, 0,1,2)(0^8)~, 
}
where we have taken into account that $\mathbf{B}_s$ do not fulfill~\eqref{eq:basisVecs}. We arrive at this form by flipping signs and adding lattice vectors. Notice that these elements are related to the standard embedding choices~\eqref{StandardEmbedding}  as $V_s = V_s^\text{SE} + \sfrac12\, (1^8)(0^8)$. 

The separation of the twists in two bunches of eight entries is possible because we have the element $\Bgx_2$ which distinguishes the second eight entries from the first eight. Notice that in this case, the gauge shifts are not in the standard embedding, hence, the number of SO(10) generations does not necessarily correspond to the Hodge numbers.

In the new basis, $\big\{ \mathbf{S}, \mathbf{B}_1,\mathbf{B}_2, \Bgb, \Bgx_2\big\}$, the generalized GSO matrix~\eqref{GSOphasesNAHE} takes the form 
\equ{ 
C[^{\mathbf{B}_a}_{\mathbf{B}_b}] = 
\scalebox{.7}{$
\begin{blockarray}{cccccc}
{\mathbf{B}_a} \backslash {\mathbf{B}_b}\!\!\!\! &  \mathbf{S}&\mathbf{B}_1&\mathbf{B}_2 & \Bgb & \Bgx_2 \\ 
\begin{block}{l(rrrrr)}
\mathbf{S} 	&  -1 & -1 & -1 & -1 &-1  \\
\mathbf{B}_1 	&  \color{gray}1  &  -1  & -1 & -1 & 1 \\
\mathbf{B}_2 	&  \color{gray}1 &  \color{gray}-1  & -1 & -1 & 1 \\
\Bgb 			&  \color{gray}1 &  \color{gray}-1  &  \color{gray}-1 &  -1 & 1 \\
\Bgx_2		& \color{gray}1 &  \color{gray}-1 & \color{gray}-1 & \color{gray}-1 & 1   \\
 \end{block}
\end{blockarray}
$}
}
from which all the orbifold phases can be read using~\eqref{eq:IdentifiedPhasesOrbi->FFF}.

The final step is to identify the Narain moduli, which are given in the third row of Table \ref{tb:NarainModels}. Note that even though the vectors $\Bgx_1$ and $\Bgx_2$ do not both appear, we can still consider the E$_8\times$E$_8$ model as the starting point of the construction because of the appearance of $\ga^{\;}_{8\times 8}$. The particular values of the rest of the moduli then place this model at a point of enhanced symmetry in the moduli space, where the lattice between the 6d and the gauge degrees of freedom is not fully factorized anymore.

\subsection{Semi-realistic free fermionic classification of $\boldsymbol{\Intr_2\times\Intr_2}$ fermionic models}

In \cite{Faraggi:2004rq} a class of free fermionic models is considered. The twelve defining basis vectors are 
\equ{
\Big\{ 
\mathbf{1}, \mathbf{S}, \mathbf{e}_1, \ldots, \mathbf{e}_6, \mathbf{B}_1, \mathbf{B}_2, \mathbf{z}_1, \mathbf{z}_2
\Big\}
}
where the first eight were defined in Table~\ref{tb:BasisVectors}; the remaining read 
\equ{
\mathbf{B}_1 = \mathbf{b}_1 + \Bgx_1~, 
\quad 
\mathbf{B}_{2} = \mathbf{b}_{2} + \mathbf{e}_{56} + \Bgx_1~, 
\qquad
\mathbf{z}_2 = \Bgx_2 - \mathbf{z}~, 
\quad 
\mathbf{z}_1 = \mathbf{z} = \big\{\overline{\gf}^{1\ldots4}\big\}~. 
}
This set spans the same additive set as our standard choice 
\equ{
\Big\{
\mathbf{S}, \mathbf{b}_1, \mathbf{b}_2, \mathbf{e}_1,\ldots,\mathbf{e}_6, \Bgx_1, \Bgx_2, \mathbf{z}
\Big\}~. 
}
Since we have all the elements $\mathbf{e}_i$ separately, we know that we have moved away from the special free fermionic point with enhanced gauge symmetry. Since we have the basis vectors $\mathbf{b}_s$, these models can be interpreted as the (0-1) orbifold for the standard choice of phases, like in~\eqref{PanosChoice}. 

The new ingredient in this model is the basis vector $\mathbf{z}$. Note that it can be combined with any of the $ \mathbf{e}_i$'s of the model to be interpreted as a Wilson line. Its effect does indeed reduce the gauge symmetry of the model.

\subsection{Free fermonic MSSM-like constructions}

In this subsection, we consider some more complicated free fermionic models that were constructed in the past, and have a rich phenomenology.

\subsubsection*{An MSSM model with a symmetric orbifold interpretation}

One of the earliest MSSM-like constructions in string theory was the model constructed in~
\cite{Faraggi:1989ka} (closely related MSSM-like models were constructed in~\cite{Faraggi1992a}). This free fermionic model is an extension of the NAHE model discussed in Subsection~\ref{sc:NAHEmodel} with three additional basis elements: 
\begin{subequations}
\equa{ 
\mathbf{b}'_4 &= \big\{ \psi^\mu, \chi^{12}, y^{36},w^{45} \,|\, \byy^{36}, \bw^{45}; \bgps^{1\ldots 5}, \bget^1 \big\}~, 
\\[1ex]  
\Bga &= \big\{ \psi^\mu, \chi^{56}, y^2, w^{134} \,|\, \byy^{1236}, \bw^{46}; \bgps^{123}, \bget^{12}, \bgf^{1\ldots4} \big\}~, 
\\[1ex] 
\Bgb &= \big\{ \psi^\mu, \chi^{34}, y^{15}, w^{26} \,|\, \byy^{15},\bw^{26}; \sfrac12 \bgps^{1\ldots5},\sfrac 12\bget^{123}, \bgf^{34}, \sfrac 12\bgf^{1567} \big\}~. 
}
\end{subequations}
We notice that these three elements can be modified to $\mathbf{e}_{45} = \mathbf{b}'_4 - \mathbf{b}'_1$ and 
\begin{subequations} 
\equa{ 
\Bga' &= \Bga + \mathbf{b}'_3 = 
\big\{ 
y^2w^2 \,|\, \byy^{1236},\bw^{1236}; \bgps^{45}, \bget^{123}, \bgf^{1\ldots4}
\big\}~, 
\\[1ex] 
\Bgb' &= \Bgb + \mathbf{b}'_2 = 
\big\{
y^{25}, w^{25} \,|\, \byy^{25},\bw^{25}; \sfrac 32 \bgps^{1\ldots5}, \sfrac 12 \bget^{13}\, \sfrac 32 \bget^2,\bgf^{34}, \sfrac 12 \bgf^{1567} 
\big\}~, 
}
\end{subequations} 
which all are Narain-like elements. Hence we see that this model admits a symmetric orbifold interpretation, in the sense that the orbifold actions act symmetrically. On the other hand, we see that the basis vectors $\Bga' $ and $\Bgb'$ are asymmetric shifts, accompanied by Wilson lines. The machinery we have developed should also apply to such models. Nevertheless, even though we can use the basis vectors above to read off the generalized vielbein $E$, this is one of the cases discussed in Section \ref{sc:IdentifyNarainData} for which it is not straightforward to bring it to a basis in which it will have the form \eqref{eq:genV}.

\subsubsection*{A non-geometric MSSM model}

Another free fermionic MSSM-like realization was constructed in~\cite{Faraggi:1991jr}. This model also starts from the NAHE set and adds 
\begin{subequations}
\equa{ 
\Bga &= \big\{y^{36}, w^{36} \,|\, \byy^{1}, \bw^{23456}; \bgps^{123},  \bgf^{1\ldots4} \big\}~, 
\\[1ex] 
\Bgb &= \big\{y^{15}, w^{15} \,|\, \byy^{356},\bw^{124};  \bgps^{123},\sfrac 12\bget^{123}, \bgf^{1\ldots4} \big\}~, 
\\[1ex] 
\Bgg &= \big\{ y^{24},w^{24} \,|\, \byy^{12346}, \bw^{4}; \sfrac12 \bgps^{1\ldots 5}, \sfrac 12 \bget^{123}, \sfrac 12 \bgf^{1567}, \bgf^{34} \big\}~. 
}
\end{subequations}
All three elements are shift elements on the right-moving side: the fermions $y^i$ and $w^i$ appear in pairs. From the left-moving side these elements act as twists and roto-translations with twist parts that act in all six torus directions: All three elements either have only $\byy^i$ or only $\bw^i$ for each of the six directions. In fact, the differences $\Bgb - \Bga$ and $\Bgg - \Bga$ are ordinary Narain-like elements. They can be understood as modifying the Narain moduli of the underlying torus compactification. Hence, there is really only one element, say $\Bga$, that does not admit a symmetric orbifold interpretation; this model corresponds to an asymmetric orbifold and is therefore beyond the scope of this paper.

\subsection{The Blaszczyk model at the free fermionic point}

Our final example considers an interesting MSSM-like model construction on a $\Intr_2\times\Intr_2$ orbifold of the E$_8\times$E$_8$ string, the so-called Blaszczyk model~\cite{Blaszczyk:2009in}. This model was defined in two steps: 
\enums{
\item A six generation GUT model was constructed on the standard $\Intr_2\times\Intr_2$ orbifold with a specific choice of gauge shifts $V_s$ and  discrete Wilson lines $A_i$ in the six torus directions. 
\item By a freely acting $\Intr_2$ shift, in all three two-tori simultaneously, with an accompanying Wilson line $A$, the GUT group was broken to the SM group and the number of generations halved. 
} 

\subsubsection*{Upstairs model matching}

In detail, the upstairs model was defined by the gauge shifts 
\begin{subequations} 
\equa{ 
V_1 &= \big(\sfrac54,-\sfrac34,-\sfrac74, \sfrac14, \sfrac14,-\sfrac34,-\sfrac34, \sfrac14\big)\big(0,1,1,0,1,0,0,-1\big)~, 
\\[1ex] 
V_2 &= \big(-\sfrac12,-\sfrac12,-\sfrac12,\sfrac12, -\sfrac12,-\sfrac12,-\sfrac12,-\sfrac12\big)\big(\sfrac12,\sfrac12,0,0,0,0,0,4\big)~, 
}
\end{subequations} 
and the discrete Wilson lines 
\begin{subequations} 
\equa{
A_1 &= \big(0^8\big)\big(0^8\big)~, 
\\[1ex] 
A_{2k} &= \big(\sfrac54,\sfrac14,\sfrac34,-\sfrac14,-\sfrac14,\sfrac34,\sfrac34,\sfrac34\big)\big(\!-\!\sfrac14,\sfrac34,\sfrac54,\sfrac54,\sfrac14,\sfrac14,\sfrac14,\sfrac14\big)~, 
\\[1ex]
A_3 &= \big(\!-\!\sfrac34,-\sfrac14,\sfrac14,\sfrac74,-\sfrac14,-\sfrac14,-\sfrac14,-\sfrac14\big)\big(\sfrac14,\sfrac14,\sfrac14,\sfrac54,-\sfrac34,\sfrac14,-\sfrac34,\sfrac14\big)~,
\\[1ex]
A_5 &= \big(\!-\!\sfrac12,-\sfrac12,\sfrac12,-\sfrac12, \sfrac12,-\sfrac12,\sfrac12,\sfrac12\big)\big(\sfrac12,\sfrac12,0,0,0,0,-\sfrac12,-\sfrac12\big)~, 
}
\end{subequations} 
with $k=1,2,3$. 

To translate this model into the free fermionic language, we begin by observing that it is an orbifold of the E$_8\times$E$_8$ theory on the standard orthogonal lattice, hence the free fermionic analogue has to have the basis vectors: 
\(
\{ \mathbf{S},\mathbf{e}_1, \ldots, \mathbf{e}_6, \Bgx_1,\Bgx_2 \}\,.
\)
Since the $\Intr_2\times\Intr_2$ orbifold actions do not involve any roto-translations, we augment the standard pure twist basis vectors $\mathbf{b}_1$ and $\mathbf{b}_2$ of Table~\ref{tb:BasisVectors} with $2V_1$ and $2V_2$: 
\begin{subequations} 
\equa{
\mathbf{\widetilde{b}}_1 &= \big\{\chi^{34},-\chi^{56}; y^{34}, y^{56} \,|\, \byy^{34},\byy^{56}\big\} \big( 2V_1 \big)~, 
\\[1ex] 
\mathbf{\widetilde{b}}_2 &= \big\{-\chi^{12},\chi^{56}; y^{12}, w^{56} \,|\, \byy^{12},\bw^{56}\big\} \big( 2V_2 \big)~,
\\[1ex] 
\Bgb_i &= \big\{ \sfrac 12 y^i, \sfrac 12 w^i \,|\, \sfrac 12 \byy^i, \sfrac 12 \bw^i \big\} \big( 2 A_i \big)~, 
\quad i=1,\ldots, 6~. 
}
\end{subequations} 
Note that we have included some minus signs in front of some of the $\chi^i$ to ensure that we satisfy the conditions~\eqref{eq:ABnorms}, as they then precisely correspond to the orbifold consistency conditions~\eqref{ModInvConditions}. 

There are no discrete torsion phases turned on in the orbifold description of this model, so we can make the standard choice \eqref{PanosChoice} for the resulting free fermionic model. 
The only subtlety here is that in the free fermionic language not all of the above basis vectors are independent (mod 2), because $2\mathbf{\widetilde{b}}_1=\Bgx_1$. This is easily rectified by removing $\Bgx_1$ from the set of basis vectors, to get a minimal set.

\subsubsection*{Downstairs model matching}

The downstairs model is obtained by modding out a freely acting Wilson line, which acts in the $\sfrac12(e_2+e_4+e_6)$ direction with 
\equ{ 
A = \sfrac12\,\big(A_2+A_4+A_6\big)~. 
} 
Before the freely acting shift, the model lives on the (0~-~1) geometry; after the freely acting element is applied, the underlying geometry is (1~-~1). Similarly, in the free fermionic language we have to include the element
\equ{ 
\Bgb = \sfrac12\,\big(\Bgb_2+\Bgb_4+\Bgb_6\big)~, 
} 
and then select an appropriate minimal set of independent vectors.

\section{Conclusion}
\label{sc:Conclusion}

\subsection*{Summary}

In this paper we developed a detailed dictionary between the free fermionic models and symmetric $\Intr_2\times\Intr_2$ orbifold models. To this end, we first gave a detailed summary of the heterotic string constructions in both formulations:  

A free fermionic model is fully specified by a set of basis vectors and a choice of generalized GSO phases. 
An orbifold model is characterized by a torus lattice on which an orbifold acts. 
The orbifold action may be simple twists or composite roto-translations. In addition, the orbifold elements may act on the gauge degrees of freedom of the E$_8\times$E$_8$ or Spin(32)$/\Intr_2$ theories; we assumed that this action can always be described by gauge shift vectors. 
Moreover, the translations that define this lattice can have accompanying actions in the form of discrete Wilson lines. 
The geometry, background B-field and the Wilson lines can be conveniently combined in the Narain description of heterotic toroidal orbifolds. 
Finally, it is possible to switch on generalized discrete torsion phases in heterotic orbifold constructions. 

To translate the input data for a free fermionic model to the orbifold language, we first determine linear combinations of the basis vectors to facilitate their interpretations. We distinguished between the following types of basis vectors: the target space supersymmetry generator; the twist-like elements; and the Narain-like elements. 
If there are Narain-like elements that do not act on the geometry, then we can often decide whether the free fermionic theory can be most naturally thought of as an orbifold of the E$_8\times$E$_8$ or of the Spin(32)$/\Intr_2$ theory. 
In any case, we can extract from the Narain-like basis elements the corresponding Narain torus compactification up to discrete O(6,22;$\Intr$) T-duality transformations. 
We gave a criterion to decide whether the twist-like elements are associated with twist and/or roto-translations and described how to read off the gauge shift vectors. 
Finally, we derived formulae that associate most generalized GSO phases with  the symmetric and asymmetric generalized discrete torsion phases. 
For certain generalized GSO phases this was not possible, because in the orbifold literature, standard choices for the corresponding phases are always assumed.

\subsection*{Outlook}

The main application of our detailed dictionary between orbifold and free fermionic models is that it allows the study of the same theory in different regimes of its moduli space. For example, this dictionary allows us to study $\Intr_2\times\Intr_2$ orbifold models at special points of enhanced symmetry with radii at the order of the string scale and investigate consequences of special choices for the B-field. 

The connection between singular orbifolds and the models that result on their resolutions has been studied in great detail in the past~\cite{Denef:2005mm,Reffert:2006du,Nibbelink:2007rd,Nibbelink:2009sp,Blaszczyk:2010db,Blaszczyk:2011hs}. By establishing a dictionary between free fermionic and $\Intr_2\times\Intr_2$ orbifolds, one may also study what happens if vacuum expectation values of various scalars in free fermionic models are switched on that deform the theory away from the free fermionic point, as well as the orbifold locus. 

Moreover, the phenomenological studies of string constructions do not stop at the construction of interesting string vacua, they begin there: For example, one can look for stringy doublet--triplet splitting mechanisms~\cite{Faraggi1994b,Faraggi2001a} and study Yukawa coupling selection rules. In particular, for the latter, there has been quite some controversy in recent literature~\cite{Kobayashi:2011cw,Nilles:2013lda,Bizet:2013wha}, thus, perhaps studying selection rules in the free fermionic context~\cite{Faraggi:1991mu} could help settle this debate. 

In this paper, we have focused on finding a dictionary between free fermionic models and bosonic constructions that can be interpreted as symmetric orbifolds. Recently there has been a revived interest in asymmetric orbifolds~\cite{Beye:2013moa,Beye:2013ola} and their connection to non-geometry and non-geometric fluxes~\cite{Condeescu:2012sp,Condeescu:2013yma}. It would, therefore, be very interesting to extend our dictionary to include asymmetric cases in both the fermionic and bosonic language. Moreover, as we have seen in this work it is very natural and useful to formulate the underlying lattice properties of symmetric and asymmetric orbifolds using the Narain formulation. Thus, it would be very useful to have a formulation of (a)symmetric orbifolds as Narain orbifolds. Work in this direction is underway in~\cite{GN_V}.

Moreover, in light of the absence of a signal for supersymmetry, there has been some recent effort in constructing non-supersymmetric models directly in string theory. These investigations have been done both in the context of orbifold theories~\cite{Blaszczyk:2014qoa,Nibbelink:2015ena} 
and free fermionic models~\cite{Abel:2015oxa,Ashfaque:2015vta}. 
Furthermore, there have been some interesting studies of some of their properties, like threshold corrections~\cite{Angelantonj:2014dia,Faraggi:2014eoa,Angelantonj:2015nfa}. 
It would certainly be useful if the dictionary that is presented here for supersymmetric string vacua can also be extended to incorporate non-supersymmetric constructions. 
Given that in both formulations some generalized GSO or torsion phases need to be chosen very carefully in order to preserve supersymmetry, it is very likely that also a dictionary between various non-supersymmetric orbifold and free fermionic constructions can be established.

\subsection*{Acknowledgements}

We are indebted to Orestis Loukas and Patrick K.S.\ Vaudrevange for many helpful discussions to test many of our ideas on symmetric orbifolds and the use of the {\tt orbifolder} package. 
We have also benefited from communications with Maximilian Fischer about the classification of orbifold geometries and the use of {\tt CARAT} program. 
In addition, we would like to thank 
Ioannis Florakis
and
Fabian Ruehle 
for very valuable discussions. 
AEF is supported, in part, by the STFC (ST/L000431/1) and VMM by the Transregio TR33 ``The Dark Universe'' project.

\bibliographystyle{paper}
{\small
\bibliography{paper}
}
\end{document}